\documentclass{aa}  

\usepackage{graphicx}
\usepackage{txfonts}
\usepackage{hyperref}
\usepackage{booktabs}
\usepackage{bm}
\usepackage[normalem]{ulem}
\usepackage[dvipsnames]{xcolor}
\usepackage{nicefrac}
\usepackage{mathtools}

\usepackage{hyperref}
\hypersetup{
	colorlinks,
	citecolor=blue,
	linkcolor=blue,
	urlcolor=blue}

\usepackage{amsmath}
\usepackage{yhmath}

\DeclarePairedDelimiter\floor{\lfloor}{\rfloor}

\newcommand{\mach}[0]{\mathcal{M}}
\newcommand{\mrm}[1]{\mathrm{#1}}
\newcommand{\LR}[0]{{\mathrm{L},\mathrm{R}}}

\newcommand{\code}[1]{\texttt{#1}}

\begin{document}

	\title{Performance of high-order Godunov-type methods in simulations of astrophysical low Mach number flows}
	\titlerunning{Performance of high-order Godunov methods}
	
	\author{
		G.~Leidi\inst{1}\and
		R.~Andrassy\inst{1}\and
		W.~Barsukow\inst{2}\and
		J.~Higl\inst{1,3}\and
		P.~V.~F.~Edelmann\inst{4}\and
		F.~K.~R{\"o}pke\inst{1,5}
	}
	\institute{
		Heidelberger Institut f{\"u}r Theoretische Studien,
		Schloss-Wolfsbrunnenweg 35, D-69118 Heidelberg, Germany\\
		\email{giovanni.leidi@h-its.org}
		\and
		Bordeaux Institute of Mathematics, Bordeaux University and CNRS/UMR5251, Talence, 33405 France
		\and
		High-Performance Computing Center Stuttgart,
		Nobelstraße 19, 70569 Stuttgart, Germany
		\and
		Computer, Computational and Statistical Sciences (CCS) Division and Center for Theoretical
		Astrophysics (CTA), Los Alamos National Laboratory, Los Alamos, PO Box 1663, NM
		87545, USA
		\and
		Zentrum f\"ur Astronomie der Universit\"at Heidelberg, Institut f\"ur
		Theoretische Astrophysik, Philosophenweg 12, D-69120 Heidelberg, Germany
	}
	
	\date{Received 8 December 2023 / Accepted 23 February 2024}
	
	
	\abstract
	{High-order Godunov methods for gas dynamics have become a standard tool for simulating different classes of astrophysical flows. Their accuracy is mostly determined by the spatial interpolant used to reconstruct the pair of Riemann states at cell interfaces and by the Riemann solver that computes the interface fluxes. In most Godunov-type methods, these two steps can be treated independently, so that many different schemes can in principle be built from the same numerical framework. Because astrophysical simulations often test out the limits of what is feasible with the computational resources available, it is essential to find the scheme that produces the numerical solution with the desired accuracy at the lowest computational cost.  However, establishing the best combination of numerical options in a Godunov-type method to be used for simulating a complex hydrodynamic problem is a nontrivial task. In fact, formally more accurate schemes do not always outperform simpler and more diffusive methods, especially if sharp gradients are present in the flow. In this work, we use our fully compressible \textsc{Seven-League Hydro} (\code{SLH}) code to test the accuracy of six  reconstruction methods and three approximate Riemann solvers on two- and three-dimensional (2D and 3D) problems involving subsonic flows only. We consider Mach numbers in the range from $10^{-3}$ to $10^{-1}$, which are characteristic of many stellar and geophysical flows. In particular, we consider a well-posed, 2D, Kelvin--Helmholtz instability problem and a 3D turbulent convection zone that excites internal gravity waves in an overlying stable layer. Although the different combinations of numerical methods converge to the same solution with increasing grid resolution for most of the quantities analyzed here, we find that (i) there is a spread of almost four orders of magnitude in computational cost per fixed accuracy between the methods tested in this study, with the most performant method being a combination of a ``low-dissipation'' Riemann solver and a sextic reconstruction scheme, (ii) the low-dissipation solver always outperforms conventional Riemann solvers on a fixed grid when the reconstruction scheme is kept the same, (iii) in simulations of  turbulent flows, increasing the order of spatial reconstruction reduces the characteristic dissipation length scale achieved on a given grid even if the overall scheme is only second order accurate, (iv) reconstruction methods based on slope-limiting techniques tend to generate artificial, high-frequency acoustic waves during the evolution of the flow, (v) unlimited reconstruction methods introduce oscillations in the thermal stratification near the convective boundary, where the entropy gradient is steep.}

	
	\maketitle
	%
	
	\section{Introduction}
	\label{sec:introduction}

	High-resolution schemes for gas dynamics \citep[see, e.g.,][]{vanleer1979,colella1984a,harten1987,colella1990,liu1994,jiang1996,colella2008a,toro2009a,balsara2017} are routinely used for modeling a broad variety of astrophysical flow phenomena. Their popularity derives from their conservation properties and robustness, which allow them to accurately capture both smooth and discontinuous solutions on the same computational grid without sacrificing numerical stability. 
	
	These schemes are based on higher-order extensions of the original first-order accurate method of \cite{godunov1959} and their time-integration algorithm is typically carried out in three steps. First, a pair of Riemann states is reconstructed at each grid cell interface by applying high-order monotonic interpolants to a set of cell-averaged hydrodynamic quantities. Second, the resulting Riemann problems are solved (either exactly or approximately) to obtain fluxes across every cell boundary. Finally, the cell surface integral of the fluxes is evaluated, allowing the cell-volume-averaged state quantities to be advanced in time\footnote{High-resolution schemes for gas dynamics can be fully discrete, where the system of equations is discretized both in space and time, or semi-discrete, where spatial discretization is performed first while leaving the problem continuous in time. In the latter approach, state quantities are then advanced in time using any standard numerical solver for systems of ordinary differential equations.}. 
	
	In most high-order Godunov schemes, the solution strategy of the Riemann problem is independent of  the spatial interpolant used for reconstructing the Riemann states. Therefore, many different schemes can be built from the same numerical framework. The choices made in the construction of a particular scheme, however, do have a strong effect on its accuracy, that is the difference between the numerical ($U$) and the true ($u$) solution, 
	\begin{equation}\label{eq:accuracy}
		\| U-u \| = \mathcal{O}({\Delta x}^m) + \mathcal{O}({\Delta t}^n),
	\end{equation}
	computed in some norm $\| \cdot \|$ \citep[see, e.g.,][]{leveque2002}. Here, $\Delta x$ is the width of the grid cell and $\Delta t$ is the time step. Although the formal order of the spatial and temporal accuracy of a Godunov-like scheme, ``$m$'' and ``$n$'' in Eq.~(\ref{eq:accuracy}), can be derived for smooth flows, they do not give any information about the magnitude of the numerical errors generated on a given grid, which is problem-dependent. The convergence rates can also be significantly lower than the formal order of accuracy of the scheme for problems that admit non-smooth solutions. Consequently, formally higher-than-second-order interpolants do not always outperform simpler linear spatial reconstruction schemes when large gradients or discontinuities are present in the flow \citep{greenough2003}. Moreover, if the flow is stochastic or chaotic, like in the case of turbulence, convergence may not be achieved in the sense of Eq.~(\ref{eq:accuracy}), but rather the quality of the numerical results can only be judged in terms of global or ensemble-averaged quantities that characterize the flow and its evolution. 
	
	Given these considerations, it is impossible to generalize the convergence properties of a certain combination of numerical methods in a Godunov-type scheme, so they have to be explored by running numerical tests. Such tests, to be significant, have to be challenging enough and close to the actual application case. The performance of the numerical scheme is another crucial aspect to be considered alongside its accuracy, especially in astrophysical simulations, which often test out the limits of what is feasible nowadays with available computational resources. Therefore, the question arises of what combination of different ingredients in a Godunov-type scheme should be used to produce the desired solution at a minimal computational cost. 
	
	Several comparison studies have been presented in the literature with the aim of shedding light on the behavior of different high-resolution schemes in simulations of complex hydrodynamic phenomena, such as forced turbulence \citep{klingenberg2007, kritsuk2011,san2015,radice2015,seo2023}, convection \citep{muller2020}, jet evolution \citep{beckwith2011,musoke2020}, magneto-rotational instabilities in accretion disks \citep{flock2010}, Richtmyer-Meshkov instabilities \citep{latini2007}, and shear instabilities \citep{mcnally2012,lecoanet2016a}. These numerical experiments focused on supersonic or mildly subsonic flow regimes, for which Godunov-type methods are highly optimized \citep[see, e.g.,][]{leveque2002,toro2009a}. Nonetheless, high-resolution schemes have been proven to be a powerful tool also for modeling regimes of low Mach numbers ($\mathcal{M}\coloneqq|\bm{V}|/c\,{\lesssim}\,0.1$, where $\bm{V}$ is the fluid velocity and $c$ is the sound speed), especially in simulations of terrestrial \citep[see, e.g.,][]{day2000,klein2009,dumbser2009,motheau2018} and stellar \citep[see, e.g.,][]{meakin2007,muthsam2010,woodward2015,goffrey2017a,muller2020,horst2021a,cuissa2022} flows. To our knowledge, other than idealized tests, no extensive work along the line of the aforementioned comparison studies has been presented for low-Mach-number flows yet. Only a few studies evaluated the impact of the order of the spatial and temporal discretization in the numerical scheme on the properties of highly subsonic turbulent flows, but they kept the Riemann solver fixed \citep{wongwathanarat2016,teissier2023}.
	
	In this work, we use our fully compressible \textsc{Seven-League Hydro} (\code{SLH}) code to test 18 different combinations of spatial reconstruction schemes and Riemann solvers on two test problems in which flows are highly subsonic ($10^{-3}\,{\lesssim}\,\mathcal{M}\,{\lesssim}\,10^{-1}$). In particular, we consider a two-dimensional (2D) Kelvin--Helmholtz instability with smooth initial conditions and a 3D, turbulent convection zone that entrains material from an upper, stably stratified layer, where internal waves are free to propagate. The initial conditions of the latter test are adopted from the work of \cite{andrassy2022}. Here, we opt to reduce the strength of the heat source driving the convection in order to achieve lower convective speeds than those obtained by \cite{andrassy2022}. Also, contrary to that work, we provide performance measurements for all the methods tested in our study. In both tests, we conduct a resolution study to analyze the convergence of the numerical results obtained by each method.

	The paper is structured as follows: in Sect.~\ref{sec:methods}, we provide a detailed description of the equations solved and the numerical methods included in this study. In Sect.~\ref{sec:convergence-properties}, we measure the convergence properties of different Godunov-type schemes for the  2D, Kelvin--Helmholtz instability test problem (see Sect.~\ref{sec:results_kh}) and for the 3D setup involving turbulent convection, convective boundary mixing, and wave excitation (see Sect.~\ref{sec:code-comparison}). In Sect.~\ref{sec:performance}, we use the kinetic energy spectrum of the convective flows simulated in the latter test, which is close to a real astrophysical application, to provide measurements of the computational cost per fixed accuracy for each method. Finally, in Sect.~\ref{sec:conclusions}, we summarize the main results and we give some guidance on which methods to use for specific applications.

	\section{Methods}
	\label{sec:methods}
	
	\subsection{Governing equations}

	We solve the fully compressible, inviscid Euler equations with a source term $\bm{S}$ in the integral form 
	\begin{equation}\label{eq:euler-int}
		\frac{1}{|\Omega|}\frac{\partial}{\partial t} \Bigg( \int_\Omega \bm{U}\ \mathrm{d}\Omega \Bigg) + \frac{1}{|\Omega|}\oint_{\partial \Omega}\mathbb{T}\cdot\bm{n}\ \mathrm{d}A =  \frac{1}{|\Omega|}\int_\Omega \bm{S}\ \mathrm{d}\Omega,
	\end{equation}
	where
	\begin{align}
		\bm{U} =  \begin{bmatrix}
			\rho \\
			\rho u\\
			\rho v\\
			\rho w\\
			\rho e_\mathrm{tot} \\
			\rho X\\
		\end{bmatrix}
	\end{align}
	is the set of conserved quantities, $|\Omega|$ is the volume of a fluid element $\Omega$ enclosed by a surface $\partial \Omega$, $\bm{n}$ is the outward normal vector to the surface, and $\mathbb{T}=[\bm{F}|\bm{G}|\bm{H}]$ is a tensor defined by the flux vectors
	\begin{align}\label{eq:fluxes}
		\bm{F} = \begin{bmatrix}
			\rho u \\
			\rho u^2+p \\
			\rho u v \\
			\rho u w \\
			( \rho e_\mathrm{tot} + p )u \\
			\rho X u
		\end{bmatrix},
		\ \bm{G} = 
		\begin{bmatrix}
			\rho v \\
			\rho u v \\
			\rho v^2 + p \\
			\rho v w \\
			( \rho e_\mathrm{tot} + p )v \\
			\rho X v
		\end{bmatrix},
		\ \bm{H} = 
		\begin{bmatrix}
			\rho w \\
			\rho u w \\
			\rho v w \\
			\rho w^2 + p \\
			( \rho e_\mathrm{tot} + p )w \\
			\rho X w
		\end{bmatrix}.
	\end{align}
	Here, $\rho$ denotes the mass density, $\bm{V}=( u,  v,  w)$ the velocity field, $ e_\mathrm{tot}= e_\mathrm{int} + \frac{1}{2}|\bm{V}|^2$ the total energy per unit mass, $e_\mathrm{int}$ the specific internal energy,  and $X$ the mass fraction of a passive scalar used as a tracer advected with the fluid. The system represented by Eq.~(\ref{eq:euler-int}) is closed by an equation of state (EoS), which gives the gas pressure as a function of the density and internal energy, 
	\begin{equation}
		p=p(\rho,e_\mathrm{int}).
	\end{equation}
	In this work, we only consider a perfect gas with a given adiabatic index $\gamma$, for which
	\begin{equation}
		p(\rho,e_\mathrm{int}) = (\gamma-1)\rho e_\mathrm{int}.
	\end{equation}
	
	For modeling the test problem introduced in Sect.~\ref{sec:code-comparison}, which involves the presence of a gravitational field, $\bm{g}\,{=}\,(g_x,g_y,g_z)$, the corresponding source term,
	\begin{align}
		\bm{S}_\mathrm{grav} =  \begin{bmatrix}
			0 \\
			\rho g_x\\
			\rho g_y\\
			\rho g_z\\
			\rho \bm{g}\cdot\bm{V} \\
			0\\
		\end{bmatrix},
	\end{align}
	must be included in the right-hand-side term of Eq.~(\ref{eq:euler-int}). Because we assume $\bm{g}$ to be time-independent, we opt to solve an equivalent form of Eq.~(\ref{eq:euler-int}), in which the gravitational potential $\phi$ is directly added to $e_\mathrm{tot}$, eliminating the $\rho \bm{g}\cdot\bm{V}$ source term from the energy equation,
	\begin{align}
		\quad e_\mathrm{tot} \mapsto e_\mathrm{tot} + \phi, \quad
		\bm{S}_\mathrm{grav} =  \begin{bmatrix}
			0 \\
			\rho g_x\\
			\rho g_y\\
			\rho g_z\\
			\rho \bm{g}\cdot\bm{V} \\
			0\\
		\end{bmatrix}\mapsto 
		\begin{bmatrix}
			0 \\
			\rho g_x\\
			\rho g_y\\
			\rho g_z\\
			0 \\
			0\\
		\end{bmatrix}.
	\end{align}
	Numerical schemes that solve this form of Eq.~(\ref{eq:euler-int}) are capable of conserving the total energy of the system over time exactly. Such a property is  particularly important in low-Mach-number hydrodynamics, where even small energy conservation errors can become comparable to the kinetic energy content of the flows \citep{muller2020,edelmann2021a}. 
	
	\subsection{Seven-League Hydro code}\label{sec:slh}
	In our study, Eq. (\ref{eq:euler-int}) is solved numerically using the \textsc{Seven-League Hydro} code \citep[\code{SLH}, ][]{miczek2013a, edelmann2014a}, which was originally developed to model the broad variety of hydrodynamic processes that characterize the  deep interiors of stars, such as shear instabilities \citep{edelmann2017a}, excitation of internal waves \citep{horst2020a}, convective boundary mixing \citep{horst2021a, andrassy2022,andrassy2023}, and turbulent dynamos \citep{leidi2022,leidi2023}. \code{SLH} makes use of the finite-volume discretization on an arbitrarily curvilinear, but logically rectangular, Eulerian grid to retain the conservation properties of the fluid-dynamics equations. Numerical solutions to Eq.~(\ref{eq:euler-int}) are computed by means of Godunov-type methods based on the definition of Riemann problems at cell interfaces. The code is parallelized using the Message Passing Interface (MPI) and it has been proven to scale up to several hundred thousand processes \citep{edelmann2016b}. 
	
	\code{SLH} allows the user to choose among many different numerical options at compile time, which makes this code perfectly suited to run the comparison study introduced in Sect.~\ref{sec:introduction}. In particular, in addition to the well-known Rusanov \citep{rusanov1962}, Roe \citep{roe1981}, and Harten-Lax-van Leer-Contact \citep[HLLC,][]{Toro1994} approximate Riemann solvers, \code{SLH} adopts special low-Mach-number methods \citep{Liou2006, miczek2015, minoshima2021} to reduce the excessive numerical dissipation introduced by shock-capturing schemes at low Mach numbers (see Sect.~\ref{sec:numerical-fluxes}). A wide spectrum of spatial reconstruction methods can be used to generate a pair or Riemann states at each grid cell interface, ranging from  (first-order accurate) constant reconstruction to very high order methods, some of which are described in Sect.~\ref{sec:spatial-reconstruction}. In problems involving the presence of a gravitational field, the deviation well-balancing method \citep{berberich2019, edelmann2021a} is used to preserve hydrostatic solutions and to reduce the strength of spurious flows generated by grid discretization errors in strongly stratified media. 
	
	In \code{SLH}, the cell-volume-averaged source term and the cell-surface-averaged fluxes are approximated using the midpoint method, making the code at best second-order accurate in space. On a 3D, evenly spaced Cartesian grid, the final expressions for these integrals read
	\begin{align}
		\label{eq:s}
		\frac{1}{\Delta V}\int_{\Omega_{i,j,k}} \bm{S}\ \mathrm{d} x \mathrm{d} y \mathrm{d} z = & \  \bm{\mathcal{S}}_{i,j,k} 
		+ \mathcal{O}(\Delta x^2), \\
		\label{eq:fx}
		\frac{1}{\Delta A} \int_{\Delta A_{i+1/2,j,k}} \bm{F}\ \mathrm{d}y \mathrm{d}z = & \  \bm{\mathcal{F}}_{i+1/2,j,k} + \mathcal{O}(\Delta x^2), \\
		\label{eq:fy}
		\frac{1}{\Delta A} \int_{\Delta A_{i,j+1/2,k}} \bm{G}\ \mathrm{d}x \mathrm{d}z = & \  \bm{\mathcal{G}}_{i,j+1/2,k} + \mathcal{O}(\Delta x^2), \\
		\label{eq:fz}
		\frac{1}{\Delta A} \int_{\Delta A_{i,j,k+1/2}} \bm{H}\ \mathrm{d}x \mathrm{d}y = & \  \bm{\mathcal{H}}_{i,j,k+1/2} + \mathcal{O}(\Delta x^2),
	\end{align}
	where $\bm{\mathcal{S}}_{i,j,k}$ is the point value of the source term $\bm{S}$ at the center of the cell represented by the set of indices $(i,j,k)$ and vector quantities such as $\bm{\mathcal{F}}_{i+1/2,j,k}$ refer to the face-centered value of the flux at the boundary between two adjacent cells, in this case $(i,j,k)$ and $(i+1,j,k)$. The volume of the cell and the surface of a cell face are $\Delta V=\Delta x^3$ and $\Delta A=\Delta x^2$, respectively. The discretized source term and fluxes in Eqs.~(\ref{eq:s})--(\ref{eq:fz}) are used to build a semi-discrete version of Eq.~(\ref{eq:euler-int}) which leaves the problem continuous in time, following the method of lines \citep[see, e.g.,][]{leveque2002},
	\begin{equation}\label{eq:semi-discrete}
		\begin{split}
			\frac{\partial \overline{\bm{U}}_{i,j,k}}{\partial t}  =  -&\frac{1}{\Delta x}\Big(\bm{\mathcal{F}}_{i+1/2,j,k}-\bm{\mathcal{F}}_{i-1/2,j,k}  \\
			+ & \bm{\mathcal{G}}_{i,j+1/2,k}-\bm{\mathcal{G}}_{i,j-1/2,k}  \\  
			+ & \bm{\mathcal{H}}_{i,j,k+1/2}-\bm{\mathcal{H}}_{i,j,k-1/2}\Big)  \\ 
			+ & \bm{\mathcal{S}}_{i,j,k}.
		\end{split}
	\end{equation}
	The time update on the cell-volume-averaged conserved variables, $\overline{\bm{U}}_{i,j,k}$, is then carried out in a dimensionally unsplit fashion using explicit or implicit time stepping. Here, we only consider a limited set of all of the numerical methods available in \code{SLH} to avoid constructing a too large parameter space. In the following sections, we provide a detailed description of the algorithms that we use for running the tests presented in Sect.~\ref{sec:results_kh} and \ref{sec:code-comparison}.

	\subsection{Spatial reconstruction methods}
	\label{sec:spatial-reconstruction}
	
	We use six reconstruction methods as summarized in Table~\ref{tab:rec_methods}. We do not attempt to be exhaustive in the choice of our methods, which is why, for example, essentially non-oscillatory (ENO) and weighted ENO (WENO) schemes \citep[see, e.g.,][]{liu1994,jiang1996,shu2009} are left out from our comparison study\footnote{The influence of the parameter that occurs in the smoothness indicators of the compact third order WENO scheme of \cite{kolb2014} on the generation of sound waves by turbulent flows is discussed in Appendix~\ref{sec:sound-weno}.}. We also exclude multidimensional reconstruction methods. However, the methods we do include cover a wide range in complexity and formal order of accuracy and several of them are often used in stellar hydrodynamics.
	
	\begin{table*}
		\caption{Overview of spatial reconstruction methods used in this work.}
		\label{tab:rec_methods}
		\centering
		\begin{tabular}{lllll}
			\toprule
			ID & Theoretical & Practical & Number of & Description \\
			& order & order & ghost cells & \\
			\midrule
			LIN & 2 & 2.0 & 2 & piecewise linear, no limiter \\
			LIN+VL & 2 & 2.0 & 2 & piecewise linear, van Leer limiter \\
			PAR & 3 & 3.0 & 2 & piecewise parabolic, no limiter \\
			PPM84 & 4 & 2.3, 1.6 & 3 & PPM method of \citet{colella1984a} \\
			PPM08 & 6 & 6.0 & 4 & PPM method of \citet{colella2008a} \\
			PSH & 7 & 7.0 & 4 & hybrid piecewise sextic with no limiter for dynamic \\
			& & & & variables and PPM08 for passive scalars \\
			\bottomrule
		\end{tabular}
		\vspace{0.5em}
		\tablefoot{ID: identifier; Theoretical order: formal order of accuracy for smooth and monotonic solutions in 1D; Practical order: asymptotic order obtained in the 1D experiments reported in Appendix~\ref{sec:1D_test_cases} (smooth but non-monotonic solutions). In the case of PPM84, the higher order of accuracy corresponds to linear advection and the lower one to the propagation of a sound wave using the LHLLC flux function, see Appendix~\ref{sec:1D_test_cases} for details.}
	\end{table*}
	
	The simplest are unlimited linear (LIN) and parabolic (PAR) methods, which have proved to be well-behaved in implicit simulations of slow flows using our \code{SLH} code \citep[e.g.][]{horst2020a, horst2021a, andrassy2023}. Their main disadvantage -- the generation of artificial oscillations around steep gradients -- can be eliminated using slope limiters. Out of a wide spectrum of limiters available, we have decided to include the popular van Leer limiter \citep{vanleer1974a} in combination with linear reconstruction (LIN+VL). This limiter has the total-variation-diminishing (TVD) property and eliminates the oscillations completely.\footnote{This formally holds for linear advection in one spatial dimension. In practice, we do not observe oscillations even when the LIN+VL method is applied to a multi-dimensional system of conservation laws, see Sect.~\ref{sec:results_kh}.} As examples of higher-order methods with limiters, we include two versions of the widely used piecewise-parabolic method (PPM): \citet[][CW84 hereinafter]{colella1984a} and \citet[][CS08 hereinafter]{colella2008a}, for which we introduce the acronyms PPM84 and PPM08, respectively. We also construct a hybrid method, which we name piecewise sextic hybrid (PSH). It combines unlimited sextic reconstruction for dynamic variables ($\rho$, $\bm{V}$, $p$) with the PPM08 method for passive scalars. We provide lower-order alternatives to the PSH method in Appendix~\ref{sec:PPH_PQH}, although we do not test them in this study.
	
	In the remainder of this section, we provide detailed descriptions of all the six reconstruction methods in unified notation. We do this to (i) maximise the reproducibility of our work, (ii) simplify the methods' more general original forms (e.g.\ for non-uniform grids), (iii) make clear what choices we make if several options are available, and (iv) to point out a few typographic mistakes in the original works.
	
	We refer to the variable being reconstructed as $a$ in Sections~\ref{sec:method_LIN}--\ref{sec:method_PSH}. Although reconstruction can be performed in physical coordinates, we chose to do it in logical coordinates defined by the cell index $i$, which leads to much simpler expressions. This choice does not influence our results in any way since we use uniform Cartesian grids. In some cases, we introduce the continuous linear logical coordinate $\zeta$ such that the coordinate of the centre of cell $i$ is $\zeta_i = i$ and the left and right cell interfaces are located at $\zeta_{i-1/2} = i - \frac{1}{2}$ and $\zeta_{i+1/2} = i + \frac{1}{2}$, respectively. The reconstruction is usually discontinuous at the interfaces. We use the notation $a_{i+1/2,\mrm{L}}$ and $a_{i+1/2,\mrm{R}}$ to denote the reconstructed states on the left and right side of the interface at $\zeta_{i+1/2}$, respectively. Reconstruction is performed along each spatial axis separately, i.e.\ the reconstruction procedure is always one-dimensional. To highlight the difference between cell averages and point values, we use the notation $\overline{a}_i$ for the average value of $a$ within cell $i$.

	In this work, we always reconstruct the set of cell-volume-averaged primitive variables $\overline{\bm{q}}$,
	where $\bm{q}=(\rho,\ \bm{V},\ p,\ X)$, because it helps reducing oscillations near discontinuities as compared to reconstructing cell-volume-averaged conserved quantities $\overline{\bm{U}}$. In SLH, transformations between $\overline{\bm{q}}$ and $\overline{\bm{U}}$ are performed using  2\textsuperscript{nd}-order approximations,
	\begin{equation}
		\begin{split}
			\overline{\bm{q}} & = \bm{\xi}\Big({\overline{\bm{U}}}\Big) + \mathcal{O}(\Delta x^2), \\
			\overline{\bm{U}} & = \bm{\xi}^{-1}\Big({\overline{\bm{q}}}\Big) + \mathcal{O}(\Delta x^2), \\
		\end{split}
	\end{equation}
	where $\bm{\xi}$ is a nonlinear, invertible transformation,
	\begin{equation}
		\bm{\xi}(\bm{U}) = 
		\begin{bmatrix}
			U_1 \\
			U_2/U_1 \\
			U_3/U_1 \\
			U_4/U_1 \\
			(\gamma-1)\Bigg(U_5 - \frac{1}{2U_1}\Big(U_2^2+U_3^2+U_4^2\Big)\Bigg) \\
			U_6/U_1
		\end{bmatrix},
	\end{equation}
	and $U_i$ is the $\mathrm{i}$-th component of $\bm{U}$.
	
	\subsubsection{The LIN method}
	\label{sec:method_LIN}
	
	The unlimited linear reconstruction method is based on a linear approximation to the underlying function $a(\zeta)$. Its slope $\delta_i$ (in cell-index coordinates) is estimated using the central difference (the Fromm method)
	\begin{align}
		\delta_i = \frac{\overline{a}_{i+1} - \overline{a}_{i-1}}{2}  = \Delta x \, \partial_x a|_{x_i} + O(\Delta x^3), \label{eq:LIN_sigma}
	\end{align}
	using which we obtain the reconstructed states
	\begin{align}
		a_{i-1/2,\mrm{R}} &= \overline{a}_i - \frac{\delta_i}{2} + O(\Delta x^2), \label{eq:LIN_aimhR} \\
		a_{i+1/2,\mrm{L}} &= \overline{a}_i + \frac{\delta_i}{2} + O(\Delta x^2). \label{eq:LIN_aiphL}
	\end{align}
	The LIN method is exact wherever $a(\zeta)$ is locally linear, \mbox{2\textsuperscript{nd}-order} accurate for general but smooth functions $a(\zeta)$, and it requires two ghost cells at domain boundaries.\footnote{Although the method only uses the cells $i-1$, $i$, and $i+1$, reconstruction must also be performed in the first ghost cell to fully define the Riemann problem at the domain boundary.}
	
	\subsubsection{The LIN+VL method}
	\label{sec:method_LINVL}
	
	The van-Leer-limited linear reconstruction method is particularly easy to describe in terms of the 2\textsuperscript{nd}-order-accurate, interface-centred slopes
	\begin{align}
		\delta_{i-1/2} = \overline{a}_i - \overline{a}_{i-1}, \label{eq:LIN_VL_dimh} \\
		\delta_{i+1/2} = \overline{a}_{i+1} - \overline{a}_i. \label{eq:LIN_VL_diph}
	\end{align}
	
	The final slope $\delta_{i,\mrm{lim}}$ is then obtained by applying the limiter of \citet{vanleer1974a},
	\begin{align}
		\delta_{i,\mrm{lim}} = 
		\begin{cases}
			\frac{2 \delta_{i-1/2}\, \delta_{i+1/2}}{\delta_{i-1/2} + \delta_{i+1/2}}\quad &\text{if}\ \delta_{i-1/2}\, \delta_{i+1/2} > 0, \\
			0\quad &\text{otherwise,}
		\end{cases}
	\end{align}
	which gives the reconstructed states
	\begin{align}
		a_{i-1/2,\mrm{R}} &= \overline{a}_i - \frac{\delta_{i,\mrm{lim}}}{2} + O(\Delta x^2), \label{eq:LIN_VL_aimhR} \\
		a_{i+1/2,\mrm{L}} &= \overline{a}_i + \frac{\delta_{i,\mrm{lim}}}{2} + O(\Delta x^2). \label{eq:LIN_VL_aiphL}
	\end{align}
	The limiter makes the slope less steep where $a(\zeta)$ is strongly curved and flat at local extrema. With a smooth and monotonic function $a(\zeta)$, the effect of the limiter weakens upon grid refinement and the left and right states converge to those provided by the LIN method. This makes the LIN+VL method formally \mbox{2\textsuperscript{nd}-order} accurate  away from any extrema. Two ghost cells are required at domain boundaries.
	
	\subsubsection{The PAR method}
	\label{sec:method_PAR}
	
	The unlimited parabolic method assumes that $a(\zeta)$ can within cell $i$ be described using the parabola
	\begin{align}
		a(\zeta) = \sum_{n=0}^{2} c_n (\zeta - \zeta_i)^n. \label{eq:PAR_a}
	\end{align}
	The three coefficients $c_n$ are uniquely determined by the requirement that the averages of $a(\zeta)$ in cells $i-1$, $i$, and $i+1$ equal $\overline{a}_{i-1}$, $\overline{a}_i$, and $\overline{a}_{i+1}$, respectively. The reconstructed states are then obtained by evaluating Eq.~(\ref{eq:PAR_a}) at $\zeta_{i-1/2}$ and $\zeta_{i+1/2}$, respectively. The resulting expressions are
	\begin{align}
		a_{i-1/2,\mrm{R}} &= \frac{1}{6}\left( 2\overline{a}_{i-1} + 5\overline{a}_i - \overline{a}_{i+1} \right) + O(\Delta x^3), \label{eq:PAR_aimhR} \\
		a_{i+1/2,\mrm{L}} &= \frac{1}{6}\left( -\overline{a}_{i-1} + 5\overline{a}_i + 2\overline{a}_{i+1} \right) + O(\Delta x^3). \label{eq:PAR_aiphL}
	\end{align}
	The PAR method is exact wherever $a(\zeta)$ is locally parabolic, \mbox{3\textsuperscript{rd}-order} accurate for general but smooth functions $a(\zeta)$, and it requires two ghost cells at domain boundaries.
	
	\subsubsection{The PPM84 method}
	\label{sec:method_PPM84}
	
	The piecewise parabolic reconstruction of CW84 is a two-step process. The first is based on the 4\textsuperscript{th}-order-accurate interpolation formula
	\begin{align}
		a_{i+1/2} = \frac{1}{2}\left( \overline{a}_i + \overline{a}_{i+1} \right) - \frac{1}{6}\left( \delta_{i+1} - \delta_i \right) = a|_{x_{i+1/2}} + O(\Delta x^4).
	\end{align}
	This expression is the equivalent of Eq.~1.6 of CW84 in the special case of a uniform grid. CW84 then replace $\delta_i$ by the limited value
	\begin{align}
		\delta_{i,\mrm{lim}} = 
		\begin{cases}
			\mrm{sgn}(\delta_i) \min\left( |\delta_i|,\ 2|\delta_{i-1/2}|,\ 2|\delta_{i+1/2}| \right) &\text{if}\ \delta_\mrm{i-1/2}\, \delta_{i+1/2} > 0 \\
			0 &\text{otherwise}, \label{eq:PPM84_limiter_1}
		\end{cases}
	\end{align}
	which is the monotonised central limiter of \citet{vanleer1977a}. In our implementation, we do not set $\delta_{i,\mrm{lim}} = 0$ if $\delta_\mrm{i-1/2}\, \delta_{i+1/2} \le 0$ (i.e.\ at local extrema). We have tested that, thanks to the presence of another limiter in PPM84 (see below), this modification has essentially no influence on the results. However, it makes the code faster because it removes three conditional expressions per reconstruction step (we need $\delta_{i-1,\mrm{lim}}$, $\delta_{i,\mrm{lim}}$, and $\delta_{i+1,\mrm{lim}}$ to obtain $a_{i-1/2}$ and $a_{i+1/2}$).
	
	The interpolated value $a_{i+1/2}$ is initially assigned to both $a_{i+1/2,\mrm{L}}$ and $a_{i+1/2,\mrm{R}}$, i.e.\ there is no discontinuity at the interface. However, CW84 approximate the distribution of variable $a$ in cell $i$ by the parabola uniquely defined by $a_{i-1/2,\mrm{R}}$, $a_{i+1/2,\mrm{L}}$, and the cell average $a_i$. This parabola may in some cases take on values outside of the interval defined by $a_{i-1/2,\mrm{R}}$ and $a_{i+1/2,\mrm{L}}$, i.e.\ overshoots may appear within the cell. To prevent this, a second limiting step is introduced. We express it in terms of the differences
	\begin{align}
		a_i^- &= a_{i-1/2,\mrm{R}} - \overline{a}_i, \label{eq:PPM84_aim} \\
		a_i^+ &= a_{i+1/2,\mrm{L}} - \overline{a}_i. \label{eq:PPM84_aip}
	\end{align}
	The limiter is then defined by the variable replacements
	\begin{align}
		a_i^- \mapsto 0,\ a_i^+ \mapsto 0\quad &\text{if}\quad a_i^- a_i^+ \ge 0,\nonumber \\
		a_i^- \mapsto -2 a_i^+\quad &\text{if}\quad |a_i^-| > 2|a_i^+|, \label{eq:PPM84_limiter_2} \\
		a_i^+ \mapsto -2 a_i^-\quad &\text{if}\quad |a_i^+| > 2|a_i^-|.\nonumber
	\end{align}
	The reconstructed states $a_{i-1/2,\mrm{R}}$ and $a_{i+1/2,\mrm{L}}$ are recovered using Eqs.~(\ref{eq:PPM84_aim}) and (\ref{eq:PPM84_aip}). Equation~(\ref{eq:PPM84_limiter_2}) is equivalent to Eq.~1.10 of CW84. This second limiter introduces discontinuities at cell interfaces in regions where the gradient of $a(\zeta)$ changes rapidly and it flattens the assumed parabola at local extrema.
	
	Unlike CW84, we do not use the parabolic model of $a(\zeta)$ inside the cell in any way. We only need the states at the two sides of each interface to construct Riemann problems and time integration is done using a Runge-Kutta scheme (see Sect.~\ref{sec:time-update} for details). However, the second limiter (Eq.~(\ref{eq:PPM84_limiter_2})), which is based on the parabolic model, is still needed to remove oscillations and to introduce dissipation where necessary.
	
	Although the interpolation formula that PPM84 starts with is 4\textsuperscript{th}-order accurate, the slope flattening introduced at all extrema reduces the practically attainable order of accuracy substantially for non-monotonic solutions even if they are smooth. Our 1D experiment in Sect.~\ref{sec:linear_advection} gives the order of $2.3$ whereas \citet{colella2008a} reach the order of $2.6$ in a similar advection experiment with a Gaussian-shaped profile and PPM84 reconstruction. The PPM84 method requires three ghost cells at domain boundaries.
	
	\subsubsection{The PPM08 method}
	\label{sec:method_PPM08}
	
	The piecewise parabolic method of CS08 is based on ideas similar to those of CW84 in constructing the PPM84 scheme and PPM08 also contains two limiters. However, the limiters are modified such that PPM08 models smooth extrema instead of flattening them.
	
	In PPM08, the first estimate of $a_{i+1/2}$ is obtained using the 6\textsuperscript{th}-order-accurate interpolation formula (c.f.\ Eq.~17 of CS08)
	\begin{align}
		a_{i+1/2} &= \frac{37}{60} \left( \overline{a}_i + \overline{a}_{i+1} \right) - \frac{2}{15} \left( \overline{a}_{i-1} + \overline{a}_{i+2} \right) + \frac{1}{60} \left( \overline{a}_{i-2} + \overline{a}_{i+3} \right) \\&=a|_{x_{i+1/2}} + O(\Delta x^6).
	\end{align}
	If $a_{i+1/2}$ does not satisfy the condition (c.f.\ Eq.~13 of CS08)
	\begin{align}
		\min(\overline{a}_i,\, \overline{a}_{i+1}) \le a_{i+1/2} \le \max(\overline{a}_i,\, \overline{a}_{i+1})
	\end{align}
	a limiter is applied. It is based on three 2\textsuperscript{nd}-order-accurate second derivatives,
	\begin{alignat}{2}
		&\left(\mrm{D}^2 a\right)_{i+1/2} &&= 3\left( \overline{a}_i - 2a_{i+1/2} + \overline{a}_{i+1} \right) = \partial_x^2 a|_{x_{i+1/2}} \Delta x^2 + O(\Delta x^4), \label{eq:d2a}\\
		&\left(\mrm{D}^2 a\right)_i &&= \overline{a}_{i-1} - 2\overline{a}_i + \overline{a}_{i+1}= \partial_x^2 a|_{x_{i}} \Delta x^2 + O(\Delta x^4), \\
		&\left(\mrm{D}^2 a\right)_{i+1} &&= \overline{a}_i - 2\overline{a}_{i+1} + \overline{a}_{i+2}= \partial_x^2 a|_{x_{i+1}} \Delta x^2 + O(\Delta x^4).
	\end{alignat}
	In Eq.~(\ref{eq:d2a}), the more common finite difference formula with a prefactor 4 would arise if the involved quantities were of the same kind, i.e. all three point values, or all three averages. These derivatives are combined to obtain a limited derivative $\left(\mrm{D}^2 a\right)_{i+1/2,\mrm{lim}}$ such that
	\begin{align}
		\left(\mrm{D}^2 a\right)_{i+1/2,\mrm{lim}} =\ &\mrm{sgn}\left[ \left(\mrm{D}^2 a\right)_{i+1/2} \right] \min\Bigg( \left|\left(\mrm{D}^2 a\right)_{i+1/2}\right|, \nonumber\\
		&C\left|\left(\mrm{D}^2 a\right)_i\right|, C\left|\left(\mrm{D}^2 a\right)_{i+1}\right| \Bigg) \label{eq:PPM08_limiter_1}
	\end{align}
	if all three derivatives have the same sign and
	\begin{align}
		\left(\mrm{D}^2 a\right)_{i+1/2,\mrm{lim}} = 0
	\end{align}
	otherwise. We use $C = 1.25$ in Eq.~(\ref{eq:PPM08_limiter_1}). The limited derivative is then used to modify the value of $a_{i+1/2}$,
	\begin{align}
		a_{i+1/2} \mapsto \frac{1}{2}\left( \overline{a}_i + \overline{a}_{i+1} \right) - \frac{1}{6} \left(\mrm{D}^2 a\right)_{i+1/2,\mrm{lim}}. \label{eq:ppm08_limiter_1}
	\end{align}
	Equation~19 of CS08 is equivalent to our Eq.~(\ref{eq:ppm08_limiter_1}) except for the factor in front of the second term, which is $\frac{1}{3}$ in CS08. This is likely a typographical error because the factor of $\frac{1}{6}$ is needed to obtain the original, unlimited value of $a_{i+1/2}$ when the solution is so smooth that the limiter does not change the second derivative significantly.
	
	Just like in the PPM84 method, the interpolated value $a_{i+1/2}$ is initially assigned to both $a_{i+1/2,\mrm{L}}$ and $a_{i+1/2,\mrm{R}}$, i.e.\ there is no discontinuity at the interface. The second limiting step depends on whether cell $i$ is in the vicinity of a local extremum or not. If (c.f.\ Eq.~20 of CS08)
	\begin{align}
		(\overline{a}_i - a_{i-1/2,\mrm{R}})(a_{i+1/2,\mrm{L}} - \overline{a}_i) &\le 0\quad \text{or} \nonumber\\
		(\overline{a}_i - \overline{a}_{i-1})(\overline{a}_{i+1} - \overline{a}_i) &\le 0, \label{eq:PPM08_extremum_condition}
	\end{align}
	cell $i$ is close to a local extremum, which should be preserved if smooth enough. The second derivatives
	\begin{alignat}{2}
		&\left(\mrm{D}^2 a\right)_i^* &&= 6(a_{i-1/2,\mrm{R}} - 2\overline{a}_i + a_{i+1/2,\mrm{L}}), \\
		&\left(\mrm{D}^2 a\right)_{i-1} &&= \overline{a}_{i-2} - 2\overline{a}_{i-1} + \overline{a}_i, \label{eq:PPM08_typo_2}\\
		&\left(\mrm{D}^2 a\right)_i &&= \overline{a}_{i-1} - 2\overline{a}_i + \overline{a}_{i+1}, \\
		&\left(\mrm{D}^2 a\right)_{i+1} &&= \overline{a}_i - 2\overline{a}_{i+1} + \overline{a}_{i+2}.
	\end{alignat}
	are combined to judge the solution's smoothness (CS08 have one wrong index in their equivalent of our Eq.~(\ref{eq:PPM08_typo_2}), c.f.\ their Eq.~21). We then set
	\begin{align}
		\left(\mrm{D}^2 a\right)_{i,\mrm{lim}}^* =\ &\mrm{sgn}\left[ \left(\mrm{D}^2 a\right)_i^* \right] \min\Bigg( \left|\left(\mrm{D}^2 a\right)_i^*\right|, \nonumber\\
		&C\left|\left(\mrm{D}^2 a\right)_{i-1}\right|, C\left|\left(\mrm{D}^2 a\right)_i\right|, C\left|\left(\mrm{D}^2 a\right)_{i+1}\right| \Bigg) \label{eq:PPM08_limiter_2}
	\end{align}
	if all four second derivatives have the same sign and
	\begin{align}
		\left(\mrm{D}^2 a\right)_{i,\mrm{lim}}^* = 0
	\end{align}
	otherwise. Finally, the reconstructed states are updated,
	\begin{align}
		a_{i-1/2,\mrm{R}} \mapsto \overline{a}_i + (a_{i-1/2,\mrm{R}} - \overline{a}_i)\frac{\left(\mrm{D}^2 a\right)_{i,\mrm{lim}}^*}{\left(\mrm{D}^2 a\right)_i^*}, \label{eq:PPM08_update_2_imhR} \\
		a_{i+1/2,\mrm{L}} \mapsto \overline{a}_i + (a_{i+1/2,\mrm{L}} - \overline{a}_i)\frac{\left(\mrm{D}^2 a\right)_{i,\mrm{lim}}^*}{\left(\mrm{D}^2 a\right)_i^*}. \label{eq:PPM08_update_2_iphL}
	\end{align}
	If $\left|\left(\mrm{D}^2 a\right)_i^*\right| < \varepsilon = 10^{-12}$ we do not modify the reconstructed states in this step. Equations (\ref{eq:PPM08_update_2_imhR}) and (\ref{eq:PPM08_update_2_iphL}) are equivalent to Eq.~23 of CS08.
	
	If the condition in Eq.~(\ref{eq:PPM08_extremum_condition}) is not satisfied, i.e.\ cell $i$ is not in the vicinity of a local extremum, we use Eqs.~(\ref{eq:PPM84_aim})-(\ref{eq:PPM84_limiter_2}) instead of Eqs.~(\ref{eq:PPM08_update_2_imhR}) and (\ref{eq:PPM08_update_2_iphL}) to limit the reconstructed states. CS08 propose to use a slightly less restrictive limiter away from extrema (their Eq.~26), but that limiter produces oscillations with our time-discretization scheme and we do not use it.
	
	Just as we do in the case of PPM84, we only use the reconstructed states and not the assumed parabolic model of $a(\zeta)$ within the cell, see Sect.~\ref{sec:method_PPM84} for details. The PPM08 method is 6\textsuperscript{th}-order accurate for smooth functions $a(\zeta)$ even if they are not monotonic. Our 1D experiment in Sect.~\ref{sec:linear_advection} confirms this. The PPM08 method requires four ghost cells at domain boundaries.
	
	\subsubsection{The PSH method}
	\label{sec:method_PSH}
	
	Whereas all of the previous reconstruction methods are applied to all variables in the same way, the PSH method is hybrid: it is an unlimited piecewise-sextic method for all dynamic variables combined with PPM08 for passive scalars. This allows us to eliminate certain issues that occur with methods containing limiters when applied to slow flows (see Sect.~\ref{sec:results_kh} for details) while essentially eliminating oscillations in the passive scalars, which could represent mass fractions. The piecewise sextic reconstruction assumes that within cell $i$ $a(\zeta)$ can be described by the sextic polynomial
	\begin{align}
		a(\zeta) = \sum_{n=0}^{6} c_n (\zeta - \zeta_i)^n. \label{eq:PSH_a}
	\end{align}
	The seven coefficients $c_n$ are uniquely determined by the requirement that the averages of $a(\zeta)$ in cells $i-3+n$ equal $\overline{a}_{i-3+n}$ for $n = 0, 1, \dots, 6$. The reconstructed states are then obtained by evaluating Eq.~(\ref{eq:PSH_a}) at $\zeta_{i-1/2}$ and $\zeta_{i+1/2}$, respectively. The resulting expressions are
	\begin{align}
		\begin{split}\label{eq:PSH_aL}
			a_{i-1/2,\mrm{R}} = \frac{1}{420}\bigg({}& 4\overline{a}_{i-3} - 38\overline{a}_{i-2} + 214\overline{a}_{i-1} + 319\overline{a}_i \\
			&- 101\overline{a}_{i+1} + 25\overline{a}_{i+2} - 3\overline{a}_{i+3}\bigg) + O(\Delta x^7),
		\end{split}\\
		\begin{split}\label{eq:PSH_aR}
			a_{i+1/2,\mrm{L}} = \frac{1}{420}\bigg({}& -3\overline{a}_{i-3} + 25\overline{a}_{i-2} - 101\overline{a}_{i-1} + 319\overline{a}_i \\
			&+ 214\overline{a}_{i+1} - 38\overline{a}_{i+2} + 4\overline{a}_{i+3}\bigg) + O(\Delta x^7),
		\end{split}
	\end{align}
	The method is exact wherever $a(\zeta)$ is locally a sextic polynomial, 7\textsuperscript{th}-order accurate for general but smooth functions $a(\zeta)$, and it requires four ghost cells at domain boundaries.
	
	\subsection{Approximate Riemann solvers}
	\label{sec:numerical-fluxes}
	
	The reconstructed pair of primitive state quantities, $\bm{q}_{i+1/2,\LR}$, defines a Riemann problem at the cell interface $i\,{+}\,1/2$, which \code{SLH} solves by means of 1D approximate Riemann solvers to obtain the face-centered value of the fluxes $\bm{\mathcal{F}}_{i+1/2}$. We run the tests presented in Sect.~\ref{sec:convergence-properties} using two widely popular flux functions, namely the RUSANOV and HLLC solvers. Because in this work we only focus on simulations of subsonic flows, 
	for comparison we also build a low-dissipation version of HLLC following the approach of \cite{minoshima2021}, who modified the Harten-Lax-van Leer-Discontinuities \citep[HLLD,][]{miyoshi2005} scheme for magnetohydrodynamics to diminish the magnitude of the numerical dissipation for low-Mach-number flows. The authors called this low-dissipation flux ``LHLLD'', so, for consistency, we will refer to the low-dissipation HLLC solver as ``LHLLC'' throughout the text\footnote{Although \code{SLH} was already equipped with several low-Mach solvers, like AUSM$^+$-up \citep{Liou2006} and Miczek-Roe \citep{miczek2015}, we decide not to use them in this study because they all suffer from a more restrictive stability criterion on the time step than LHLLC when used in combination with explicit time steppers, whilst all of these fluxes provide very similar results in terms of accuracy.}. In the rest of this section, we summarize the main aspects of each of these solvers and provide the implementation details whenever several choices can be made for specifying the value of a certain quantity that is needed to evaluate the numerical flux. 
	
	\subsubsection{RUSANOV}
	The RUSANOV flux is computed by adding an upwind, numerical diffusive term proportional to the maximum wave speed at the cell interface, $S_\mathrm{max}$, to every component of the central flux. The final expression for the numerical flux reads\footnote{For sake of clarity, here we assume that the fluxes are computed in the $x$ direction and dropped the indices, but analogous expressions can be obtained for the $y$ and $z$ directions.}
	\begin{equation}
		\label{eq:RUSANOV}
		\bm{\mathcal{F}}(\bm{U}_\mathrm{L},\bm{U}_\mathrm{R}) = \frac{1}{2}\left[ \bm{F}(\bm{U}_\mathrm{L})+\bm{F}(\bm{U}_\mathrm{R})\right] - \frac{1}{2} S_\mathrm{max}(\bm{U}_\mathrm{R}-\bm{U}_\mathrm{L}),
	\end{equation}
	where $\bm{U}_\mathrm{L,R}$ are the left and right sets of conserved quantities, respectively. In \code{SLH}, $S_\mathrm{max}$ is estimated as
	\begin{equation}
		S_\mathrm{max} = \max \Big(|u_\mathrm{L}|+c_\mathrm{L},|u_\mathrm{R}|+c_\mathrm{R}\Big),
	\end{equation}
	where $c\,{=}\,(\gamma p /\rho)^{1/2}$ is the sound speed. The diffusive term in Eq.~(\ref{eq:RUSANOV}) scales with the Mach number of the flow $\mathcal{M}$ and allows the scheme to achieve numerical stability by smearing out any discontinuity that may arise in the vector of state quantities $\bm{U}$.
	
	The RUSANOV solver is one of the simplest schemes that can be used to approximate the fluxes at grid cell interfaces, which makes it very efficient in terms of Floating Point Operations per Second. However, it does not take into account the complex structure of the solution arising from the Riemann problem of gas dynamics \citep[see, e.g.,][]{toro2009a}, so the states between the two outer waves in the Riemann fan are averaged out. For this reason, this flux function is particularly diffusive for transporting contact and shear waves, which lack the self-steepening property of sound waves.
	
	\subsubsection{HLLC}
	\label{sec:hllc}
	\begin{figure}
		\includegraphics[width=0.5\textwidth]{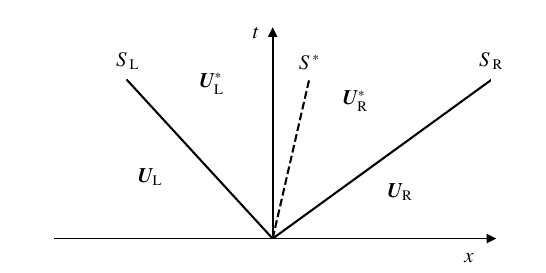}
		\caption{Space-time diagram showing the wave structure of the HLLC solver for a Riemann problem of gas dynamics. $S_\LR$ are the two outer sonic waves (solid lines), while $S^*$ is the speed of the linearly degenerate contact and shear waves (dashed line) that separate 
			the two intermediate ``star'' states, $\bm{U}^*_\LR$.  }
		\label{fig:Riemann-fan}
	\end{figure}
	Different from RUSANOV, the HLLC solver restores the linearly degenerate contact and shear waves back to the structure of the solution of the Riemann problem  (see Fig.~\ref{fig:Riemann-fan}). In this method, the numerical flux is chosen according to the sign of the wave speeds in the Riemann fan,
	\begin{align}
		\label{flux-hllc}
		\bm{\mathcal{F}}(\bm{U}_\mathrm{L},\bm{U}_\mathrm{R}) = \left\{
		\begin{array}{@{}l@{\thinspace}l}
			\bm{F}_\mathrm{L}    & \text{ if  } 0 \le  S_\mathrm{L},\\
			\bm{F}^*_\mathrm{L}  & \text{ if  } S_\mathrm{L} < 0 \le S^*,\\
			\bm{F}^*_\mathrm{R}  & \text{ if  } S^* <  0 \le S_\mathrm{R},\\
			\bm{F}_\mathrm{R}    & \text{ if  } S_\mathrm{R} < 0.
		\end{array}
		\right.
	\end{align}
	While the computation of the physical fluxes $\bm{F}_\mathrm{L}=\bm{F}(\bm{U}_\mrm{L})$ and $\bm{F}_\mathrm{R}=\bm{F}(\bm{U}_\mrm{R})$ is trivial, the  fluxes in the intermediate regions, $\bm{F}^*_\mathrm{L}$ and $\bm{F}^*_\mathrm{R}$, are obtained by solving the Rankine--Hugoniot jump conditions across the two outer sonic waves,
	\begin{align}
		\label{eq:UsL}
		S_\mathrm{L}\bm{U}^*_\mathrm{L}-\bm{F}^*_\mathrm{L} &= S_\mathrm{L}\bm{U}_\mathrm{L}-\bm{F}_\mathrm{L},\\
		\label{eq:UsR}
		S_\mathrm{R}\bm{U}^*_\mathrm{R}-\bm{F}^*_\mathrm{R} &= S_\mathrm{R}\bm{U}_\mathrm{R}-\bm{F}_\mathrm{R}.
	\end{align}
	Here, $\bm{U}^*_\mathrm{L}$ and $\bm{U}^*_\mathrm{R}$ represent the state quantities in the star regions. In order to solve Eqs.~(\ref{eq:UsL})--(\ref{eq:UsR}), proper estimates of the wave speeds $S_\mathrm{L}$ and $S_\mathrm{R}$ must be provided beforehand. \code{SLH} computes these wave speeds as
	\begin{align}
		S_\mathrm{L} &= \min(u_\mathrm{L},u_\mathrm{R}) - \max(c_\mathrm{L},c_\mathrm{R}), \\
		S_\mathrm{R} &= \max(u_\mathrm{L},u_\mathrm{R}) + \max(c_\mathrm{L},c_\mathrm{R}).
	\end{align}
	After some assumptions and algebraic manipulations \citep[see][]{toro2009a}, these estimates allow Eqs.~(\ref{eq:UsL})--(\ref{eq:UsR}) to be solved for the star states $\bm{U}^*_\LR$ ,
	\begin{align}
		(\rho)^*_\LR & = \tilde{\alpha}_\LR, \\
		(\rho u)^*_\LR   & = \tilde{\alpha}_\LR S^*, \\
		(\rho v)^*_\LR   & = \tilde{\alpha}_\LR v_\LR, \\
		(\rho w)^*_\LR   & =  \tilde{\alpha}_\LR w_\LR, \\
		\begin{split}
			(\rho e_\mathrm{tot})^*_\LR & = \tilde{\alpha}_\LR\frac{(\rho e_{\mathrm{tot}})_\LR}{\rho_\LR} \\ &+\tilde{\alpha}_\LR(S^*-u_\LR) \Bigg[  S^* + \frac{p_\LR}{\rho_\LR(S_\LR-u_\LR)} \Bigg],
		\end{split}\\
		(\rho X)^*_\LR   & =  \tilde{\alpha}_\LR X_\LR, 
	\end{align}
	with 
	\begin{equation}
		\tilde{\alpha}_\LR = \rho_\LR\Bigg(\frac{S_\LR-u_\LR}{S_\LR-S^*}\Bigg),
	\end{equation}
	and $S^*$ being the speed of the intermediate wave,
	\begin{equation}
		S^* = \frac{p_\mathrm{R}-p_\mathrm{L}+\rho_\mathrm{L}u_\mathrm{L}(S_\mathrm{L}-u_\mathrm{L})-\rho_\mathrm{R}u_\mathrm{R}(S_\mathrm{R}-u_\mathrm{R})}{\rho_\mathrm{L}(S_\mathrm{L}-u_\mathrm{L})-\rho_\mathrm{R}(S_\mathrm{R}-u_\mathrm{R})}.
	\end{equation}
	The gas pressure is preserved across the middle wave and takes the value
	\begin{equation}
		\begin{split}
			p^* = &  \frac{1}{2}\Big[ p_\mathrm{L} + p_\mathrm{R} + \rho_\mathrm{L}(S_\mathrm{L}-u_\mathrm{L})(S^*-u_\mathrm{L})  \\
			& +\rho_\mathrm{R}(S_\mathrm{R}-u_\mathrm{R})(S^*-u_\mathrm{R})  \Big].
		\end{split}
	\end{equation}
	These expressions are then inserted back into Eqs.~(\ref{eq:UsL})--(\ref{eq:UsR}) to compute $\bm{F}^*_\mathrm{L}$ and $\bm{F}^*_\mathrm{R}$. Finally, the interface flux is chosen according to Eq.~(\ref{flux-hllc}). 
	
	In our study, we use a variant of the original HLLC solver of \cite{Toro1994}, which allows the low-Mach correction presented in the next section to be implemented trivially into the solver. In particular, we directly evaluate the physical fluxes in the selected state of the Riemann fan,
	\begin{align}
		\label{flux-hllc-slh}
		\bm{\mathcal{F}}(\bm{U}_\mathrm{L},\bm{U}_\mathrm{R}) = \left\{
		\begin{array}{@{}l@{\thinspace}l}
			\bm{F}_\mathrm{L}    & \text{ if  } 0 \le  S_\mathrm{L},\\
			\bm{F}(\bm{U}^*_\mathrm{L})  & \text{ if  } S_\mathrm{L} < 0 \le S^*,\\
			\bm{F}(\bm{U}^*_\mathrm{R})  & \text{ if  } S^* <  0 \le S_\mathrm{R}, \\
			\bm{F}_\mathrm{R}    & \text{ if  } S_\mathrm{R} < 0,
		\end{array}
		\right.
	\end{align}
	and we compute $p^*$ using a linearized Riemann solver for the equations of gas dynamics \citep[see, e.g.,][]{toro1991}, 
	\begin{equation}
		\label{eq:star-pressure}
		p^*=\frac{1}{2}(p_\mathrm{L}+p_\mathrm{R})-\frac{1}{2}\tilde{\rho}\tilde{c}(u_\mathrm{R}-u_\mathrm{L}),
	\end{equation}
	where $\tilde{c}=(c_\mathrm{L}+c_\mathrm{R})/2$ and $\tilde{\rho}=(\rho_\mathrm{L}+\rho_\mathrm{R})/2$. The system in Eq.~(\ref{flux-hllc-slh}) is consistent with the physical fluxes in the sense that
	\begin{equation}
		\bm{\mathcal{F}}(\bm{U},\bm{U}) = \bm{F}(\bm{U})
	\end{equation}
	and it satisfies the Rankine--Hugoniot jump conditions across the contact wave $S^*$ as the original solver,
	\begin{equation}\label{eq:contact-star}
		S^*\bm{U}^*_\mathrm{R}-\bm{F}^*_\mathrm{R} = S^*\bm{U}^*_\mathrm{L}-\bm{F}^*_\mathrm{L}.
	\end{equation}
	Diagnostic tests of the Kelvin--Helmholtz instability problem, described in Sect.~\ref{sec:results_kh}, show that the numerical solutions computed with our modified version and the original solver of \cite{Toro1994} are virtually indistinguishable for subsonic flows.
	We stress, however, that the fluxes in Eq.~(\ref{flux-hllc-slh}) do not satisfy the jump conditions across the sonic waves $S_\mathrm{L}$ and $S_\mathrm{R}$ (see Eqs.~(\ref{eq:UsL}) and (\ref{eq:UsR})). Therefore, there is no guarantee that the resulting scheme preserves positivity of density and internal energy when the flow is nearly transonic, in which case effects of compressibility and nonlinearities can become dominant. Such a flow regime, however, is not considered in this study.
	
	\subsubsection{LHLLC}
	\label{sec:LHLLC}
	As discussed in Sect.~\ref{sec:hllc}, HLLC restores the intermediate, linearly degenerate waves, so it is generally more accurate than two-wave solvers like RUSANOV or HLL \citep{harten1983} in simulations involving the presence of material interfaces or the propagation of entropy waves. However, the effects of the numerical dissipation introduced by HLLC on the evolution of the flow become progressively more dominant as $\mathcal{M}\rightarrow0$, thus producing unnecessarily large diffusive errors in highly subsonic velocity regimes \citep[see, e.g.,][]{fleischmann2020}. In our variant of HLLC, this behavior is caused by the upwind term in the expression for $p^*$ (see Eq.~(\ref{eq:star-pressure})),
	\begin{equation}
		\label{eq:diffusion-star-pressure}
		D(p^*)=-\frac{1}{2}\tilde{\rho}\tilde{c}(u_\mathrm{R}-u_\mathrm{L}).
	\end{equation}
	This term scales with $\mathcal{M}$, which is inconsistent with the scaling of pressure fluctuations dynamically generated by subsonic flows. In fact, in the asymptotic limit $\mathcal{M}\rightarrow0$, the solution to the compressible Euler equations approaches the incompressible regime \citep{guillard1999}, in which the gas pressure is homogeneous in space except for fluctuations proportional to $\mathcal{M}^2$. At low Mach numbers, the numerical term in Eq.~(\ref{eq:diffusion-star-pressure}) can eventually become larger than the physical pressure fluctuation at the cell interface, thus leading to an highly inaccurate pressure flux estimation.

	In order to correct for the flawed scaling of the numerical dissipation introduced by HLLC-like methods, we here follow the approach described in \cite{minoshima2021}, who proposed to multiply the diffusive term in Eq.~(\ref{eq:diffusion-star-pressure}) by a factor $\phi$ proportional to the local Mach number of the flow\footnote{Other low-Mach corrections for the HLLC Riemann solver can be found, e.g., in \cite{thornber2008,rieper2011,xie2019,chen2020,fleischmann2020}.}. Such a correction was originally applied to the  magnetohydrodynamic solver HLLD, but it can easily be used in HLLC by setting all magnetic field components to zero, resulting in
	\begin{equation}\label{eq:low-mach-fix}
		\phi = \chi(2-\chi), 
	\end{equation}
	with
	\begin{equation}
		\chi = \min \Bigg\{ 1, \max \Bigg( \frac{|\bm{V}_\mathrm{L}|}{c_\mathrm{L}},\frac{|\bm{V}_\mathrm{R}|}{c_\mathrm{R}} \Bigg) \Bigg\}.
	\end{equation}
	The final expression for $p^*$ then reads
	\begin{equation}\label{eq:pressure-low-mach-fix}
		p^*=\frac{1}{2}(p_\mathrm{L}+p_\mathrm{R})-\phi\frac{1}{2}\tilde{\rho}\tilde{c}(u_\mathrm{R}-u_\mathrm{L}).
	\end{equation}
	The resulting upwind term in this ``low-dissipation'' version of the HLLC flux (LHLLC) scales with $\mathcal{M}^2$ when the flow is subsonic, so the ratio of the numerical diffusive term to the amplitude of pressure fluctuations is independent of $\mathcal{M}$. 
	
	We note that the same fix cannot equally be applied to the RUSANOV flux without sacrificing numerical stability. In particular, a diffusive coefficient proportional to $\mathcal{M}^2$ would result in too little dissipation for sound waves. This is not the case for LHLLC, in which the complex upwinding performed in Eq.~(\ref{flux-hllc-slh}) guarantees that the scheme remains stable for the propagation of sound waves (see also Appendix \ref{sec:linear_advection}).
	
	\subsection{Time discretization}
	\label{sec:time-update}

	Because the acoustic Courant-Friedrichs-Lewy \citep[CFL,][]{courant1928} criterion on the time step becomes excessively strict in regimes of very low Mach numbers, implicit time discretization techniques are typically better suited for simulating the evolution of such slow flows \citep[see, e.g.,][]{viallet2011,miczek2015,dumbser2019}. However, we recognize that most hydrodynamic codes nowadays do not have 
	time-implicit integration capabilities, whose implementation requires a considerable effort from code developers. Thus, to make our study easily reproducible, we decide to target in our test setups, see Sect.~\ref{sec:results_kh} and \ref{sec:code-comparison}, flows with  Mach numbers in the range $10^{-3}\,{\lesssim}\,\mathcal{M}\,{\lesssim}\,10^{-1}$, where simple time-explicit marching schemes are still competitive with implicit ones. In this work, explicit time integration is performed in a semi-discrete fashion, in which the cell-surface integral of the fluxes and the cell-volume integral of the source terms in Eq.~(\ref{eq:euler-int}) are first separately discretized in space whilst the system is left continuous in time according to the method of lines (see Sect.~\ref{sec:slh}). The resulting system of ordinary differential equations (see Eq.~(\ref{eq:semi-discrete})), 
	is then solved numerically to advance the cell-volume-averaged state quantities in time. To solve Eq.~(\ref{eq:semi-discrete}), we use the third-order accurate, strong stability preserving  (SSP) RK3 method of \cite{shu1988}, in which the update on $\overline{\bm{U}}_{i,j,k}^{(n)}$ from time $t_n$ to time $t_{n+1} = t_n + \Delta t$ is performed in three stages,
	\begin{align}
		\bm{\overline{U}}^{(1)}_{i,j,k} &= \bm{\overline{U}}^{(n)}_{i,j,k} -  \bm{\mathcal{R}}\Big(\bm{\overline{U}}^{(n)}_{i,j,k}\Big)\Delta t,
		\\
		\bm{\overline{U}}^{(2)}_{i,j,k} &= \frac{3}{4}\bm{\overline{U}}^{(n)}_{i,j,k} + \frac{1}{4}\bm{\overline{U}}^{(1)}_{i,j,k} -  \frac{1}{4}\bm{\mathcal{R}}\Big(\bm{\overline{U}}^{(1)}_{i,j,k}\Big)\Delta t,
		\\
		\bm{\overline{U}}^{(n+1)}_{i,j,k} &= \frac{1}{3}\bm{\overline{U}}^{(n)}_{i,j,k} + \frac{2}{3}\bm{\overline{U}}^{(2)}_{i,j,k} -  \frac{2}{3}\bm{\mathcal{R}}\Big(\bm{\overline{U}}^{(2)}_{i,j,k}\Big)\Delta t.
	\end{align}
	In particular, we compute the spatial residuals at stage $s$,
	\begin{equation}
		\begin{split}
			\bm{\mathcal{R}}\Big(\bm{\overline{U}}^{(s)}_{i,j,k}\Big) = & \frac{1}{\Delta x}\Big(\bm{\mathcal{F}}^{(s)}_{i+1/2,j,k}-\bm{\mathcal{F}}^{(s)}_{i-1/2,j,k}  \\
			+ &  \bm{\mathcal{G}}^{(s)}_{i,j+1/2,k}-\bm{\mathcal{G}}^{(s)}_{i,j-1/2,k}   \\    
			+ &  \bm{\mathcal{H}}^{(s)}_{i,j,k+1/2}-\bm{\mathcal{H}}^{(s)}_{i,j,k-1/2}\Big)  \\ 
			- & \bm{\mathcal{S}}^{(s)}_{i,j,k},
		\end{split}
	\end{equation}
	using the numerical techniques described in Sects.~\ref{sec:spatial-reconstruction} and \ref{sec:numerical-fluxes}. Finally, in order to achieve numerical stability, we limit the time step according to
	\begin{equation}
		\Delta t = \frac{\mathrm{CFL}}{N_\mathrm{dim}}\min_{i,j,k} \Bigg(\frac{\Delta x}{|\bm{V}|_{i,j,k}+c_{i,j,k}} \Bigg),
	\end{equation}
	where $N_\mathrm{dim}$ is the number of spatial dimensions. In all the tests presented in Sect.~\ref{sec:convergence-properties}, we always adopt $\mathrm{CFL}\,{=}\,0.8$. We prefer to use a third-order accurate time stepper over less computationally expensive (but more inaccurate) methods, such as the ``Midpoint rule'' or SSP-RK2 \citep{shu1988}, so that the spatial instead of the temporal discretization would contribute most to the building up of global truncation errors.
	
	\section{Convergence properties of different Godunov-type methods}
	\label{sec:convergence-properties}

	In this section, we check if the methods included in our study (described in Sect.~\ref{sec:methods}) converge to the same numerical solution for several physical quantities of interest. In particular, we test a Kelvin--Helmholtz instability and a more complex setup characterized by the presence of turbulent convective flows, turbulent entrainment, and wave excitation. The results of the latter set of simulations allow us to estimate the computational cost per fixed accuracy for any given scheme, which we show in Sect.~\ref{sec:performance}.
	
	\subsection{Kelvin--Helmholtz instability}
	\label{sec:results_kh}
	
	We first test all of our 18 combinations of numerical schemes as described in Sects.~\ref{sec:spatial-reconstruction} and \ref{sec:numerical-fluxes} on a 2D Kelvin--Helmholtz problem with the initial condition
	\begin{align}
		\rho &= \gamma, \\
		u &= \mach_0 \left[1 - 2\eta(y) \right], \\
		v &= \frac{\mach_0}{10} \sin(2\pi x), \label{eq:vpert}\\
		p &= 1, \\
		X &= \eta(y),
	\end{align}
	where $\gamma = 1.4$ and
	\begin{align}
		\eta(y) = 
		\begin{cases}
			\frac{1}{2} \big\{ 1+\sin \left[ 16\pi(y+0.25)\right] \big\}, &\text{for}\ y > -\frac{9}{32}\ \text{and}\ y < -\frac{7}{32}, \\
			1, &\text{for}\ y \ge -\frac{7}{32}\ \text{and}\ y \le \frac{7}{32}, \\
			\frac{1}{2}\big\{1-\sin\left[16\pi(y-0.25) \right] \big\}, &\text{for}\ y > \frac{7}{32}\ \text{and}\ y < \frac{9}{32}, \\
			0,  &\text{otherwise}.\label{eq:eta}
		\end{cases}
	\end{align}
	The smooth function $\eta(y)$ provides a resolvable transition between layers moving in opposite horizontal directions.  The initial speed of sound is unity, so $\mach_0$ is a tunable initial Mach number of the shear flow. We discuss solutions with $\mach_0 = 10^{-1}$, $ 10^{-2}$, and $10^{-3}$. In Eq.~(\ref{eq:vpert}), a smooth initial perturbation with an amplitude of $\mach_0/10$ is included as a velocity component perpendicular to the shear flow. The computational domain, assumed to be periodic in both $x$ and $y$, spans $0 \le x \le 2$, $-0.5 \le y \le 0.5$.
	
\begin{figure}
		\includegraphics[width=\linewidth]{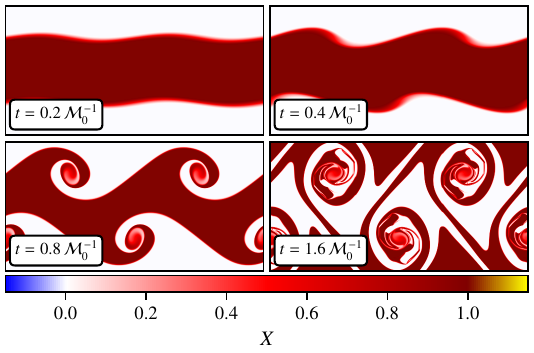}
		\caption{Reference solution to the Kelvin--Helmholtz problem with initial Mach number $\mach_0 = 10^{-2}$. The solution was computed using PSH reconstruction and the LHLLC flux function on an $8192 \,{\times}\, 4096$ grid. The mass fraction $X$ of the passive scalar is shown at four points in time: late in the linear growth of the instability ($t = 0.2\,\mach_0^{\,-1}$), at an early stage of non-linear evolution ($t = 0.4\,\mach_0^{\,-1}$), at a stage when the primary vortices have fully formed ($t = 0.8\,\mach_0^{\,-1}$; the final time for all of our other Kelvin--Helmholtz simulations), and at a late stage when fine threads have formed inside the primary vortices ($t = 1.6\,\mach_0^{\,-1}$). We use the same colour scale as in Figs.~\ref{fig:kh2d-machx-1.000e-02-128x64-ps}, \ref{fig:kh2d-machx-1.000e-03-128x64-ps}, and \ref{fig:kh2d-machx-1.000e-01-128x64-ps}, although $0 \leq X \leq 1$ in the reference solution.}
		\label{fig:kh2d-ref}
	\end{figure}

\begin{figure}
		\includegraphics[width=\linewidth]{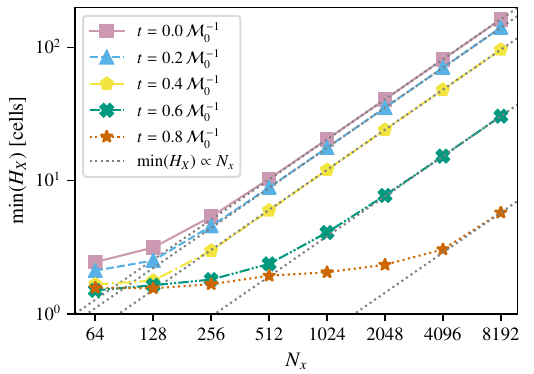}
		\caption{Resolution dependence of the minimum scale height $\min(H_X)$ of the passive scalar in the Kelvin--Helmholtz problem expressed in units of the computational cell width and shown at five points in time. Grid resolution is given by the number $N_x$ of computational cells along the $x$ axis. The initial Mach number is $\mach_0 = 10^{-2}$ and we use PSH reconstruction and the LHLLC flux function in this series of simulations. Once the steepest gradients in $X$ become resolved the minimum scale height starts to follow the linear scaling relations shown.}
		\label{fig:ps-scale-height}
	\end{figure}

	The fact that the transition function $\eta(y)$ between the shearing layers is smooth\footnote{$\eta(y)$ and its first derivative are continuous but the second derivative is not.} allows us to compute numerically converged solutions even in the absence of physical viscosity as long as the simulations are stopped before the flow field becomes chaotic \citep[see also][]{robertson2010a, mcnally2012, lecoanet2016a, berlok2019a}. Each of the two transitions spans only $\nicefrac{1}{16}$ of the domain height and is poorly resolved on the coarser grids used in our tests. Therefore, we improve the accuracy of the initial cell averages that involve $\eta(y)$ by averaging $\eta(y)$ over $100$ points uniformly distributed in the $y$-range covered by each cell. We measure numerical errors with respect to a reference solution computed using PSH reconstruction and the LHLLC flux function on a $8192 \times 4096$ grid. The solution for $\mach_0 = 10^{-2}$ is shown in Fig.~\ref{fig:kh2d-ref} at four points in time\footnote{We give the time in units of $\mach_0^{-1}$ such that the same numerical value corresponds to the same evolutionary stage of the instability at all three initial Mach numbers we use.}. As the instability grows in amplitude, the sinusoidal initial perturbation is rolled up into a series of vortices.  Parts of the initial shear layers are stretched and become trapped in the centres of the vortices. Other parts of the shear layers become substantially narrower. We quantify this phenomenon by computing the minimum scale height $\min(H_X) \equiv 1 / \max\left( |\bm{\nabla} X| \right)$ of the passive scalar $X$. Figure~\ref{fig:ps-scale-height} shows that this quantity drops by as much as a factor of $28$ between $t = 0$ and $t = 0.8 \mach_0^{-1}$. At the latter point in time, the minimum scale height is only $5.75$ computational cells on the $8192 \times 4096$ reference grid. Extremely thin and difficult-to-resolve filaments appear at even later times (see Fig.~\ref{fig:kh2d-ref}). Therefore, we compare the solutions at $t = 0.8 \mach_0^{-1}$, making the problem non-linear and challenging enough but not computationally prohibitive. The maximum Mach number in the flow field is $1.8 \mach_0$ at this point in time.
	
\begin{figure*}[p]
		\includegraphics[width=\linewidth]{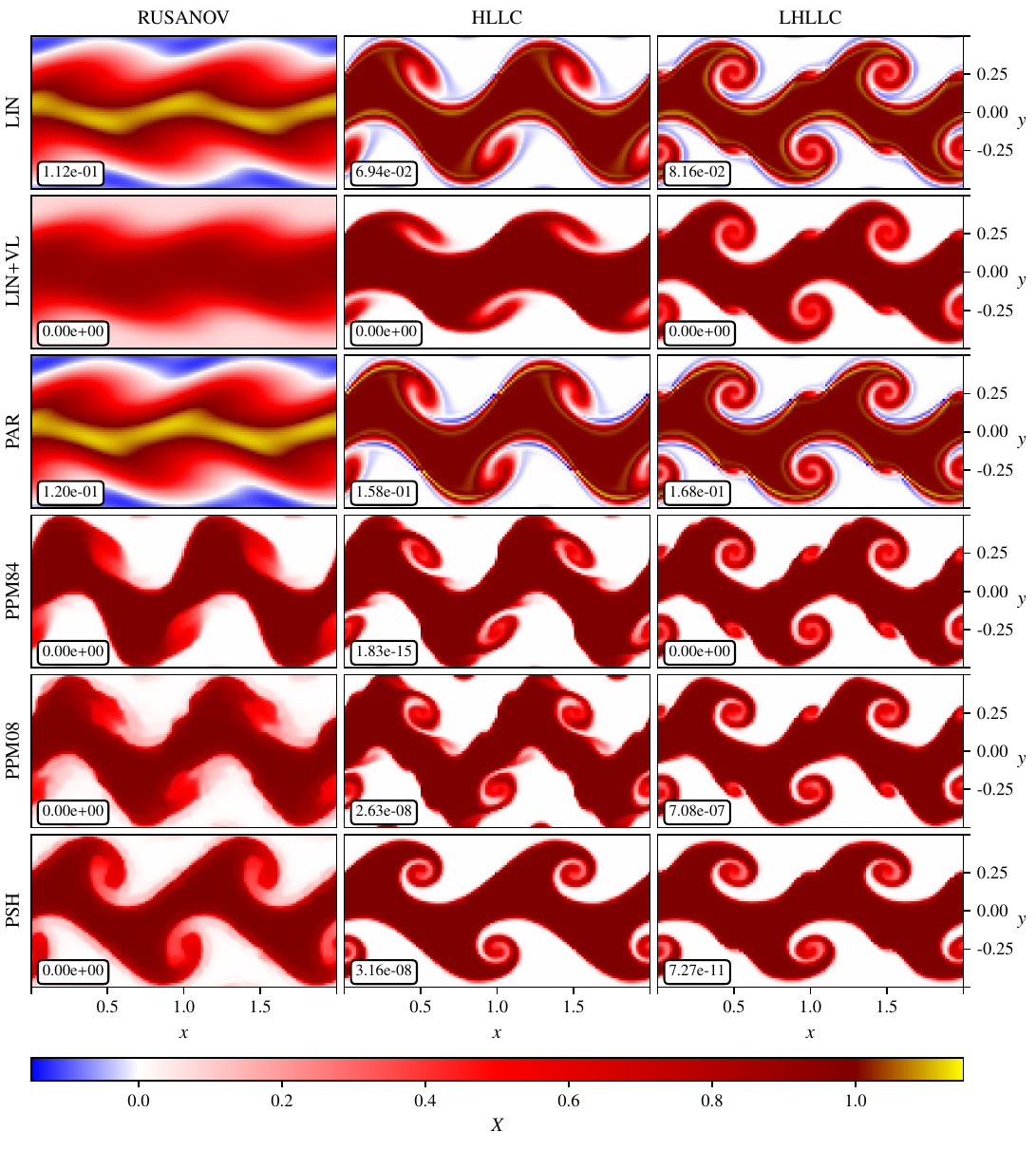}
		\caption{Distributions of the mass fraction $X$ of the passive scalar in simulations of the Kelvin--Helmholtz instability with the initial Mach number $\mach_0 = 10^{-2}$ on a $128 \times 64$ grid. The results are shown at $t = 0.8\,\mach_0^{\,-1}$. Rows and columns show different reconstruction schemes and numerical flux functions, respectively. The colour scale intentionally shows values $X < 0$ and $X > 1$ to highlight the overshoots that some of the schemes produce. The absolute value of the largest overshoot outside of this range is given in each panel.}
		\label{fig:kh2d-machx-1.000e-02-128x64-ps}
	\end{figure*}

	Figure~\ref{fig:kh2d-machx-1.000e-02-128x64-ps} compares the distributions of the passive scalar in simulations with $\mach_0 = 10^{-2}$ computed on a $128 \times 64$ grid. The steepest gradients are strongly under-resolved on this grid (see Fig.~\ref{fig:ps-scale-height}), which increases the amplitude and visibility of small-scale artefacts produced by different methods. All of the six reconstruction functions lead to extremely diffusive solutions with the RUSANOV flux, although high-order methods with limiters (PPM84, PPM08, PSH) still preserve steep gradients at some places. Numerical diffusion is strongly suppressed with the HLLC flux function owing to the explicit treatment of the contact wave in HLLC. The three highest-order methods (PPM84, PPM08, PSH) reproduce the structure of the primary vortices (c.f.\ Fig.~\ref{fig:kh2d-ref}) much more closely than the lower-order methods (LIN, LIN+VL, PAR). However, the two PPM methods develop secondary instabilities around the primary vortices. This effect, not present in the reference solution, occurs also with the LHLLC flux function \citep[see also][]{mcnally2012}. The secondary instabilities become the dominant source of numerical errors. We find that these instabilities tend to grow when excessive velocity shear is generated at the grid scale and there is not enough numerical dissipation to suppress their growth. Therefore, the excitation of artificial, short wavelength Kelvin--Helmholtz instabilities is favored on coarser grids, which generate larger shear at the grid scale across the poorly resolved slip line, and by less dissipative Riemann solvers. For the same reason, simulations run with the HLLC solver are more prone to developing secondary instabilities at higher rather than lower Mach numbers, as HLLC introduces less numerical diffusion into the system when modeling faster flows (see, e.g., the panels for PPM84+HLLC in Fig.~\ref{fig:kh2d-machx-1.000e-02-128x64-ps} and Fig.~\ref{fig:kh2d-machx-1.000e-01-128x64-ps}). As the grid is refined, the shear layers are progressively better resolved thus reducing grid scale shear and suppressing the growth of the secondary instabilities. Not surprisingly, the smooth interiors of the primary vortices are best represented with the highest-order method PSH (c.f.\ Fig.~\ref{fig:kh2d-ref}). Figures~\ref{fig:kh2d-machx-1.000e-01-128x64-ps} and \ref{fig:kh2d-machx-1.000e-03-128x64-ps}, respectively, show that the differences between the three flux functions become smaller with $\mach_0 = 10^{-1}$ and much larger with $\mach_0 = 10^{-3}$. This is expected because the amount of numerical dissipation (relative to the flow of interest) introduced by the RUSANOV and HLLC flux functions increases with decreasing Mach number of the flow, see Sect.~\ref{sec:numerical-fluxes}. With $\mach_0 = 10^{-3}$, only the PSH method reproduces the basic structure of the primary vortices when combined with the HLLC flux. Comparing Fig.~\ref{fig:kh2d-machx-1.000e-02-128x64-ps} with Fig.~\ref{fig:kh2d-machx-1.000e-03-128x64-ps}, we see that the LHLLC flux produces results independent of the initial Mach number $\mathcal{M}_0$ because all of the flows are considerably subsonic. All of the methods we test converge to the same flow pattern with $\mach_0 = 10^{-2}$, see Fig.~\ref{fig:kh2d-machx-1.000e-02-4096x2048-ps} for solutions computed on the $4096 \times 2048$ grid\footnote{Similar plots for all initial Mach numbers and all computational grids are available on Zenodo (\url{https://zenodo.org/doi/10.5281/zenodo.10280900}).}.

	The magnitude of the largest overshoot (or undershoot) in the passive scalar is shown in the insets in Figs.~\ref{fig:kh2d-machx-1.000e-02-128x64-ps}, \ref{fig:kh2d-machx-1.000e-01-128x64-ps}, and \ref{fig:kh2d-machx-1.000e-03-128x64-ps}. Thanks to their use of limiters, the LIN+VL, PPM84, PPM08, and PSH methods largely eliminate the overshoots, making the methods useful for the advection of mass fractions. We observe only some accumulation of round-off-level overshoots (up to ${\approx}\,10^{-14}$) with LIN+VL, independently of the initial Mach number or grid resolution. The PPM84, PPM08, and PSH methods produce small but finite overshoots. At $t = 0.8 \mach_0^{-1}$, the magnitude of the largest overshoots is ${\approx}\,10^{-4}$ with PPM84, ${\approx}\,10^{-2}$ with PPM08, and ${\approx}\,10^{-6}$ with PSH across all grids and all initial Mach numbers. However, they only occur in a few cells or groups of cells scattered across the computational grid, so the total mass affected by this effect is negligibly small. Moreover, the amplitude of such overshoots, drops to the round-off level in many of our simulations as the resolution is increased. Unlimited methods produce substantially larger overshoots, reaching $8\%$ and $17\%$ for LIN and PAR, respectively. They are the result of the dispersion errors characteristic of linear schemes of $2$\textsuperscript{nd} or higher order, so they are nearly always present around sharp structures and affect a much larger amount of mass than the sporadic and isolated overshoots produced by PPM84, PPM08, and PSH. The benefits of the hybrid approach in PSH thus become obvious -- secondary instabilities are suppressed (as compared with PPM84 and PPM08) thanks to the absence of limiters for dynamic variables while mass fractions, for which limiters are used, are well behaved.
	%
	%
	%
	%
	
	We quantify numerical convergence of all of our methods by measuring $L_1$ errors with respect to the reference solution in the density $\rho$, kinetic energy $E_{\mrm{k},y}$ associated with motions perpendicular to the initial shear flow, and mass fraction $X$ of the passive scalar. We define the relative $L_1$ error for any quantity $q$ as
	\begin{align}
		L_1 = \frac{1}{\sigma^\mrm{ref}} \frac{\sum_{i=1}^{N_x} \sum_{j=1}^{N_y} \left| q_{i,j} - q_{i,j}^\mrm{ref} \right|}{N_x N_y},
	\end{align}
	where $q^\mrm{ref}$ is the reference solution and the sums run over the whole computational grid of $N_x \times N_y$ cells. The error is normalised using the standard deviation $\sigma^\mrm{ref}$ of $q^\mrm{ref}$. This choice is motivated by the fact that density fluctuations are much smaller than the mean density in our setup. The reference solution is always re-binned from its original, $8192 \times 4096$ grid, by repeatedly averaging groups of $2 \times 2$ neighbouring cells until the desired grid resolution is achieved.
	
\begin{figure*}
		\includegraphics[width=\linewidth]{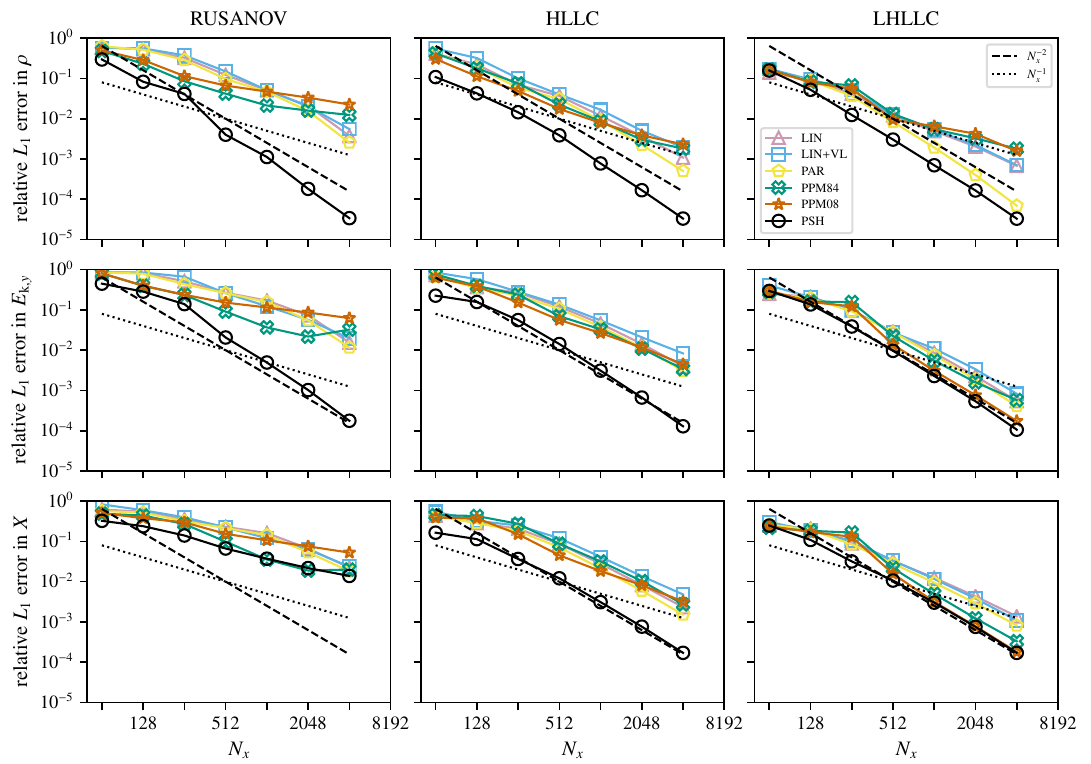}
		\caption{Relative $L_1$ errors for different variables (rows), flux functions (columns), and reconstruction methods (legend) as functions of grid resolution in the simulations of the Kelvin--Helmholtz instability with the initial Mach number $\mach_0 = 10^{-2}$. The dashed and dotted lines, which are at the same locations in all of the panels, show the $1$\textsuperscript{st}- and $2$\textsuperscript{nd}-order scalings to guide the eye.}
		\label{fig:kh2d-machx-1.000e-02-L1}
	\end{figure*}

	The $L_1$ errors for the set of simulations with the initial Mach number $\mach_0 = 10^{-2}$ are shown in Fig.~\ref{fig:kh2d-machx-1.000e-02-L1}. We first focus on the kinetic energy $E_{\mrm{k},y}$ and mass fraction $X$. The solutions computed using the LHLLC flux approach $2$\textsuperscript{nd}-order convergence, as expected. Only the PPM84 reconstruction method gives a slight decrease in the convergence rate in $E_{\mrm{k},y}$ on the finest of our grids. The same effect is observed in the sets of simulations with $\mach_0 = 10^{-1}$ (Fig.~\ref{fig:kh2d-machx-1.000e-01-L1}) and $\mach_0 = 10^{-3}$ (Fig.~\ref{fig:kh2d-machx-1.000e-03-L1}). The convergence rate of PPM08 also decreases on very fine grids in the latter case. Apart from this, the convergence curves are nearly independent of the initial Mach number with the LHLLC flux (c.f.\ Figs.~\ref{fig:kh2d-machx-1.000e-02-L1}, \ref{fig:kh2d-machx-1.000e-01-L1}, and \ref{fig:kh2d-machx-1.000e-03-L1}), confirming its low-Mach property. The highest-order reconstruction methods usually (but not always) produce the smallest errors on a given and sufficiently fine grid. The errors differ by up to one order of magnitude, although the overall 2D scheme is $2$\textsuperscript{nd}-order accurate in all of the cases.
	
	The magnitude of numerical errors strongly increases when we decrease initial Mach number with the non-low-Mach flux functions RUSANOV and HLLC. As described above, the spatial structure of errors produced by linear methods differs from that produced by methods with limiters. Figures~\ref{fig:kh2d-machx-1.000e-02-L1}, \ref{fig:kh2d-machx-1.000e-01-L1}, and \ref{fig:kh2d-machx-1.000e-03-L1} show that the linear methods LIN, PAR, and for dynamic variables also PSH approach $2$\textsuperscript{nd}-order convergence on sufficiently fine grids even if the magnitude of the errors is much larger than what we obtain with the LHLLC flux function.
	
\begin{figure*}
		\includegraphics[width=\linewidth]{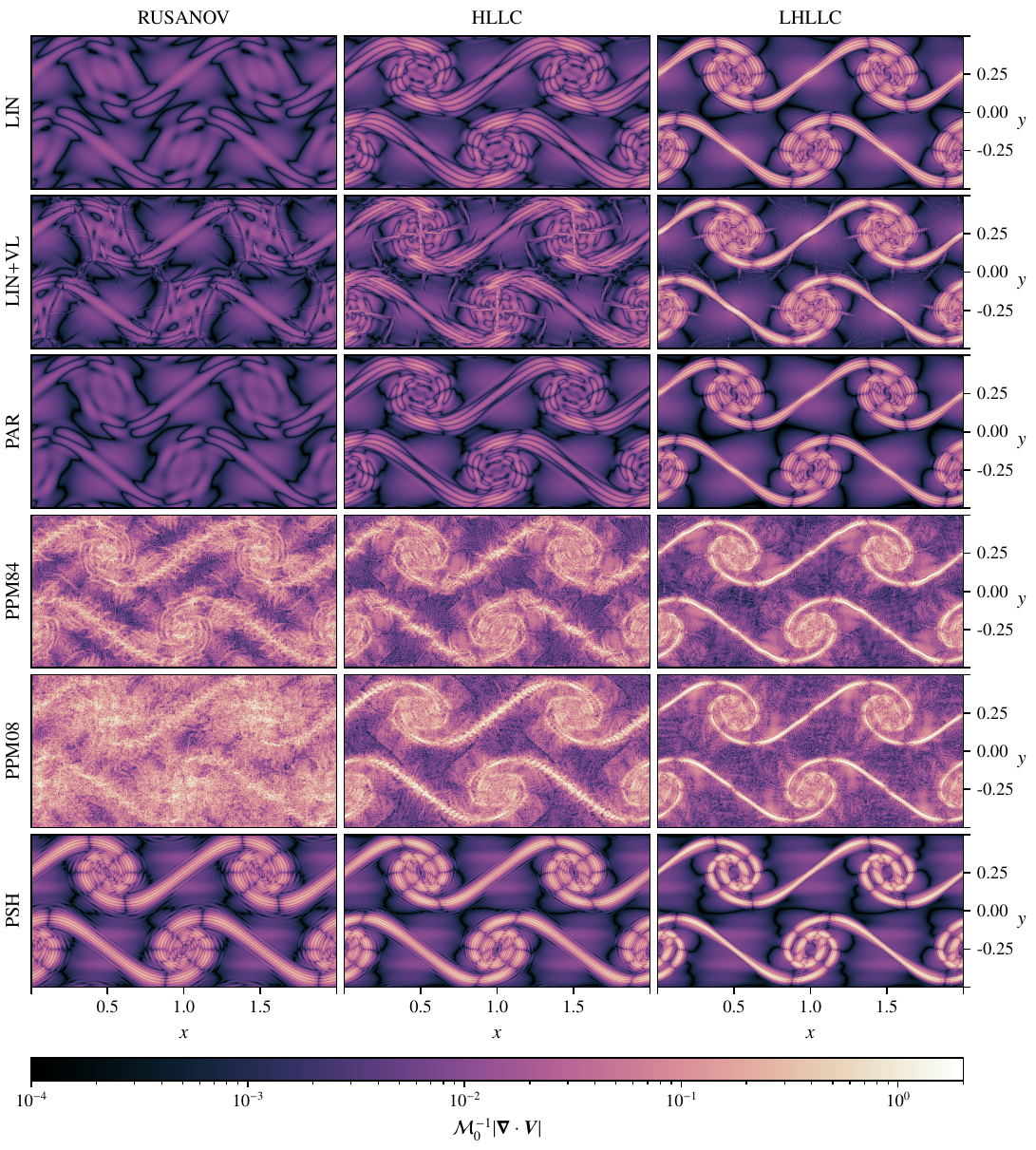}
		\caption{Magnitude of velocity divergence in simulations of the Kelvin--Helmholtz instability with the initial Mach number $\mach_0 = 10^{-2}$ on a $512 \,{\times}\, 256$ grid. The results are shown at $t = 0.8\,\mach_0^{\,-1}$. Rows and columns show different reconstruction schemes and numerical flux functions, respectively. The small-scale structure produced by the LIN+VL, PPM84, and PPM08 schemes, which does not occur with the LIN, PAR, and PSH schemes, is caused by the use of limiters.}
		\label{fig:kh2d-machx-1.000e-02-512x256-abs_div_u}
	\end{figure*}

	The two PPM methods converge poorly in $\rho$ at low Mach numbers (Figs.~\ref{fig:kh2d-machx-1.000e-02-L1} and \ref{fig:kh2d-machx-1.000e-03-L1}). Because the initial density is constant everywhere, all of the density fluctuations are the integrated effects of the divergence (or convergence) of the velocity field in the continuity equation. The magnitude of the velocity divergence in simulations computed on the $512 \times 256$ grid is shown in Figs.~\ref{fig:kh2d-machx-1.000e-02-512x256-abs_div_u}, \ref{fig:kh2d-machx-1.000e-01-512x256-abs_div_u}, and \ref{fig:kh2d-machx-1.000e-03-512x256-abs_div_u}. The solutions computed using the LIN, PAR, and PSH methods clearly show the structure of the primary vortices with some oscillations in the shear layers and a background of relatively weak, large-scale sound waves. On the other hand, the solutions computed using the PPM84 and PPM08 methods show a large amount of small-scale ``numerical noise''. After inspecting the time dependence of these artificial structures\footnote{See an animation available on Zenodo (\url{https://zenodo.org/doi/10.5281/zenodo.10280900}).}, we conclude that some of these structures travel with the flow while others have the character of small-scale sound waves. We believe that both originate from cumulative effects of the limiters switching their local state many times over the time span of the simulation\footnote{The animations show high values of the velocity divergence developing early on where the second derivative of the transition function $\eta(y)$ (Eq.~(\ref{eq:eta})) is discontinuous. This higher-order discontinuity may influence the behaviour of limiters.}. The switching can result in rapid changes in the magnitude of the discontinuities at cell faces and, consequently, in the amount of dissipation applied in the Riemann solver (i.e. the numerical flux function). This effect is weakest with the low-dissipation flux function LHLLC but it is still clearly present on the $4096 \times 2048$ grid, see Fig.~\ref{fig:kh2d-machx-1.000e-02-4096x2048-abs_div_u}. The structures disappear when we disable the limiters in PPM84 and PPM08 (not shown in the figures). We do not observe the fast-propagating, small-scale sound waves with LIN+VL but the method does produce thin structures in the velocity divergence around the primary vortices, which follow the flow.
	
	Overall, the PSH method produces by far the smallest errors with the RUSANOV and HLLC flux functions and initial Mach numbers $\mach_0 \le 10^{-2}$ (Figs.~\ref{fig:kh2d-machx-1.000e-02-L1} and \ref{fig:kh2d-machx-1.000e-03-L1}). The only exception is the poor convergence (or even divergence) in the passive tracer $X$ observed with the PSH+RUSANOV combination in simulations with $\mach_0 \le 10^{-2}$ performed on very fine grids. This effect is also likely related to the limiters switching their local state many times over the time span of the simulation but we did not investigate it further.
	
	\subsection{3D simulations of convection, turbulent entrainment, and wave excitation}
	\label{sec:code-comparison}
	
	\begin{figure*}
		\centering
		\includegraphics[width=0.5\textwidth]{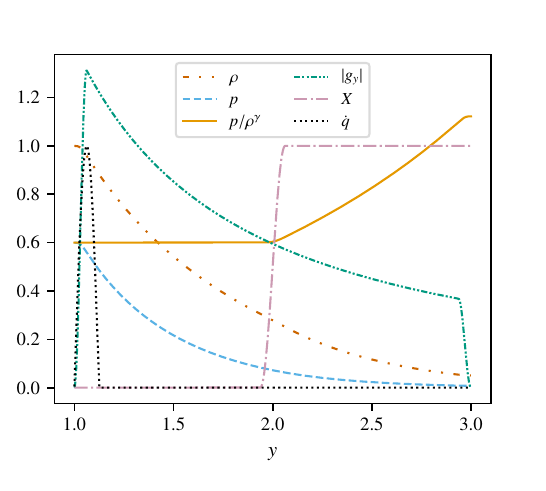}
		\caption{Hydrostatic stratification of density, gas pressure, pseudo-entropy ($p/\rho^\gamma$), and gravity at $t=0$ in the problem of turbulent convection and wave excitation at a convective boundary, based on the initial conditions provided in \cite{andrassy2022}. The distribution of the mass fractional abundance of a passive tracer ($X$) and the heat source  ($\dot{q}$) are also shown, here represented by a dash-dotted pink and dashed black line, respectively. In this setup, the gravitational acceleration smoothly turns to zero at the bottom and top boundaries of the spatial domain to make the problem consistent with the usage of reflecting, stress free boundary conditions, which for the gas pressure imply $\partial p / \partial y = 0$.}
		\label{fig:cc-stratification}
	\end{figure*}

	In this section, we consider a test problem in which a variety of complex hydrodynamic phenomena can be captured on the same computational grid. These include convective transport of energy, turbulent mixing at a convective boundary, and the propagation of internal waves in a stably stratified layer, all of which are often encountered in simulations of geophysical and stellar hydrodynamics. The initial conditions for this test are adopted from the work of \cite{andrassy2022} and they represent a hydrostatic solution of Eq.~(\ref{eq:euler-int}). In particular, the thermodynamic conditions in this setup are similar to those found in an oxygen-burning shell of a massive star. We map the hydrostatic stratification (see Fig.~\ref{fig:cc-stratification}) on an evenly spaced, 3D, Cartesian grid defined by $(x,y,z) \in [-1,1]\times [1,3] \times [-1,1]$. The gravitational acceleration points downward in the $y$-direction,
	\begin{equation}
		g_y = g_0 f_\mathrm{g}(y)y^{-5/4},
	\end{equation}
	where $g_0=-1.414870$ and
	\begin{equation}
		f_\mathrm{g}(y) = 
		\begin{cases}
			\frac{1}{2} \Big\{ 1 + \mathrm{sin} \Big[ 16\pi \Big( y-\frac{1}{32} \Big) \Big] \Big\}, & \mathrm{for}\ 1 \le y < 1+\frac{1}{16},  \\
			1, & \mathrm{for}\ 1+\frac{1}{16} \le y \le 3-\frac{1}{16} \\
			\frac{1}{2} \Big\{ 1 + \mathrm{sin} \Big[ 16\pi \Big( y-\frac{1}{32} \Big) \Big] \Big\}, & \mathrm{for}\ 3-\frac{1}{16} < y \le 3.
		\end{cases}
	\end{equation}
	As in \cite{andrassy2022}, we work with rescaled quantities such that the speed of sound and the density at the base of the box are taken as units of velocity and density, and the thickness of the convective oxygen shell  as unit of length. A detailed list of units with the associated values can be found in Table 1 of \cite{andrassy2022}. 
	
	The initially hydrostatic stratification is described by a piecewise-polytropic relation in the form
	\begin{equation}
		\frac{\partial\  \mathrm{ln}\  p}{\partial\  \mathrm{ln}\  \rho} = 
		\begin{cases}
			\gamma_0, & \mathrm{for}\ 1 \le y < 2-\frac{1}{16},  \\
			\gamma_0 + \eta(y)(\gamma_1-\gamma_0), & \mathrm{for}\  2-\frac{1}{16} \le y \le 2+\frac{1}{16},  \\
			\gamma_1, & \mathrm{for}\ 2+\frac{1}{16} < y \le 3,  \\
		\end{cases}
	\end{equation}
	where $\gamma_0=5/3$, $\gamma_1=1.3$, and $\eta(y)$ is a smooth function,
	\begin{equation}
		\eta(y) = 
		\begin{cases}
			0, & \mathrm{for}\ 1 \le y < 2-\frac{1}{16},  \\
			\frac{1}{2}[ 1 + \mathrm{sin}(8\pi y) ], & \mathrm{for}\  2-\frac{1}{16} \le y \le 2+\frac{1}{16},  \\
			1, & \mathrm{for}\ 2+\frac{1}{16} < y \le 3.  \\
		\end{cases}
	\end{equation}
	The adiabatic index is $\gamma\,{=}\,\gamma_0$. In this work, we assume periodic boundary conditions in the horizontal $x$- and $z$-direction, whereas reflecting, stress-free boundaries are used in the vertical $y$-direction. This problem is set up such that turbulent convective flows develop in the lower half of the domain, which is initially adiabatic, while internal gravity waves are free to propagate in the upper, stably stratified layer. To keep track of the position of the upper convective boundary, at $t\,{=}\,0$ we fill the stable layer with a passive scalar whose abundance smoothly turns to 0 across the upper boundary of the adiabatic region according to $X\,{=}\,\eta(y)$.
	
	In order to drive the convection, we include a time-independent heat source that continuously injects energy into the system close to the base of the box. The  rate of energy released per unit volume,
	\begin{equation}
		\dot{q}(y) = 
		\begin{cases}
			\dot{q}_0\mathrm{sin}(8\pi y)\dfrac{\mathrm{sin}(4\pi\Delta x)}{(4\pi\Delta x)}, & \mathrm{for}\ 1 \le y < 1+\frac{1}{8},  \\
			0, & \mathrm{for}\ 1+\frac{1}{8} \le y \le 3,  \\
		\end{cases}
	\end{equation}
	is added to the right-hand side of Eq.~(\ref{eq:euler-int}) as 
	\begin{equation}
		\bm{S}\mapsto \bm{S} +
		\begin{bmatrix}
			0 \\
			0\\
			0\\
			0\\
			\dot{q} \\
			0\\
		\end{bmatrix}.
	\end{equation}
	In the work of \cite{andrassy2022}, the amplitude of the heat source was $\dot{q}_0=3.795720\times 10^{-4}$, which gave rise to convective flows with a root-mean-square Mach number $\mathcal{M}_\mathrm{rms}\,{\approx}\,0.04$. To make the problem more challenging, here we decrease $\dot{q}$ by a factor of ten, so $\dot{q}_0=3.795720\times 10^{-5}$. The lower heating rate, according to the well-established  $\mathcal{M} \propto \dot{q}_0^{1/3}$ relation  \citep[see, e.g.,][]{woodward2015, kapyla2021, horst2021a}, should drive convection at $\mathcal{M}_\mathrm{rms}\,{\approx}\, 0.02$. 
	
	In this setup, internal gravity waves (IGWs) are excited by the interaction of the convective flows with the bottom boundary of the stably stratified layer. The wavelength of IGWs in the direction of gravity becomes shorter when the waves are excited at progressively lower temporal frequencies \citep{sutherland2010}.  At the heating rate we consider, the most prominent IGWs that originate at the convective boundary are only barely spatially resolved on the coarsest of our grids with $128^3$ cells. Therefore, we decide not to decrease $\dot{q}_0$ even further for these simulations because it would give rise to convective flows with lower characteristic frequencies and lead to the generation of unresolved IGWs in the stable layer. Due to the fully compressible nature of \code{SLH}, we also expect short-wavelength sound waves to be generated, although at much lower amplitudes than those of IGWs at the typical Mach numbers encountered in this test problem \citep{lighthill1952}.

	As done for the test described in Sect.~\ref{sec:results_kh}, here we run simulations for each of the 18 considered combinations of Riemann solvers and spatial reconstruction schemes. To judge the numerical convergence of our results, each combination of methods is run on grids with $128^3$ and $256^3$ cells. Additionally, we run a single simulation on a $512^3$ grid using the LHLLC Riemann solver and the PAR reconstruction scheme, which we consider the reference solution for this test problem. However, because of the chaotic nature of the turbulent flows that arise in the convective layer, convergence is not expected in the exact flow morphology, so we do not compute $L_1$ errors as done in Sect.~\ref{sec:results_kh}. Instead, we analyze the convergence of the numerical results in terms of ensemble-averaged quantities that are representative of the dynamical properties of the system, such as kinetic energy spectra computed in the convective and stable layers.
	
\begin{figure*}
		\centering
		\includegraphics[width=0.9\textwidth]{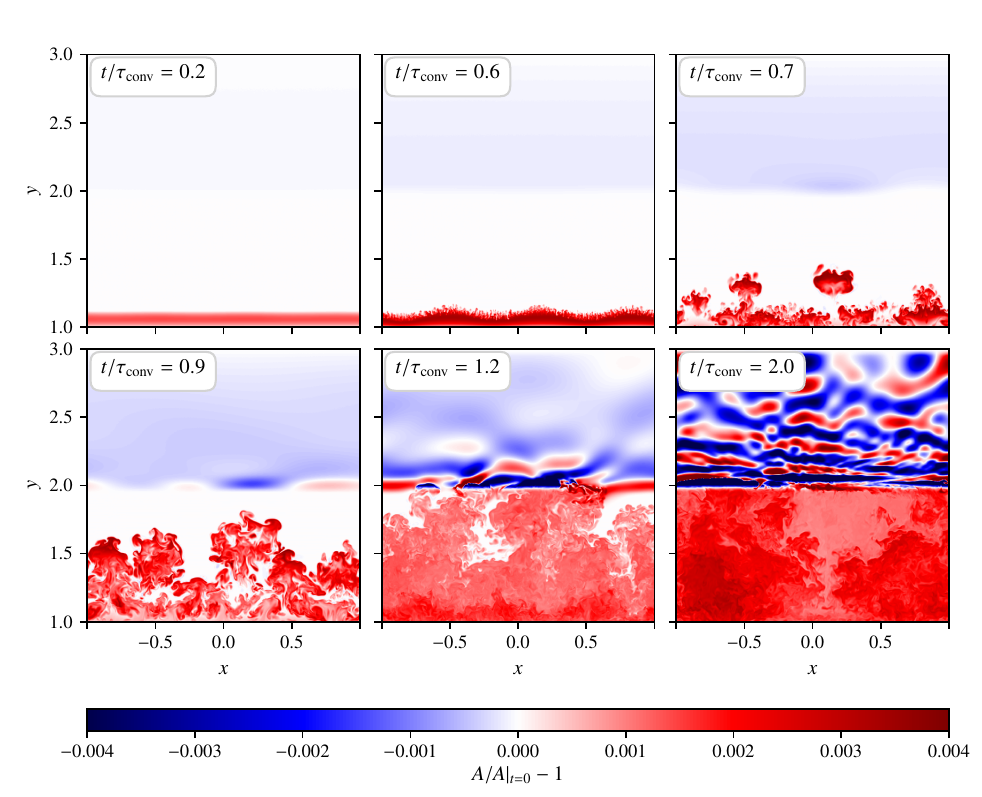}
		\caption{Pseudo-entropy ($A=p/\rho^\gamma)$ fluctuations with respect to the hydrostatic, background state computed at different times during the development of the convection, as indicated by the insets, in the problem of turbulent convection and wave excitation. Data are visualized in the $z=0$ plane. This simulation was run using the PAR method and LHLLC on a $512^3$ grid and we consider it our reference solution. Convective plumes start exciting IGWs at the upper convective boundary by $t\approx \tau_\mathrm{conv}$ (see center bottom panel).}\label{fig:initial-transient}
	\end{figure*}

	To break the initial symmetry, we add a perturbation to the hydrostatic density stratification in the form
	\begin{equation}
		\Delta \rho = 1.1\times10^{-5}\frac{\dot{q}(y)}{\dot{q}_0} [\mathrm{sin}(3\pi x)+\mathrm{cos}(\pi x)][\mathrm{sin}(3\pi z) - \mathrm{cos}(\pi z)].
	\end{equation}
	The subsequent evolution of the system and the development of convection is shown in Fig.~\ref{fig:initial-transient}. The density perturbation, alongside the action of the heat source, generate packets of fluid with higher entropy content than the adiabatic surroundings. The packets of hot and low-density material buoyantly rise in the adiabatic stratification until they reach $y\,{\approx}\,2$. At this height, the temperature stratification turns subadiabatic and the buoyant force acting on the rising plumes changes sign, forcing them to overturn. IGWs excited at the bottom boundary of the subadiabatic region propagate upward in the stratification (with characteristic Mach numbers in the range from 0.005 to 0.01) and are subsequently reflected at the top boundary of the domain. Shear instabilities break the large-scale buoyant structures that arise in the adiabatic layer and initiate the cascade of kinetic energy toward smaller scales. Turbulent convection fully develops after approximately one convective turnover time scale, $\tau_\mathrm{conv} = 133$ time units, which we define according to 
	\begin{equation}\label{eq:conv-to}
		\tau_\mathrm{conv} = \frac{2L_\mathrm{conv}}{\langle|\bm{V}|_\mathrm{rms}\rangle}.
	\end{equation}
	In Eq.~(\ref{eq:conv-to}), the root-mean-square convective speed is averaged over several convective turnover time scales and $L_\mathrm{conv}\,{=}\,1$ is taken as representative of the vertical extent of the convection zone. In our reference simulation, the root-mean-square Mach number in the convection zone is\footnote{The error bar represents one standard deviation computed over the last 20 convective turnovers.} $\mathcal{M}_\mathrm{rms}=0.019\pm0.001$, which is in agreement with the value predicted by the $\mathcal{M}_\mathrm{rms} \propto \dot{q}^{1/3}_0$ scaling relation. All simulations are run until $t_\mathrm{max}\,{=}\,32 \tau_\mathrm{conv}$ to have a proper coverage of the dynamical evolution of the system and to compute meaningful time averages needed for the following analysis.

	\subsubsection{Flow morphology and spatial kinetic energy spectra}
	
\begin{figure*}
		\centering
		\includegraphics[width=0.905\textwidth]{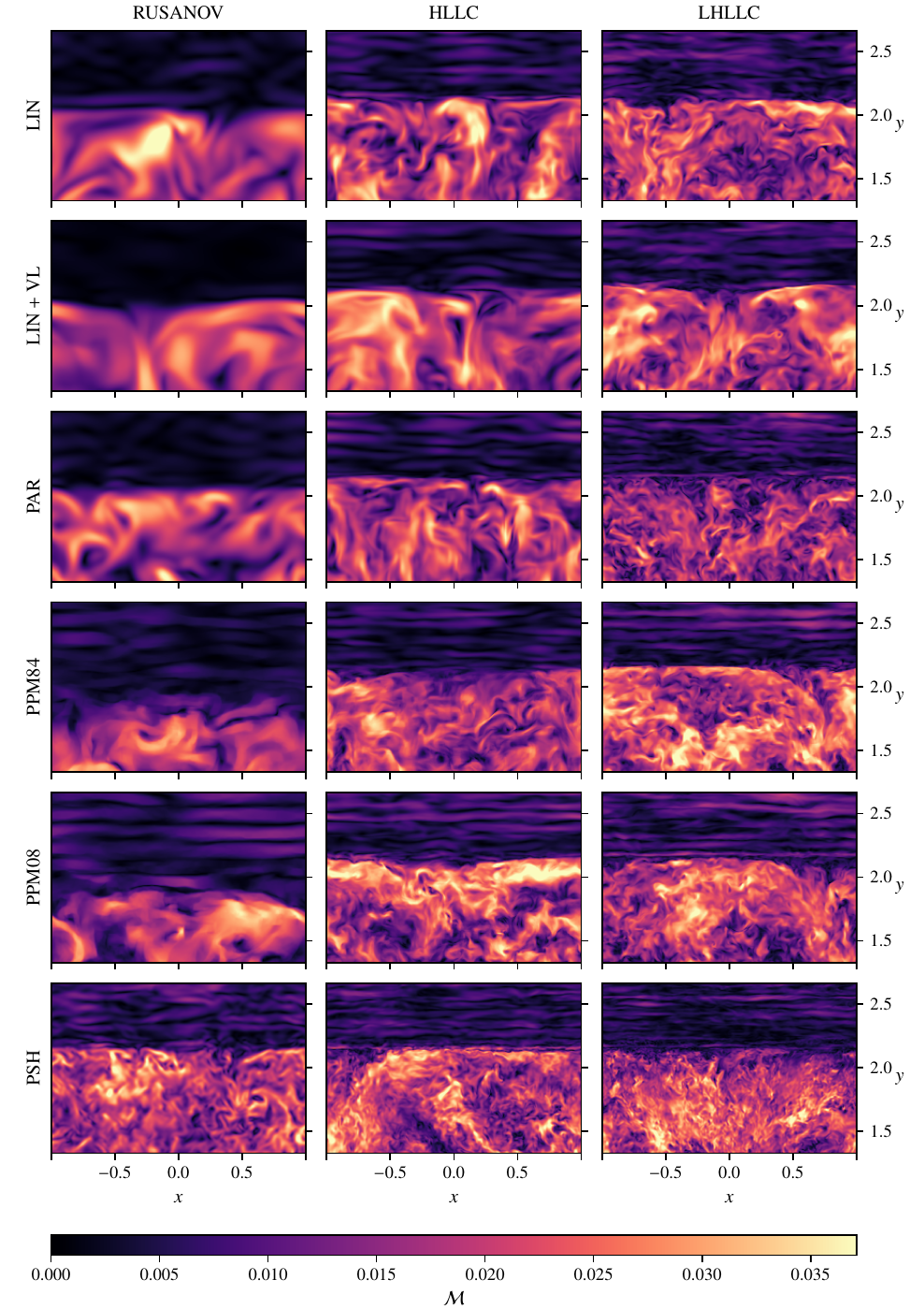}
		\caption{Snapshots obtained at the final simulation time ($t=32\tau_\mathrm{conv}$) showing the distribution of the Mach number at $z=0$ in the simulations of turbulent convection and wave propagation in a 3D box. The grid resolution is $256^3$. Each panel is cut between $y=1.3$ and $y=2.7$ for a better visualization of the flows in the proximity of the upper convective boundary.}
		\label{fig:mach-z-0.0-cuts}
	\end{figure*}

	In Fig.~\ref{fig:mach-z-0.0-cuts}, we show snapshots of the Mach number taken at the vertical plane $z\,{=}\,0$, obtained at the final time $t\,{=}\,32\tau_\mathrm{conv}$ with all the methods tested here. Although all panels show results computed using the same grid resolution ($256^3$), there is a vast difference between the methods in terms of effects induced by numerical dissipation.  With the RUSANOV flux, convection mainly happens in the form of large-scale plumes and IGWs are barely excited in the stable layer except when high-order methods such as PSH or PPM08 are used to reconstruct the Riemann states. Smaller-scale structures typical of turbulent flows become progressively more visible in the convection zone with higher-order reconstruction methods and less dissipative solvers. Internal waves with progressively shorter wavelengths also tend to be more visible in the stable layer with less dissipative methods.
	
\begin{figure*}
		\includegraphics[width=0.9\textwidth]{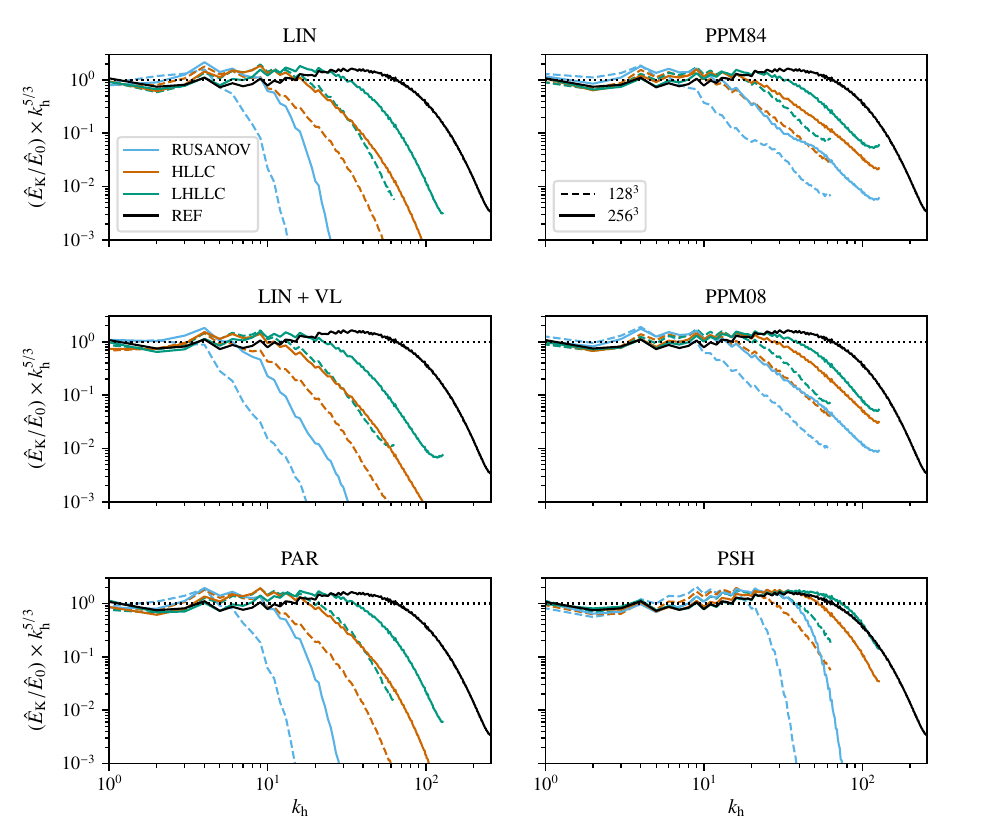}
		\caption{Kinetic energy spectra computed on the horizontal plane $y=1.5$ of the convective layer for all of our 18 combinations of Riemann solvers and spatial reconstruction schemes in the problem of turbulent convection and wave excitation. The spectra have been averaged over the time interval $t\in(10\tau_\mathrm{conv},32\tau_\mathrm{conv}$) and rescaled by the Kolmogorov law ($k_\mathrm{h}^{-5/3}$) and by the value $\hat{E}_0$ of the spectral energy density of the reference run at $k_\mathrm{h}=15$.}
		\label{fig:spectra-Y-1.5}
	\end{figure*}

\begin{figure*}
		\includegraphics[width=0.9\textwidth]{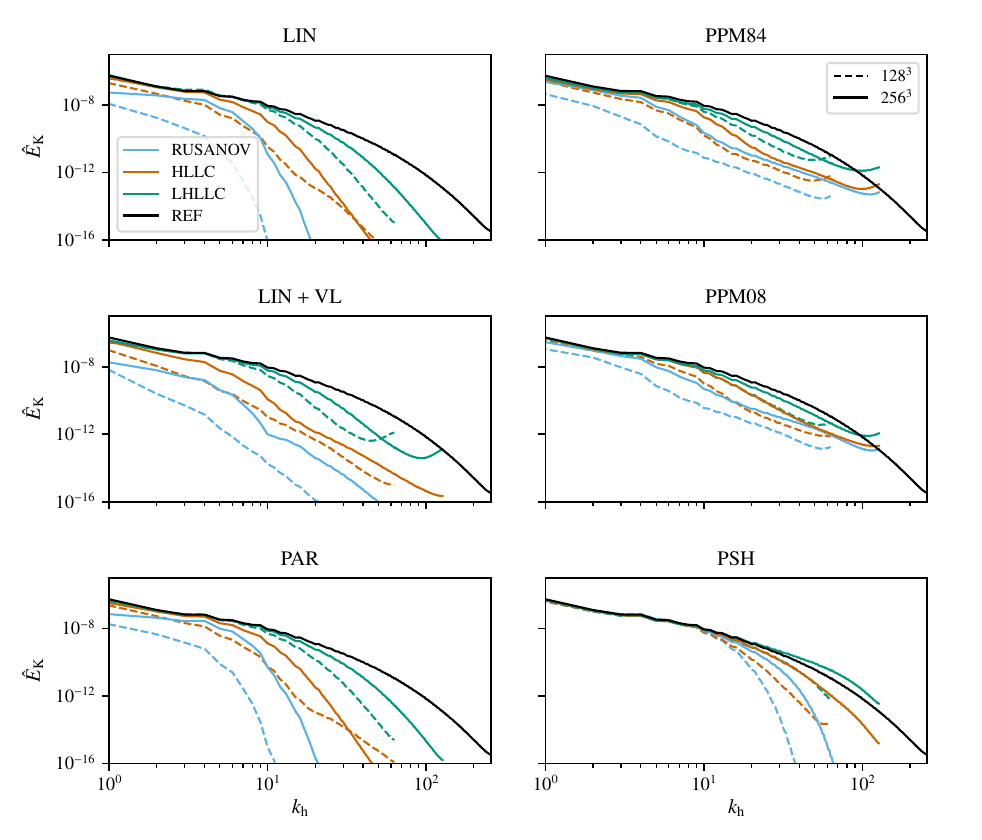}
		\caption{Kinetic energy spectra computed on the horizontal plane $y=2.5$ of the stable layer for all of our 18 combinations of Riemann solvers and spatial reconstruction schemes in the problem of turbulent convection and wave excitation. The spectra have been averaged over the time interval $t\in(10\tau_\mathrm{conv},32\tau_\mathrm{conv}$). The black solid line represents the reference run ($\mathrm{REF}$), which is computed using the $\mathrm{LHLLC}$ Riemann solver and the $\mathrm{PAR}$ reconstruction scheme on a $512^3$ grid.}
		\label{fig:spectra-Y-2.5}
	\end{figure*}

	The effects of the numerical dissipation on the properties of the flow are better quantified in Fig.~\ref{fig:spectra-Y-1.5} and \ref{fig:spectra-Y-2.5}, where we show the power spectra of kinetic energy extracted from a horizontal plane in the middle of the convection zone ($y\,{=}\,1.5$) and stable layer ($y\,{=}\,2.5$) as functions of the horizontal wavenumber, $k_\mathrm{h} \,{=}\,  \sqrt{k_x^2+k_z^2}$. We compute $k_x$ and $k_z$  as 
	\begin{align}
		k_x &= 
		\begin{cases}
			m, & \mathrm{for}\ 0 \le m \le \floor*{\frac{N_x - 1}{2}},  \\
			-N_x + m, & \floor*{\frac{N_x - 1}{2}} < m  < N_x, 
		\end{cases} \\
		k_z &= 
		\begin{cases}
			n, & \mathrm{for}\ 0 \le n \le \floor*{\frac{N_z - 1}{2}},  \\
			-N_z + n, & \floor*{\frac{N_z - 1}{2}} < n  < N_z, 
		\end{cases} 
	\end{align}
	where $\floor*{.}$ represents the floor function and $N_x$ and $N_z$ are the numbers of cells in the $x-$ and $z-$direction, respectively. The spectra are averaged over the time interval $t \,{\in}\, (10 \tau_\mathrm{conv}, 32 \tau_\mathrm{conv})$. In the convection zone, all of the curves approximately agree with the Kolmogorov scaling law ($k_\mathrm{h}^{-5/3}$) on some intermediate scales. We notice that the kinetic energy spectra shown in Fig.~\ref{fig:spectra-Y-1.5} present a pile-up of kinetic energy at the bottom of the inertial range. This phenomenon, also known as ``bottleneck effect'', is often observed both in hydrodynamic simulations \citep[see, e.g.,][]{dobler2003} and experimental studies \citep[see, e.g.,][]{kuchler2019} of turbulent flows. The extent of the inertial range of the turbulent kinetic energy spectrum greatly differs from method to method. With a fixed spatial reconstruction scheme, the kinetic energy spectrum obtained with LHLLC enters the viscous sub-range (where the  kinetic energy of the turbulent eddies is dissipated into internal energy  of the gas) at higher wavenumbers as compared to both HLLC and RUSANOV. Reconstruction schemes based on slope limiters are characterized by shallower spectra in the viscous sub-range than those generated by unlimited schemes. In the stable layer (see Fig.~\ref{fig:spectra-Y-2.5}), RUSANOV and HLLC generate much weaker IGWs than the reference run even on large scales, except when used in combination with PSH. Both in the convective and stable layers, all spectra converge to the reference solution if the order of the spatial reconstruction method or the grid resolution are increased. 
	
	\subsubsection{Turbulent entrainment at the convective boundary}
	
	The overturning of the convective flows at the upper convective boundary gives rise to a variety of hydrodynamic processes (including shear instabilities, convective overshoot, and breaking of surface waves) that lead to the entrainment of material from the upper, stably stratified layer into the convection zone. The conglomeration of these processes is known in the literature as ``convective boundary mixing'' \citep[see, e.g.,][]{meakin2007,andrassy2020,horst2021a,anders2023} and it increases the size of convective regions over time. As part of our comparison study, we quantify the impact of the choice of a specific combination of methods in Godunov-type schemes on the evolution of the convective boundary. At any given point in time, we assume that the vertical coordinate of the boundary between the convective and stable layer, $y_\mathrm{cb}$, is the position at which the horizontal average of the passive scalar $X$ takes the value
	\begin{equation}
		\tilde{X} = \frac{1}{2}(X_\mathrm{min}+X_\mathrm{max}),
	\end{equation}
	where $X_\mathrm{min}$ and $X_\mathrm{max}$ are the minimum and maximum values of $X$, respectively. Such a choice for $y_\mathrm{cb}$ is justified by the fact that the initial abundance of the passive scalar, $X$, smoothly transitions from 0 to 1 at $y\,{\approx}\,2$, which is the boundary of the initially adiabatic layer where convection sets in first. The time evolution of $y_\mathrm{cb}$ is shown in Fig.~\ref{fig:ycb} for all of our combinations of methods.  We note that PPM-based methods used in combination with $\mathrm{RUSANOV}$ do not show signs of numerical converge to the reference solution. Also, the other reconstruction schemes used with $\mathrm{RUSANOV}$ seem to accelerate the mixing at the convective boundary with respect to the reference run, especially on the $128^3$ grid. In reality, this is an artifact of the method used to estimate the position of the convective boundary $y_\mathrm{cb}$. In fact, because $\mathrm{RUSANOV}$  does not resolve shear or contact waves (see Sect.~\ref{sec:numerical-fluxes}), the initial gradient in $X$ at the convective boundary is further smoothed out by the effects of numerical dissipation. When convection fully develops, it mixes the passive scalar that has diffused inside the convection zone and homogenizes its abundance, thus shifting the formal position of $X=\tilde{X}$ outward. This effect is further enhanced with the most diffusive reconstruction methods tested here. Therefore, with the $\mathrm{RUSANOV}$ solver, entrainment of material from the stable to the convective layer is mostly determined by numerical diffusion rather than turbulent mixing and the distribution of the passive tracer is not representative of $y_\mathrm{cb}$. On the other hand, all of the other methods tested here quickly achieve numerical converge to the reference solution with increasing grid resolution, the order of the reconstruction scheme, or by using progressively less dissipative Riemann solvers. Among the data sets obtained with the six reconstruction schemes, the maximum relative deviation of $y_\mathrm{cb}$  from the reference solution at the final time is $20\%$ and $15\%$ with $\mathrm{HLLC}$ on the $128^3$ and $256^3$ grids, respectively, while with $\mathrm{LHLLC}$ it decreases from $12\%$ to $5\%$ with increasing grid resolution.
	
	\subsubsection{Shape of the convective boundary}

\begin{figure*}
		\includegraphics[width=0.96\textwidth]{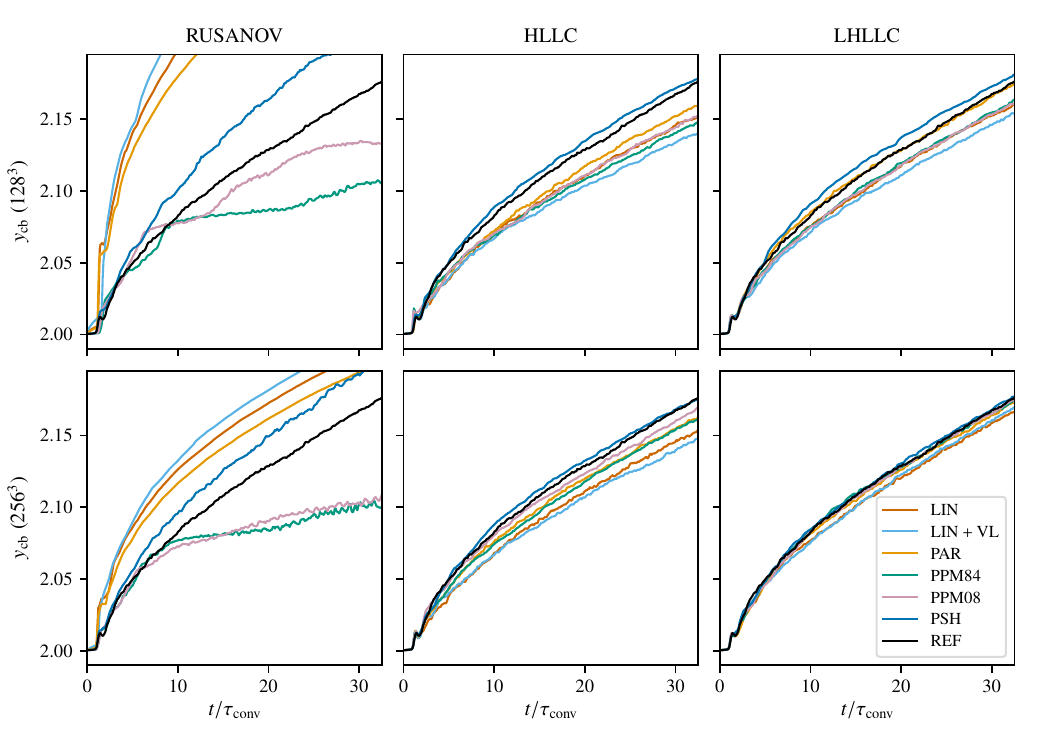}
		\caption{Time evolution of the position of the upper convective boundary $(y_\mathrm{cb})$ in the problem of turbulent convection and entrainment described in Sect.~\ref{sec:code-comparison} The top and bottom rows show results from $128^3$ and $256^3$ simulations, respectively. The black solid line represents the reference run ($\mathrm{REF}$).}
		\label{fig:ycb}
	\end{figure*}

\begin{figure*}
		\centering
		\includegraphics[width=0.84\textwidth]{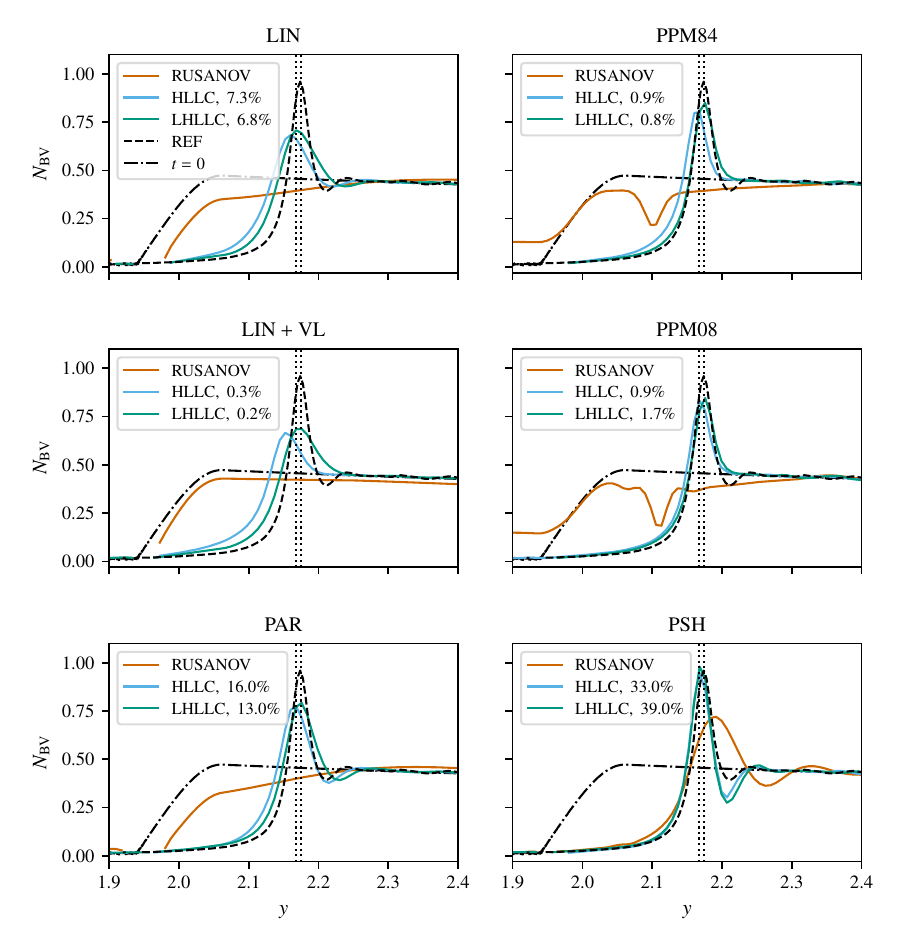}
		\caption{Vertical profiles of the Brunt-V\"ais\"al\"a frequency ($N_\mathrm{BV}$) averaged over the time interval $t\in(30\tau_\mathrm{conv},32\tau_\mathrm{conv}$) in the problem of turbulent convection and wave excitation. Each panel shows the results of simulations run using the same reconstruction scheme but different Riemann solvers. Here, we only show results from the $256^3$ grid to avoid cluttering the figures. The reference run ($\mathrm{REF}$) is represented by a black dashed line. The black dash-dotted line is the profile of the Brunt-V\"ais\"al\"a frequency at $t=0$, and the vertical black dotted lines represent the position of the convective boundary at the beginning and the end of the chosen averaging time interval in the reference run, 
			$y=2.168$ and $y=2.174$, respectively. The percentages shown in the insets for simulations run with the $\mathrm{HLLC}$ and $\mathrm{LHLLC}$ solvers represent the amplitude of the largest undershoot in $N_\mathrm{BV}$ relative to $N_{\mathrm{BV},t=0}$, in the spatial range $y \in(2.15,2.40)$. The amplitude of the largest undershoot in the reference run is $12\%$.}
		\label{fig:brunt}
	\end{figure*}

	The properties of the upper convective boundary are also studied by means of horizontal averages in the Brunt-V\"ais\"al\"a frequency, which determines both the spectrum of convectively driven IGWs \citep[see, e.g.,][]{sutherland2010} and the rate of mass entrainment from the stable layer according to the bulk-Richardson-number mixing model \citep[see, e.g.,][]{meakin2007,andrassy2020,rizzuti2023}. Here, the Brunt-V\"ais\"al\"a frequency is computed as
	\begin{equation}
		N_\mathrm{BV} = \Bigg[ -\frac{g_y}{H_p}(\nabla_\mathrm{ad}-\nabla) \Bigg]^{1/2},
	\end{equation}
	where 
	\begin{equation}
		\nabla = \frac{\mathrm{d}\ \mathrm{ln} \ T}{\mathrm{d}\ \mathrm{ln} \ p},
	\end{equation}
	$T=p/\rho$ is the temperature of the gas\footnote{Here, we assume that the gas constant is unity.}, $H_p \,{=}\, - \left(\mrm{d} \ln p / \mrm{d} y\right)^{-1}$ is the local pressure scale height, and $\nabla_\mathrm{ad}=0.4$ is the adiabatic temperature gradient. We average the profiles of $N_\mathrm{BV}$ over $t \in (30\tau_\mathrm{conv},32\tau_\mathrm{conv})$ so that the change in the position of the convective boundary is smaller than its width. At the same time, the chosen averaging time interval is long enough to partly suppress the statistical fluctuations induced by the turbulent nature of the convective flows. The results are shown in Fig.~\ref{fig:brunt}. As convection sets in and entrains material from the upper, stably stratified layer, it steepens the vertical entropy gradient $\mathrm{d}s/\mathrm{d}y$ across the convective boundary. Therefore, due to the $N^2_\mathrm{BV} \propto \mathrm{d}s/\mathrm{d}y$  proportionality  \citep[see, e.g.,][]{maeder2009}, the vertical profile of the Brunt-V\"ais\"al\"a frequency develops a spike in the region close to the convective boundary and it quickly approaches  $N_\mathrm{BV,}^{t=0}$ in the rest of the stable layer. The fact that, in this setup, $N_\mathrm{BV}$ is characterized by large vertical gradients near the convective boundary (at least with $\mathrm{HLLC}$ and $\mathrm{LHLLC}$), makes this quantity particularly suited to measure the amplitude of numerical undershoots or overshoots generated by the methods included in our study. In the regions right above the position of the spike in $N_\mathrm{BV}$, limited reconstruction methods (i.e., $\mathrm{LIN+VL, \ PPM08,}$ and $\mathrm{PPM08}$) do not generate undershoots below the curve $N_\mathrm{BV,}^{t=0}$ except for dynamical fluctuations smaller than $2\%$ induced by the flows in the stable layer. On the other hand, reconstruction schemes that do not use limiters for dynamic variables (i.e., $\mathrm{LIN, \ PAR,}$ and $\mathrm{PSH}$) produce undershoots whose amplitude is considerably larger than that of the dynamical fluctuations. The values of the largest undershoots relative to $N_\mathrm{BV,}^{t=0}$ are shown in the insets of Fig.~\ref{fig:brunt}. Overall, the amplitude of the undershoots increases with the order of the unlimited reconstruction method, and it ranges from $\approx7\%$ with $\mathrm{LIN}$ to almost $40\%$ with $\mathrm{PSH}$. The profiles of $N_\mathrm{BV}$ obtained on the $128^3$ grid, not shown here (but available on Zenodo\footnote{\url{https://zenodo.org/doi/10.5281/zenodo.10280900}}), reveal that the amplitude of the undershoots slightly increases from the $128^3$ to the $256^3$ grid for $\mathrm{LIN}$ and $\mathrm{PAR}$, although the undershoot region becomes narrower on progressively finer grids. We note that, despite the large-amplitude undershoots generated by some of our unlimited reconstruction methods, such numerical errors do not have any significant impact on the growth of the convection zone, as shown in Fig.~\ref{fig:ycb}. Nonetheless, they could still affect the frequency spectrum of IGWs in the stable layer, although the spatial spectra from the stable layer (see Fig.~\ref{fig:spectra-Y-2.5}) do show any such effect and converge to the reference solution. If numerical undershoots are largely to be avoided whilst generating a steep convective boundary, $\mathrm{PPM84+LHLLC}$ or $\mathrm{PPM08+LHLLC}$ could be the methods of choice.
	
	\subsection{Frequency power spectra}\label{sec:frequency-spectrum}
	
\begin{figure*}
		\centering
		\includegraphics[width=0.9\textwidth]{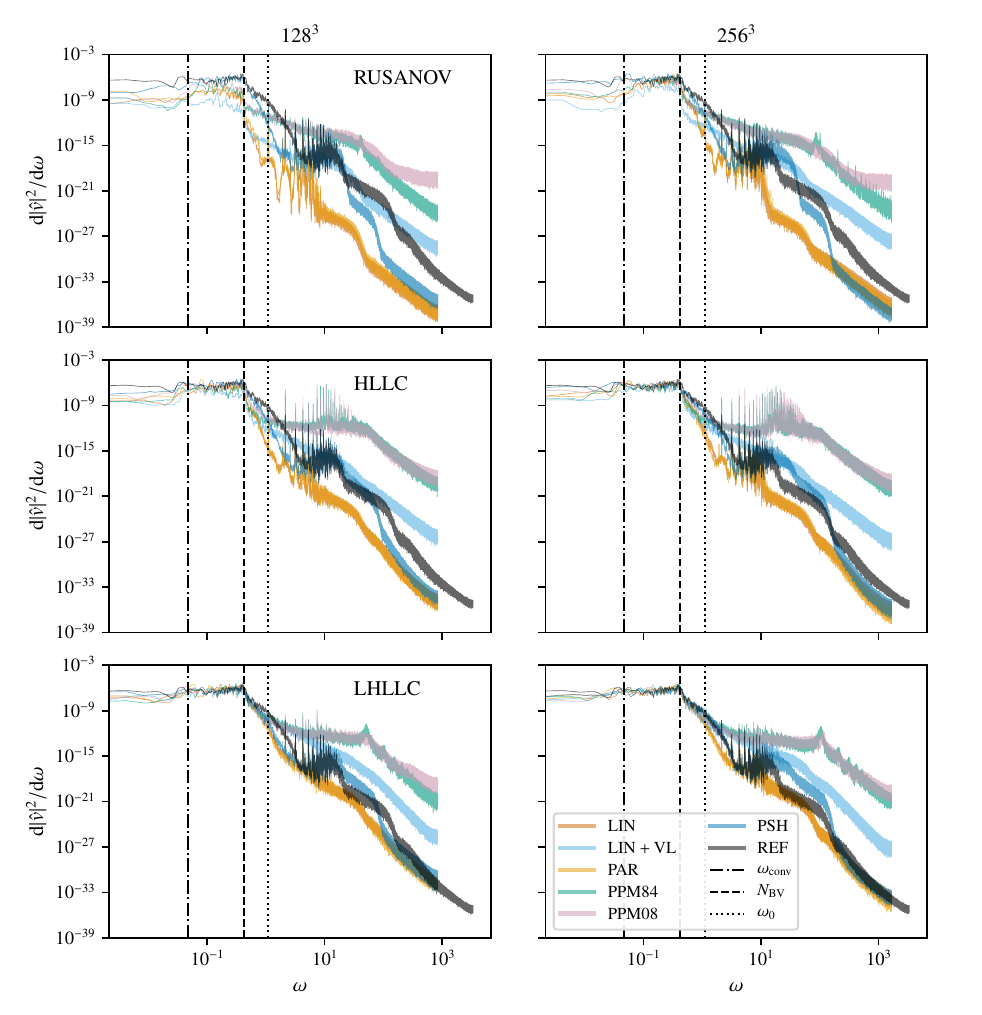}
		\caption{Frequency power spectrum of the vertical velocity component $v$ extracted in the middle of the stable layer at $(x,y,z)=(-0.2,2.5,-0.2)$ over the time series $t \in (10\tau_\mathrm{conv},32\tau_\mathrm{conv})$ in the problem of turbulent convection and wave excitation. The left and right panels show results obtained on the $128^3$ and the $256^3$ grid, respectively. Each row of panels shows the results of simulations run using 6 reconstruction schemes with the same Riemann solver. The reference solution (indicated with $\mathrm{REF}$) is the black curve. The convective turnover frequency ($\omega_\mathrm{conv}=2\pi/\tau_\mathrm{conv}$), the Brunt-V\"ais\"al\"a frequency at $(x,y,z)=(-0.2,2.5,-0.2)$ ($N_\mathrm{BV}$), and the frequency of the fundamental oscillation mode of the cavity ($\omega_0=1.1$) are represented by the black dashed-dotted, dashed, and dotted lines, respectively.}
		\label{fig:temporal spectra}
	\end{figure*}

\begin{figure*}
		\centering
		\includegraphics[width=0.5\textwidth]{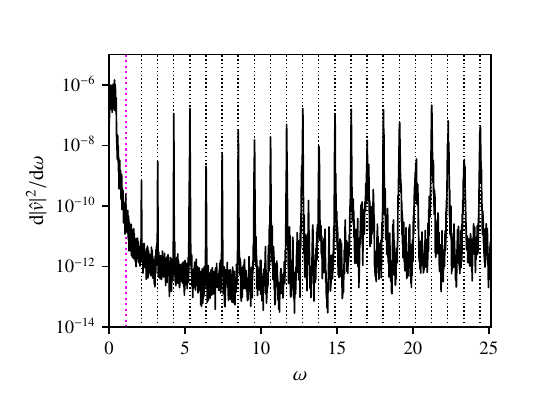}
		\caption{Frequency power spectrum of $v$ as in Fig.~\ref{fig:temporal spectra} but showing only the results of $\mathrm{PPM08+HLLC}$ on the $256^3$ grid. The vertical black dotted lines represent the resonant modes of the cavity which we derived from the theory of linear oscillations in Cowling approximation \citep[see, e.g.,][]{aerts2021}. The pink line at $\omega=1.1$ is the frequency of the fundamental oscillation mode of the cavity.}
		\label{fig:p-modes}
	\end{figure*}

	Finally, in Fig.~\ref{fig:temporal spectra}, we show the frequency power spectrum of the vertical velocity component, $v$, obtained in the middle of the stable layer for all of our 18 Godunov-type schemes. In particular, the frequency power spectra shown in Fig.~\ref{fig:temporal spectra} are Fourier projections of an array\footnote{The array of the values of $v$ obtained at each time step is multiplied by the Hanning window function to reduce the amplitude of the discontinuities in the signal at the boundaries of the time domain.} containing the value of $v$ at $(x,y,z)=(-0.2,2.5,-0.2)$ at each time step. There is a clear distinction between the frequency range corresponding to the regime of IGWs ($\omega < N_\mathrm{BV}$), where all spectra are almost flat, and the high-frequency range ($\omega > N_\mathrm{BV}$, where only sound waves are non-evanescent) in which the spectra are characterized by a steep decrease in power (although non monotonic) toward higher frequencies. The forest of lines visible at frequencies in the range from 1 to 50 correspond to the resonant pressure modes of the cavity as predicted by the linear wave theory in the Cowling approximation \citep[][see also Fig.~\ref{fig:p-modes}]{aerts2021}. In the frequency regime of IGWs, the power spectrum seems to converge to the reference solution when increasing the grid resolution or when using less dissipative Riemann solvers. A more quantitative analysis of the IGW spectrum would require substantially longer simulations, which would benefit from more time averaging as well as reach higher frequency resolution.
	
	On the other hand, the high-frequency domain is well resolved even in simulations run on the $128^3$ grid, and differences between the spectra are clearly visible. Overall, there is a large spread in power (by almost 20 orders of magnitude at the Nyquist frequency) among the different methods tested in our study, which does not decrease with grid resolution. Therefore, numerical convergence is not achieved at high frequencies. We note that the spectra with the highest power density in the high-frequency range are always those obtained with limited reconstruction methods, in particular $\mathrm{PPM84}$ and $\mathrm{PPM08}$, while unlimited reconstruction methods tend to generate much ``quiter'' spectra and that are closer to the reference solution. In the simulations of the Kelvin--Helmholtz instability shown in in Sect.~\ref{sec:results_kh}, we find that a power excess in high-frequency, short-wavelength sound waves is generated by the complex limiting procedure performed in $\mathrm{PPM84}$ and $\mathrm{PPM08}$ to reconstruct the Riemann states at cell interfaces. In this setup, a similar phenomenon is likely responsible for generating an acoustic power excess close to the Nyquist frequency, where the spread between the spectra is maximum. The power excess, however, is still large down to frequencies similar to the that of the fundamental oscillation mode of the cavity $\omega_\mathrm{0}=1.1$. Also, PPM-based methods used in combination with $\mathrm{HLLC}$ excite much stronger resonant lines than those generated with $\mathrm{LHLLC}$, and their peak power density is almost as high as the flat part of the spectrum associated with IGWs.  When used with PPM84, both RUSANOV and LHLLC produce a broad feature in the frequency spectrum whose peak is at $\omega_\mathrm{peak}\approx50$ and $100$ on the $128^3$ and $256^3$ grids, respectively. If these were sound waves, their wavelength $\lambda$ would be
	\begin{equation}
		\lambda = \frac{2 \pi}{\omega_\mathrm{peak}} c_\mathrm{sound}(y=2.5) \approx 2.7\Delta x,
	\end{equation}
	which is close to the Nyquist frequency in space. Therefore, such a feature in the spectrum may be caused by odd-even cell decoupling traveling at the local speed of sound. Additional peaks are observed at even higher frequencies than the frequency of the least resolvable sound wave on the grid (with a wavelength of two cells), so these are most likely numerical artifacts. Tests performed with lower CFL factors of 0.4 and 0.2 show that the amplitude of these peaks is slightly reduced when using shorter time steps in the simulation, but the level of the continuum in the power spectrum remains essentially unaltered.  In light of these results, we advise against using PPM-based methods in simulations of sound generation by low-Mach-number turbulence.

	\section{Performance metrics}\label{sec:performance}
	
	After proving that all of the methods tested in our study converge to the correct solution for most of the physical quantities of interest\footnote{A prominent example of a nonconvergent numerical solution is the frequency power spectrum of sound waves in the test problem of turbulent convection and mass entrainment shown in Sect.~\ref{sec:code-comparison}.}, we can now search for the most efficient way to generate a numerical solution at the desired accuracy. In principle, one could use the $L_1$ errors computed from the simulations of the Kelvin--Helmholtz instability (see Sect.~\ref{sec:results_kh}) to find 
	the resolution of the grid on which the scheme achieves a given level of accuracy. Then, the computational cost of the simulation run on such a grid can be estimated if the wall-clock time spent by the program to perform a single cell update is known.  Here, we prefer to use the results from the test problem involving turbulent convective flows and wave excitation (see Sect.~\ref{sec:code-comparison}), which is more challenging than the  Kelvin--Helmholtz instability test and much closer to a real application of stellar hydrodynamics. However, due to the chaotic nature of the turbulent flows that develop in the convection zone, convergence in the $L_1$ error norm cannot be achieved. Therefore, rather than measuring the computational cost per fixed accuracy in the sense of $L_1$ errors, we opt to measure the computational cost of a simulation run with a specific combination of numerical methods that achieves a given level of effective resolution of the turbulent flows, $N_\mathrm{eff}$. We estimate the effective resolution obtained on a given grid with $N_x^3$ cells as
	\begin{equation}
		N_\mathrm{eff}(N_x) = \frac{2L_\mathrm{conv}}{\lambda_\mathrm{vis}(N_x)},
	\end{equation}
	where $\lambda_\mathrm{vis}(N_x)$ is a characteristic dissipation length scale in the turbulent kinetic energy spectrum. In high-resolution schemes, the amount of kinematic viscosity introduced into the system is not fixed, but rather its value depends on the local dynamical properties of the flow and it is often found to be a steep function of the spatial wavenumber \citep[see, e.g.,][]{porter1994}. Therefore, $\lambda_\mathrm{vis}(N_x)$ cannot be defined uniquely. Here, we choose
	\begin{equation}
		\lambda_\mathrm{vis}(N_x) \approx \frac{2 L_\mathrm{conv}}{k_\mathrm{h,10}(N_x)}
	\end{equation}
	as a representative value for the dissipation length scale, with $k_\mathrm{h,10}(N_x)$ being the spatial wavenumber at which the kinetic power spectrum rescaled by the Kolmogorov law drops by one dex from its maximum. In this approximation, the effective resolution is simply given by
	\begin{equation}
		N_\mathrm{eff}(N_x) = k_\mathrm{h,10}(N_x).
	\end{equation}
	To perform a more precise measurement of $N_\mathrm{eff}(N_x)$, both the array of wavenumbers and of the kinetic energy spectrum are linearly interpolated on a finer grid. The values of the effective resolution (rescaled by the grid resolution $N_x$) obtained in the simulations included in our study are collected in Table \ref{tab:eff-res}. There is a clear trend toward higher effective resolution when using progressively higher-order reconstruction methods or less dissipative Riemann solvers.

	By combining the values of $N_\mathrm{eff}(N_x)$ and the average wall-clock time $\delta t(N_x)$ spent by the program to perform a single cell update, shown in Table \ref{tab:abs-perf}, it is possible to estimate the computational effort required by each combination of numerical options to achieve the effective resolution of $\mathrm{PPM08+LHLLC}$\footnote{In the following expression, we make use of the fact that the complexity of a 3D, Godunov algorithm, such as that implemented in SLH, scales with $N_x^4$, with $N_x$ being the number of grid cells per dimension.} as
	\begin{equation}\label{eq:rel-perf}
		\Theta(N_x) = \frac{\delta t(N_x)}{\delta t(N_x)_\mathrm{PPM08+LHLLC}}\Bigg(\frac{N_\mathrm{eff}(N_x)}{N_\mathrm{eff, PPM08+LHLLC}(N_x)}\Bigg)^{-4}.
	\end{equation}
	Equation~(\ref{eq:rel-perf}) is rescaled such that the cost of the simulation run with $\mathrm{PPM08+LHLLC}$ on any given grid is unity. All values of $\Theta(N_x\,{=}\,128)$ and $\Theta(N_x\,{=}\,256)$ are shown in Table~\ref{tab:rel-perf}. The spread of almost four orders of magnitude in the relative computational cost among the different methods is due to the steep dependence of $\Theta(N_x)$ on $N_\mathrm{eff}(N_x)$, which varies by as much as a factor of 10  (see Table~\ref{tab:eff-res}). On the other hand, 
	$\Theta(N_x)$ only scales linearly with $\delta t(N_x)$, which in our simulations varies at most by a factor of $\,{\approx}\,3$ (with $\mathrm{LIN+RUSANOV}$ and $\mathrm{PPM84+LHLLC}$ achieving the lowest and highest wall-clock time per cell-update, respectively). The most expensive combination of methods among those tested in our study (in terms of computational cost per fixed resolving power) is $\mathrm{LIN+VL+RUSANOV}$, $\sim$1000 times as expensive as $\mathrm{PPM08+LHLLC}$. The strong numerical dissipation generated by $\mathrm{RUSANOV}$ leads to very poor performance of the finite-volume scheme even when used in combination with PPM-based reconstruction methods. Only with the unlimited $\mathrm{PSH}$ method such a flux function is capable of achieving acceptable performance ($\Theta(N_x)\approx4$). When the reconstruction scheme is kept the same, using progressively less dissipative Riemann solvers decreases the cost of reaching the same effective resolution. At the typical Mach numbers encountered in the convection zone ($\mathcal{M}\,{\approx}\,0.02$), the combination of $\mathrm{HLLC}$ and second-order reconstruction schemes or $\mathrm{PAR}$ is considerably more expensive than $\mathrm{PPM08+LHLLC}$ ($\Theta(N_x)$ in the range from 10 to 30). The performance of the scheme increases when $\mathrm{HLLC}$ is used with PPM-based reconstruction schemes  ($\Theta(N_x)$ in the range from 2 to 6) and it is even higher than that of $\mathrm{PPM08+LHLLC}$ when coupled to $\mathrm{PSH}$. $\mathrm{PSH+LHLLC}$ is the most performant method on both grids according to the chosen metric. Overall, the computational cost of the finite-volume scheme is considerably reduced when using the low-dissipation solver (by a factor from 2 to 10 with respect to a scheme using $\mathrm{HLLC}$ and the same spatial reconstruction method). The relative performance of the scheme only varies slightly when $\mathrm{LHLLC}$ is used in combination with reconstruction methods less accurate than $\mathrm{PSH}$, with the worst performance  being achieved by $\mathrm{LIN+VL}$ ($\Theta(N_x)\approx3$). 
	
	We note that, for most methods, the value of $\Theta(N_x)$ sensibly increases (by as much as a factor of ${\approx}\,2$) from the $128^3$ to the $256^3$ grid. Such differences are due to the small but systematic decrease of the rescaled effective resolution $N_\mathrm{eff}(N_x)/N_x$ with increasing the grid resolution (this behavior is also confirmed by the reference solution computed on the $512^3$ grid, see Table~\ref{tab:eff-res}). Although the differences observed in $N_\mathrm{eff}(N_x)/N_x$ among the two grids are at most $20\%$, they are significantly amplified after applying the steep scaling relation between $\Theta(N_x)$ and  $N_\mathrm{eff}(N_x)$. However, this effect is negligible if compared to the large spread obtained in the values of $\Theta(N_x)$ on a given grid. Furthermore, we stress that our measure of $\Theta(N_x)$ is based on a crude approximation of effective resolution of the turbulent flows, so the values provided in Table~\ref{tab:rel-perf} should  only be taken as estimates of the relative performance of different Godunov-type methods in simulations of turbulent convection. The measurements of the absolute performance  provided in Table \ref{tab:abs-perf}, which are needed to estimate $\Theta(N_x)$, may also depend on the parallelization strategy and the number of cores used to run the simulations, especially for the methods that require many ghost cells (e.g., $\mathrm{PPM08}$ and $\mathrm{PSH}$) and are therefore characterized by higher communication costs.

	\begin{table}
		\caption{Effective resolution $N_\mathrm{eff}(N_x)$ as defined in in Sect.~\ref{sec:code-comparison} for all of the 18 Godunov-type methods tested in our study on grids with \mbox{$N_x=128$} and \mbox{$N_x=256$} cells per dimension. Here, all numbers are given as a percentage of the grid resolution $N_x$ for an easier comparison between different grids. In this metric, the rescaled effective resolution of the reference run ($\mathrm{PAR+LHLLC+512^3}$) is 24.43.}
		\label{tab:eff-res}
		\centering
		\begin{tabular}{llll}
			\toprule
			$128^3$  & RUSANOV & HLLC & LHLLC \\
			\midrule
			LIN & 6.21 & 14.45 & 24.68 \\ 
			LIN+VL & 5.24 & 13.53 & 21.97 \\ 
			PAR & 7.04 & 16.68 & 29.05 \\ 
			PPM84 & 10.62 & 24.15 & 31.47 \\ 
			PPM08 & 14.21 & 28.42 & 33.85 \\ 
			PSH & 20.47 & 37.24 & 49.22 \\ 
			\toprule
			$256^3$    & RUSANOV & HLLC & LHLLC \\
			\midrule
			LIN & 5.47 & 12.61 & 23.30 \\ 
			LIN+VL & 4.38 & 11.08 & 19.75 \\ 
			PAR & 6.35 & 14.28 & 26.16 \\ 
			PPM84 & 10.10 & 21.23 & 29.06 \\ 
			PPM08 & 13.55 & 27.00 & 32.02 \\ 
			PSH & 19.90 & 35.08 & 47.00 \\ 
			\bottomrule
		\end{tabular}
		\vspace{0.5em}
	\end{table}

	\begin{table}
		\caption{Mean wall-clock time  spent by the program to advance the solution by one time step with RK3 in a single cell of the computational grid, in units of $\mu \mathrm{s}$. The numbers provided in the tables are averages of five measurements, each of which is obtained by evolving the setup described in Sect.~\ref{sec:code-comparison} for 100 steps. Relative errors in the averages (in the sense of $1\sigma$) are smaller than $2\%$ in all cases. Every simulation run for this analysis is MPI parallelized using grids with $32^3$ cells per task. Computations are performed on 2.3 GHz Intel Xeon, Skylake-based processors.}
		\label{tab:abs-perf}
		\centering
		\begin{tabular}{llll}
			\toprule
			$128^3$  & RUSANOV & HLLC & LHLLC \\
			\midrule
			LIN & 1.47 & 1.68 & 1.70 \\ 
			LIN+VL & 1.57 & 1.80 & 1.89 \\ 
			PAR & 1.47 & 1.74 & 1.78 \\ 
			PPM84 & 4.12 & 4.30 & 4.31 \\ 
			PPM08 & 3.57 & 3.75 & 3.83 \\ 
			PSH & 2.01 & 2.36 & 2.41 \\ 
			\toprule
			$256^3$  & RUSANOV & HLLC & LHLLC \\
			\midrule
			LIN & 1.60 & 1.73 & 1.77 \\ 
			LIN+VL & 1.73 & 1.87 & 1.88 \\ 
			PAR & 1.60 & 1.81 & 1.86 \\ 
			PPM84 & 4.23 & 4.40 & 4.47 \\ 
			PPM08 & 3.79 & 4.03 & 4.00 \\ 
			PSH & 2.24 & 2.47 & 2.52 \\ 
			\bottomrule
		\end{tabular}
		\vspace{0.5em}
	\end{table}

	\begin{table}
		\caption{Relative computational cost of each Godunov-type method considered in this study to achieve the same effective resolution as the $\mathrm{PPM08}+\mathrm{LHLLC}$ combination ($\Theta$, see Eq.~(\ref{eq:rel-perf})) in the simulations described in Sect.~\ref{sec:code-comparison}. Here all numbers are rounded to two significant figures.}
		\label{tab:rel-perf}
		\centering
		\begin{tabular}{llll}
			\toprule
			$128^3$ & RUSANOV & HLLC & LHLLC \\
			\midrule
			LIN & 340 &13.0 &1.60 \\ 
			LIN+VL & 710 &18.0 &2.80 \\ 
			PAR & 210 &7.70 &0.86 \\ 
			PPM84 & 110 &4.30 &1.50 \\ 
			PPM08 & 30.0 &2.00 &1.00 \\ 
			PSH & 3.90 &0.42 &0.14 \\ 
			\toprule
			$256^3$ & RUSANOV & HLLC & LHLLC \\
			\midrule
			LIN & 470 & 18.0 & 1.60 \\ 
			LIN+VL & 1200 & 33.0 & 3.30 \\ 
			PAR & 260 & 11.0 & 1.00 \\ 
			PPM84 & 110 & 5.70 & 1.60 \\ 
			PPM08 & 30.0 & 2.00 & 1.00 \\ 
			PSH & 3.80 & 0.43 & 0.14 \\ 
			\bottomrule
		\end{tabular}
		\vspace{0.5em}
	\end{table}

	\section{Summary and conclusions}
	\label{sec:conclusions}
	
	High-resolution, finite-volume schemes are popular methods for simulating the behaviour of astrophysical fluids. There is a wide range of spatial-reconstruction schemes, numerical flux functions and time-integration methods that can be combined into a numerically stable and robust, finite-volume scheme. Focusing on subsonic flows, we have analysed the accuracy and computational cost of all possible combinations of six spatial-reconstruction schemes and three numerical flux functions, i.e. 18 methods in total. The numerical solution was marched in time with a semi-discrete scheme based on a third-order Runge--Kutta method. This choice was motivated by our focus on the spatial accuracy of the schemes and the related need to suppress time-stepping errors.
	
	We consider two main test problems. The first is a Kelvin--Helmholtz instability problem, in which the initial shear flows have Mach numbers of $10^{-1}$, $10^{-2}$, and $10^{-3}$. We use a smooth initial condition to make it possible to obtain numerically convergeable solutions to the inviscid Euler equations at a fixed point in time in the non-linear phase of the instability. We characterise the accuracy of the solutions by (i)
	performing a qualitative assessment of the spatial structure of the solutions and (ii) by measuring $L_1$ errors with respect to a reference solution obtained on a fine grid of $8192 \times 4096$ cells.
	
	The low-Mach flux function LHLLC is found to be much less dissipative and much more accurate than the HLLC and RUSANOV flux functions at the two lowest Mach numbers considered. Even though the overall 2D numerical scheme is \mbox{$2$\textsuperscript{nd}-order} accurate, the errors produced by different spatial reconstruction schemes span as much as an order of magnitude. Unlimited reconstruction schemes of increasing order, up to $7$\textsuperscript{th}, provide progressively more accurate results when the solution is sufficiently well resolved. However, these schemes are of limited use for the advection of mass fractions due to the oscillations and overshoots they produce around discontinuities. This issue is largely eliminated by the use of limiters in the schemes LIN+VL, PPM84, and PPM08. However, we find that the switching behaviour of the limiters introduces spurious structures and small-scale sound waves to the solutions, which severely reduce the accuracy of the methods when applied to slow flows. Our hybrid method PSH, which combines unlimited, $7$\textsuperscript{th}-order reconstruction for dynamic variables with PPM08 for mass fractions, is found to provide the most accurate solutions in nearly all of our simulations of the Kelvin--Helmholtz instability.
	
	Our second test problem, chosen to be as close as possible to practical applications in the dynamics of stellar interiors, involves stratified turbulent convection generating  waves in and entraining mass from an overlying stably stratified layer. The typical Mach number of the convection is ${\approx}\,0.02$. The turbulent nature of the convective flow makes the solutions chaotic but space- and time-averaged quantities can still be meaningfully compared between different simulations. We run the simulations on grids of $128^3$ and $256^3$ cells with one additional simulation on a $512^3$ grid serving as a reference solution.
	
	Qualitatively speaking, the solutions match the trends seen in the simulations of the Kelvin--Helmholtz instability, i.e.\ combinations of the least dissipative flux functions with the highest-order reconstruction schemes provide the highest resolving power in both the convective and stable layers. Spatial spectra of kinetic energy computed in the convective and stable layers converge toward the reference solution with any of the 18 methods but the least dissipative combinations (especially PSH+LHLLC) are much closer to the reference solution than the most dissipative ones (especially LIN+VL+RUSANOV). The spatial spectrum in both layers obtained with PSH+LHLLC on a $256^3$ grid is essentially the same as that with PAR+LHLLC on a $512^3$ grid.

	The growth rate of the convective layer converges upon grid refinement with all 6 reconstruction schemes and both the HLLC and LHLLC flux function but the convergence is significantly faster with the low-Mach flux function LHLLC. The RUSANOV flux function produces extremely viscous flows and large deviations from the reference solution in terms of the position of the upper convective boundary. 
	
	We show that schemes that do not use limiting for dynamic variables (LIN, PAR, PSH), unlike those that do (LIN+VL, PPM84, PPM08), produce overshoots in the Brunt-V\"ais\"al\"a frequency at the relatively sharp convective boundary. The magnitude of the overshoots increases with increasing order of accuracy of the method, as expected. Neither the growth rate of the convective layer, nor the temporal and spatial spectra of internal gravity waves are affected by this phenomenon in our simulations but we recommend careful testing when such high-order methods are applied to other setups or used to derive quantities not investigated here.
	
	Temporal spectra of the vertical component of velocity in the stable layer show that all six reconstruction schemes produce qualitatively  similar, flat and featureless spectra in the regime of internal gravity waves. On the other hand, the spectra of pressure (i.e. high frequency) waves are sensitive to the choice of the reconstruction scheme even with the low-Mach flux function LHLLC. The spectra also reinforce our conclusions based on the Kelvin--Helmholtz problem that methods that apply limiters to dynamic variables generate much more acoustic ``noise'' than methods that do not. This includes the popular methods PPM84 and PPM08. Although we do not know what the acoustic spectrum in our problem should look like, we show that the frequencies of vertical resonant modes agree with 1D linear wave theory.
	
	Finally, we use the steep drop-off of the spatial kinetic energy spectra in the convective layer close to the grid scale to define a measure of effective resolution. Rescaling the wall-clock time of our $128^3$ and $256^3$ simulations, we estimate the computational cost of matching the same effective resolution using our 18 methods. The cost is found to span almost four orders of magnitude. The largest cost reduction comes from choosing the least dissipative flux function, which is LHLLC in our study. We show that the reconstruction schemes PSH, PAR, and PPM08, in order of increasing cost, are the cheapest at the same effective resolution.
	
	Our study demonstrates that it is generally advisable to use low-dissipation Riemann solvers in Godunov-type finite-volume schemes for simulating low-Mach-number flows. The low-Mach fix $\phi$ in Eq.~(\ref{eq:pressure-low-mach-fix}) can easily be implemented in any $\mathrm{HLLC}$-type scheme without affecting the absolute performance of the code, and it reduces the amount of computation required to achieve the same accuracy as $\mathrm{HLLC}$ by a factor ranging from 2 to 10 at typical Mach numbers of $\approx0.01$. At Mach numbers of $\,{\approx}\,10^{-4}$ or $\,{\approx}\,10^{-3}$, like those encountered in the convective cores of main sequence stars, the usage of a low-Mach solver decreases the computational cost per fixed accuracy with respect to a conventional Riemann solver by even larger factors \citep[][]{leidi2022}. Although low-dissipation Riemann solvers such as $\mathrm{LHLLC}$ bring clear advantages in simulations of subsonic flows when used in Eulerian codes, their robustness and accuracy properties in quasi-Lagrangian, moving-mesh schemes still has to be verified. We note that such schemes hold promise for a particularly significant improvement because mesh motions aim at reducing advection errors by minimizing the velocities relative to the cell interfaces. This naturally leads to low-Mach-number flows in the comoving frame, which are better modeled by low-Mach-number Riemann solvers. We are currently testing an implementation of the $\mathrm{LHLLC/D}$ solvers of \cite{minoshima2021} in the moving-mesh MHD code \code{AREPO} \citep{springel2010}, which we will report in a forthcoming study.
	
	On the other hand, the choice of the spatial reconstruction scheme depends on the physical application of interest. Overall, second-order reconstruction methods lead to the generation of considerably more inaccurate results than higher-order schemes when the Mach number of the flow is low. If numerical overshoots have to be suppressed whilst retaining high resolving power, $\mathrm{PPM84}$ and $\mathrm{PPM08}$ should be the methods of choice. However, higher-order unlimited reconstruction methods should be preferred over $\mathrm{PPM}$-based methods in simulations of sound wave generation by subsonic turbulence because they generate much less numerical ``noise'' in the acoustic spectrum. Less oscillatory alternatives to the hybrid PSH method, such as the $3$\textsuperscript{th}-order piecewise parabolic hybrid ($\mathrm{PPH}$) or the $5$\textsuperscript{th}-order piecewise quartic hybrid ($\mathrm{PQH}$) methods described in Appendix~\ref{sec:PPH_PQH}, could offer the best compromise between the complete removal of numerical overshoots and reduction of acoustic noise.
	
	Finally, we note that there are complex astrophysical environments where both high- and low-Mach-number flows can coexist. This scenario often occurs, for instance, in accretion disks \citep[see, e.g.,][]{klessen2010}, star forming regions \citep[see, e.g.,][]{wang2024}, or outer layers of stars \citep[see, e.g.,][]{wedemeyer2017}. To model such diverse flow regimes, the simple use of unlimited higher-order reconstructions and low-dissipation solvers may give rise to numerical instabilities, especially in the proximity of shock fronts. To cure this problem, an alternative approach based on the local dynamical properties of the flow is needed \citep[see, e.g.,][]{mignone2012,fleischmann2020}. One possibility is to add a selection criterion to the function that computes the flux at a cell interface such that the LHLLC solver is used if the Mach number of the flow is lower than a conservative threshold (e.g., $\mathcal{M}\lesssim0.6$), whereas a proper shock-capturing method (e.g., HLL or HLLC) is chosen for modeling faster flows. A similar argument can be made for the choice of the spatial reconstruction scheme. To avoid introducing large oscillations in the state quantities near large, non-linear discontinuities, schemes that are more dissipative and possibly close to being TVD (such as LIN+VL, PPM84, PPM08, or WENO) should be preferred over high order unlimited methods.

	\begin{acknowledgements}
		The work of GL and FKR is supported by the German Research
		Foundation (DFG) through the grant RO 3676/3-1. We acknowledge support by the Klaus Tschira Foundation. This work is funded by the Deutsche Forschungsgemeinschaft (DFG, German Research Foundation) under Germany's Excellence Strategy EXC 2181/1 - 390900948 (the Heidelberg STRUCTURES Excellence Cluster). The authors gratefully acknowledge the Gauss Centre for Supercomputing e.V. (www.gauss-centre.eu) for funding this project by providing computing time on the GCS Supercomputer SuperMUC-NG at Leibniz Supercomputing Centre (www.lrz.de). PVFE was supported by the U.S. Department of Energy through the Los Alamos National Laboratory (LANL). LANL is operated by Triad National Security, LLC, for the National Nuclear Security Administration of the U.S. Department of Energy (Contract No. 89233218CNA000001). This work has been assigned a document release number LA-UR-23-33257.
	\end{acknowledgements}
	
	\bibliographystyle{aa}
	\bibliography{godunov_performance}

\begin{thebibliography}{88}
\expandafter\ifx\csname natexlab\endcsname\relax\def\natexlab#1{#1}\fi

\bibitem[{Aerts(2021)}]{aerts2021}
Aerts, C. 2021, Rev. Mod. Phys., 93, 015001

\bibitem[{Anders \& Pedersen(2023)}]{anders2023}
Anders, E.~H. \& Pedersen, M.~G. 2023, Galaxies, 11

\bibitem[{{Andrassy} {et~al.}(2020){Andrassy}, {Herwig}, {Woodward}, \&
  {Ritter}}]{andrassy2020}
{Andrassy}, R., {Herwig}, F., {Woodward}, P., \& {Ritter}, C. 2020, \mnras,
  491, 972

\bibitem[{{Andrassy} {et~al.}(2022){Andrassy}, {Higl}, {Mao}, {Moc{\'a}k},
  {Vlaykov}, {Arnett}, {Baraffe}, {Campbell}, {Constantino}, {Edelmann},
  {Goffrey}, {Guillet}, {Herwig}, {Hirschi}, {Horst}, {Leidi}, {Meakin},
  {Pratt}, {Rizzuti}, {R{\"o}pke}, \& {Woodward}}]{andrassy2022}
{Andrassy}, R., {Higl}, J., {Mao}, H., {et~al.} 2022, \aap, 659, A193

\bibitem[{{Andrassy} {et~al.}(2023){Andrassy}, {Leidi}, {Higl}, {Edelmann},
  {Schneider}, \& {Roepke}}]{andrassy2023}
{Andrassy}, R., {Leidi}, G., {Higl}, J., {et~al.} 2023, arXiv e-prints,
  arXiv:2307.04068

\bibitem[{{Balsara}(2017)}]{balsara2017}
{Balsara}, D.~S. 2017, Living Reviews in Computational Astrophysics, 3, 2

\bibitem[{Beckwith \& Stone(2011)}]{beckwith2011}
Beckwith, K. \& Stone, J.~M. 2011, The Astrophysical Journal Supplement Series,
  193, 6

\bibitem[{Berberich {et~al.}(2021)Berberich, Chandrashekar, \&
  Klingenberg}]{berberich2019}
Berberich, J.~P., Chandrashekar, P., \& Klingenberg, C. 2021, Computers \&
  Fluids, 219, 104858

\bibitem[{{Berlok} \& {Pfrommer}(2019)}]{berlok2019a}
{Berlok}, T. \& {Pfrommer}, C. 2019, \mnras, 485, 908

\bibitem[{{Canivete Cuissa} \& {Teyssier}(2022)}]{cuissa2022}
{Canivete Cuissa}, J.~R. \& {Teyssier}, R. 2022, \aap, 664, A24

\bibitem[{Chen {et~al.}(2020)Chen, Lin, Li, \& Yan}]{chen2020}
Chen, S., Lin, B., Li, Y., \& Yan, C. 2020, SIAM Journal on Scientific
  Computing, 42, B921

\bibitem[{Colella(1990)}]{colella1990}
Colella, P. 1990, Journal of Computational Physics, 87, 171

\bibitem[{{Colella} \& {Sekora}(2008)}]{colella2008a}
{Colella}, P. \& {Sekora}, M.~D. 2008, Journal of Computational Physics, 227,
  7069

\bibitem[{{Colella} \& {Woodward}(1984)}]{colella1984a}
{Colella}, P. \& {Woodward}, P.~R. 1984, Journal of Computational Physics, 54,
  174

\bibitem[{{Courant} {et~al.}(1928){Courant}, {Friedrichs}, \&
  {Lewy}}]{courant1928}
{Courant}, R., {Friedrichs}, K., \& {Lewy}, H. 1928, Mathematische Annalen,
  100, 32

\bibitem[{Day \& Bell(2000)}]{day2000}
Day, M.~S. \& Bell, J.~B. 2000, Combustion Theory and Modelling, 4, 535

\bibitem[{{Dobler} {et~al.}(2003){Dobler}, {Haugen}, {Yousef}, \&
  {Brandenburg}}]{dobler2003}
{Dobler}, W., {Haugen}, N.~E., {Yousef}, T.~A., \& {Brandenburg}, A. 2003,
  \pre, 68, 026304

\bibitem[{{Dumbser} {et~al.}(2019){Dumbser}, {Balsara}, {Tavelli}, \&
  {Fambri}}]{dumbser2019}
{Dumbser}, M., {Balsara}, D.~S., {Tavelli}, M., \& {Fambri}, F. 2019,
  International Journal for Numerical Methods in Fluids, 89, 16

\bibitem[{Dumbser {et~al.}(2009)Dumbser, Castro, Parés, \& Toro}]{dumbser2009}
Dumbser, M., Castro, M., Parés, C., \& Toro, E.~F. 2009, Computers \& Fluids,
  38, 1731

\bibitem[{{Edelmann}(2014)}]{edelmann2014a}
{Edelmann}, P.~V.~F. 2014, Dissertation, Technische Universit\"at M\"unchen

\bibitem[{{Edelmann} {et~al.}(2021){Edelmann}, {Horst}, {Berberich},
  {Andrassy}, {Higl}, {Leidi}, {Klingenberg}, \& {R{\"o}pke}}]{edelmann2021a}
{Edelmann}, P.~V.~F., {Horst}, L., {Berberich}, J.~P., {et~al.} 2021, \aap,
  652, A53

\bibitem[{{Edelmann} \& {R\"{o}pke}(2016)}]{edelmann2016b}
{Edelmann}, P.~V.~F. \& {R\"{o}pke}, F.~K. 2016, in {JUQUEEN} {E}xtreme
  {S}caling {W}orkshop 2016, ed. D.~Br\"ommel, W.~Frings, \& B.~J.~N. Wylie,
  JSC Internal Report No. FZJ-JSC-IB-2016-01, 63--67

\bibitem[{{Edelmann} {et~al.}(2017){Edelmann}, {R{\"o}pke}, {Hirschi},
  {Georgy}, \& {Jones}}]{edelmann2017a}
{Edelmann}, P.~V.~F., {R{\"o}pke}, F.~K., {Hirschi}, R., {Georgy}, C., \&
  {Jones}, S. 2017, \aap, 604, A25

\bibitem[{Fleischmann {et~al.}(2020)Fleischmann, Adami, \&
  Adams}]{fleischmann2020}
Fleischmann, N., Adami, S., \& Adams, N.~A. 2020, Journal of Computational
  Physics, 423, 109762

\bibitem[{{Flock} {et~al.}(2010){Flock}, {Dzyurkevich}, {Klahr}, \&
  {Mignone}}]{flock2010}
{Flock}, M., {Dzyurkevich}, N., {Klahr}, H., \& {Mignone}, A. 2010, \aap, 516,
  A26

\bibitem[{Godunov(1959)}]{godunov1959}
Godunov, S.~K. 1959, Matematicheskii Sbornik, 89, 271

\bibitem[{{Goffrey} {et~al.}(2017){Goffrey}, {Pratt}, {Viallet}, {Baraffe},
  {Popov}, {Walder}, {Folini}, {Geroux}, \& {Constantino}}]{goffrey2017a}
{Goffrey}, T., {Pratt}, J., {Viallet}, M., {et~al.} 2017, \aap, 600, A7

\bibitem[{Greenough \& Rider(2003)}]{greenough2003}
Greenough, J.~A. \& Rider, W.~J. 2003, Journal of Computational Physics, 196

\bibitem[{Guillard \& Viozat(1999)}]{guillard1999}
Guillard, H. \& Viozat, C. 1999, Computers \& Fluids, 28, 63

\bibitem[{Harten {et~al.}(1987)Harten, Engquist, Osher, \&
  Chakravarthy}]{harten1987}
Harten, A., Engquist, B., Osher, S., \& Chakravarthy, S.~R. 1987, Journal of
  Computational Physics, 71, 231

\bibitem[{Harten {et~al.}(1983)Harten, Lax, \& Leer}]{harten1983}
Harten, A., Lax, P.~D., \& Leer, B.~V. 1983, SIAM Review, 25, 35

\bibitem[{{Horst} {et~al.}(2020){Horst}, {Edelmann}, {Andr{\'a}ssy},
  {R{\"o}pke}, {Bowman}, {Aerts}, \& {Ratnasingam}}]{horst2020a}
{Horst}, L., {Edelmann}, P.~V.~F., {Andr{\'a}ssy}, R., {et~al.} 2020, \aap,
  641, A18

\bibitem[{{Horst} {et~al.}(2021){Horst}, {Hirschi}, {Edelmann}, {Andrassy}, \&
  {Roepke}}]{horst2021a}
{Horst}, L., {Hirschi}, R., {Edelmann}, P.~V.~F., {Andrassy}, R., \& {Roepke},
  F.~K. 2021, \aap, 653, A55

\bibitem[{Jiang \& Shu(1996)}]{jiang1996}
Jiang, G.-S. \& Shu, C.-W. 1996, Journal of Computational Physics, 126, 202

\bibitem[{{K{\"a}pyl{\"a}}(2021)}]{kapyla2021}
{K{\"a}pyl{\"a}}, P.~J. 2021, \aap, 651, A66

\bibitem[{{Klein}(2009)}]{klein2009}
{Klein}, R. 2009, Theoretical and Computational Fluid Dynamics, 23, 161

\bibitem[{{Klessen} \& {Hennebelle}(2010)}]{klessen2010}
{Klessen}, R.~S. \& {Hennebelle}, P. 2010, \aap, 520, A17

\bibitem[{Klingenberg {et~al.}(2007)Klingenberg, Schmidt, \&
  Waagan}]{klingenberg2007}
Klingenberg, C., Schmidt, W., \& Waagan, K. 2007, Journal of Computational
  Physics, 227, 12

\bibitem[{Kolb(2014)}]{kolb2014}
Kolb, O. 2014, SIAM Journal on Numerical Analysis, 52, 2335

\bibitem[{Kritsuk {et~al.}(2011)Kritsuk, Åke Nordlund, Collins, Padoan,
  Norman, Abel, Banerjee, Federrath, Flock, Lee, Li, Müller, Teyssier,
  Ustyugov, Vogel, \& Xu}]{kritsuk2011}
Kritsuk, A.~G., Åke Nordlund, Collins, D., {et~al.} 2011, The Astrophysical
  Journal, 737, 13

\bibitem[{{K{\"u}chler} {et~al.}(2019){K{\"u}chler}, {Bewley}, \&
  {Bodenschatz}}]{kuchler2019}
{K{\"u}chler}, C., {Bewley}, G., \& {Bodenschatz}, E. 2019, Journal of
  Statistical Physics, 175, 617

\bibitem[{Latini {et~al.}(2007)Latini, Schilling, \& Don}]{latini2007}
Latini, M., Schilling, O., \& Don, W.~S. 2007, Journal of Computational
  Physics, 221, 805

\bibitem[{{Lecoanet} {et~al.}(2017){Lecoanet}, {McCourt}, {Quataert}, {Burns},
  {Vasil}, {Oishi}, {Brown}, {Stone}, \& {O'Leary}}]{lecoanet2016a}
{Lecoanet}, D., {McCourt}, M., {Quataert}, E., {et~al.} 2017, \mnras, 455, 4274

\bibitem[{{Leidi} {et~al.}(2023){Leidi}, {Andrassy}, {Higl}, {Edelmann}, \&
  {R{\"o}pke}}]{leidi2023}
{Leidi}, G., {Andrassy}, R., {Higl}, J., {Edelmann}, P.~V.~F., \& {R{\"o}pke},
  F.~K. 2023, \aap, 679, A132

\bibitem[{{Leidi} {et~al.}(2022){Leidi}, {Birke}, {Andrassy}, {Higl},
  {Edelmann}, {Wiest}, {Klingenberg}, \& {R{\"o}pke}}]{leidi2022}
{Leidi}, G., {Birke}, C., {Andrassy}, R., {et~al.} 2022, \aap, 668, A143

\bibitem[{LeVeque(2002)}]{leveque2002}
LeVeque, R.~J. 2002, Finite Volume Methods for Hyperbolic Problems, Cambridge
  Texts in Applied Mathematics (Cambridge University Press)

\bibitem[{{Lighthill}(1952)}]{lighthill1952}
{Lighthill}, M.~J. 1952, Proceedings of the Royal Society of London Series A,
  211, 564

\bibitem[{{Liou}(2006)}]{Liou2006}
{Liou}, M.-S. 2006, Journal of Computational Physics, 214, 137

\bibitem[{Liu {et~al.}(1994)Liu, Osher, \& Chan}]{liu1994}
Liu, X.-D., Osher, S., \& Chan, T. 1994, Journal of Computational Physics, 115,
  200

\bibitem[{{Maeder}(2009)}]{maeder2009}
{Maeder}, A. 2009, {Physics, Formation and Evolution of Rotating Stars}

\bibitem[{{McNally} {et~al.}(2012){McNally}, {Lyra}, \& {Passy}}]{mcnally2012}
{McNally}, C.~P., {Lyra}, W., \& {Passy}, J.-C. 2012, \apjs, 201, 18

\bibitem[{{Meakin} \& {Arnett}(2007)}]{meakin2007}
{Meakin}, C.~A. \& {Arnett}, D. 2007, \apj, 667, 448

\bibitem[{Miczek(2013)}]{miczek2013a}
Miczek, F. 2013, Dissertation, Technische Universit\"at M\"unchen

\bibitem[{{Miczek} {et~al.}(2015){Miczek}, {R{\"o}pke}, \&
  {Edelmann}}]{miczek2015}
{Miczek}, F., {R{\"o}pke}, F.~K., \& {Edelmann}, P.~V.~F. 2015, \aap, 576, A50

\bibitem[{Mignone {et~al.}(2011)Mignone, Zanni, Tzeferacos, van Straalen,
  Colella, \& Bodo}]{mignone2012}
Mignone, A., Zanni, C., Tzeferacos, P., {et~al.} 2011, The Astrophysical
  Journal Supplement Series, 198, 7

\bibitem[{Minoshima \& Miyoshi(2021)}]{minoshima2021}
Minoshima, T. \& Miyoshi, T. 2021, Journal of Computational Physics, 446,
  110639

\bibitem[{{Miyoshi} \& {Kusano}(2005)}]{miyoshi2005}
{Miyoshi}, T. \& {Kusano}, K. 2005, Journal of Computational Physics, 208, 315

\bibitem[{Motheau {et~al.}(2018)Motheau, Duarte, Almgren, \&
  Bell}]{motheau2018}
Motheau, E., Duarte, M., Almgren, A., \& Bell, J.~B. 2018, Journal of
  Computational Physics, 372, 1027

\bibitem[{{M{\"u}ller}(2020)}]{muller2020}
{M{\"u}ller}, B. 2020, Living Reviews in Computational Astrophysics, 6, 3

\bibitem[{Musoke {et~al.}(2020)Musoke, Young, \& Birkinshaw}]{musoke2020}
Musoke, G., Young, A.~J., \& Birkinshaw, M. 2020, Monthly Notices of the Royal
  Astronomical Society, 498, 3870

\bibitem[{{Muthsam} {et~al.}(2010){Muthsam}, {Kupka}, {L{\"o}w-Baselli},
  {Obertscheider}, {Langer}, \& {Lenz}}]{muthsam2010}
{Muthsam}, H.~J., {Kupka}, F., {L{\"o}w-Baselli}, B., {et~al.} 2010, \na, 15,
  460

\bibitem[{{Porter} \& {Woodward}(1994)}]{porter1994}
{Porter}, D.~H. \& {Woodward}, P.~R. 1994, \apjs, 93, 309

\bibitem[{{Radice} {et~al.}(2015){Radice}, {Couch}, \& {Ott}}]{radice2015}
{Radice}, D., {Couch}, S.~M., \& {Ott}, C.~D. 2015, Computational Astrophysics
  and Cosmology, 2, 7

\bibitem[{Rieper(2011)}]{rieper2011}
Rieper, F. 2011, Journal of Computational Physics, 230, 5263

\bibitem[{Rizzuti {et~al.}(2023)Rizzuti, Hirschi, Arnett, Georgy, Meakin,
  Murphy, Rauscher, \& Varma}]{rizzuti2023}
Rizzuti, F., Hirschi, R., Arnett, W.~D., {et~al.} 2023, Monthly Notices of the
  Royal Astronomical Society, 523, 2317

\bibitem[{{Robertson} {et~al.}(2010){Robertson}, {Kravtsov}, {Gnedin}, {Abel},
  \& {Rudd}}]{robertson2010a}
{Robertson}, B.~E., {Kravtsov}, A.~V., {Gnedin}, N.~Y., {Abel}, T., \& {Rudd},
  D.~H. 2010, \mnras, 401, 2463

\bibitem[{{Roe}(1981)}]{roe1981}
{Roe}, P.~L. 1981, Journal of Computational Physics, 43, 357

\bibitem[{Rusanov(1962)}]{rusanov1962}
Rusanov, V. 1962, USSR Computational Mathematics and Mathematical Physics, 1,
  304

\bibitem[{San \& Kara(2015)}]{san2015}
San, O. \& Kara, K. 2015, Computers \& Fluids, 117, 24

\bibitem[{Seo \& Ryu(2023)}]{seo2023}
Seo, J. \& Ryu, D. 2023, The Astrophysical Journal, 953, 39

\bibitem[{Shu(2009)}]{shu2009}
Shu, C.-W. 2009, SIAM Review, 51, 82

\bibitem[{Shu \& Osher(1988)}]{shu1988}
Shu, C.-W. \& Osher, S. 1988, Journal of Computational Physics, 77, 439

\bibitem[{{Springel}(2010)}]{springel2010}
{Springel}, V. 2010, \mnras, 401, 791

\bibitem[{Sutherland(2010)}]{sutherland2010}
Sutherland, B.~R. 2010, Internal Gravity Waves (Cambridge University Press)

\bibitem[{{Teissier} \& {M{\"u}ller}(2023)}]{teissier2023}
{Teissier}, J.-M. \& {M{\"u}ller}, W.-C. 2023, arXiv e-prints, arXiv:2306.09856

\bibitem[{Thornber {et~al.}(2008)Thornber, Mosedale, Drikakis, Youngs, \&
  Williams}]{thornber2008}
Thornber, B., Mosedale, A., Drikakis, D., Youngs, D., \& Williams, R. 2008,
  Journal of Computational Physics, 227, 4873

\bibitem[{{Toro}(1991)}]{toro1991}
{Toro}, E.~F. 1991, Proceedings of the Royal Society of London Series A, 434,
  683

\bibitem[{Toro(2009)}]{toro2009a}
Toro, E.~F. 2009, Riemann Solvers and Numerical Methods for Fluid Dynamics: A
  Practical Introduction (Berlin Heidelberg: Springer)

\bibitem[{{Toro} {et~al.}(1994){Toro}, {Spruce}, \& {Speares}}]{Toro1994}
{Toro}, E.~F., {Spruce}, M., \& {Speares}, W. 1994, Shock Waves, 4, 25

\bibitem[{{van Leer}(1974)}]{vanleer1974a}
{van Leer}, B. 1974, Journal of Computational Physics, 14, 361

\bibitem[{{van Leer}(1977)}]{vanleer1977a}
{van Leer}, B. 1977, Journal of Computational Physics, 23, 276

\bibitem[{{van Leer}(1979)}]{vanleer1979}
{van Leer}, B. 1979, Journal of Computational Physics, 32, 101

\bibitem[{{Viallet} {et~al.}(2011){Viallet}, {Baraffe}, \&
  {Walder}}]{viallet2011}
{Viallet}, M., {Baraffe}, I., \& {Walder}, R. 2011, \aap, 531, A86

\bibitem[{{Wang} {et~al.}(2024){Wang}, {Wang}, {Xu}, {Sanhueza}, {Liu},
  {Zhang}, {Lu}, {Fontani}, {Caselli}, {Busquet}, {Tan}, {Li}, {Jackson},
  {Pillai}, {Ho}, {Guzm{\'a}n}, \& {Yue}}]{wang2024}
{Wang}, C., {Wang}, K., {Xu}, F.-W., {et~al.} 2024, \aap, 681, A51

\bibitem[{{Wedemeyer} {et~al.}(2017){Wedemeyer}, {Ku{\v{c}}inskas}, {Klevas},
  \& {Ludwig}}]{wedemeyer2017}
{Wedemeyer}, S., {Ku{\v{c}}inskas}, A., {Klevas}, J., \& {Ludwig}, H.-G. 2017,
  \aap, 606, A26

\bibitem[{{Wongwathanarat} {et~al.}(2016){Wongwathanarat}, {Grimm-Strele}, \&
  {M{\"u}ller}}]{wongwathanarat2016}
{Wongwathanarat}, A., {Grimm-Strele}, H., \& {M{\"u}ller}, E. 2016, \aap, 595,
  A41

\bibitem[{Woodward {et~al.}(2014)Woodward, Herwig, \& Lin}]{woodward2015}
Woodward, P.~R., Herwig, F., \& Lin, P.-H. 2014, The Astrophysical Journal,
  798, 49

\bibitem[{Xie {et~al.}(2019)Xie, Zhang, Lai, \& Li}]{xie2019}
Xie, W., Zhang, R., Lai, J., \& Li, H. 2019, International Journal for
  Numerical Methods in Fluids, 89, 430

\end{thebibliography}
	
	\begin{appendix}
		
		\section{The PPH and PQH methods}
		\label{sec:PPH_PQH}
		
		The idea of applying limiters to passive scalars (or mass fractions) only, which lead us to formulating the PSH method (Sect.~\ref{sec:method_PSH}), can be applied at any order of accuracy. In this section, we provide two methods that can be seen as lower-order alternatives to PSH. Their potential advantages include a lower amplitude of overshoots in dynamic quantities and the need for fewer ghost cells.
		
		We start with the unlimited parabolic method PAR (Sect.~\ref{sec:method_PAR}) and apply a two-step limiter to passive scalars. The first step is defined by the variable replacements
		\begin{align}
			a_{i-1/2,\mathrm{L}/\mathrm{R}} \mapsto \mathrm{median}(a_{i-1},a_{i-1/2,\mathrm{L}/\mathrm{R}},a_i).
		\end{align}
		These replacements are followed by the application of the PPM84 limiter defined by Eq.~(\ref{eq:PPM84_limiter_2}). We call the resulting method piecewise parabolic hybrid (PPH). The PPH method is exact wherever $a(\zeta)$ is locally parabolic, \mbox{3\textsuperscript{rd}-order} accurate for general but smooth functions $a(\zeta)$, and it requires two ghost cells at domain boundaries. 
		
		A more accurate method, which we call piecewise quartic hybrid (PQH), can be obtained by assuming that within cell $i$ $a(\zeta)$ can be described by the quartic polynomial
		\begin{align}
			a(\zeta) = \sum_{n=0}^{4} c_n (\zeta - \zeta_i)^n.\label{eq:PQH_a}
		\end{align}
		The five coefficients $c_n$ are uniquely determined by the requirement that the averages of $a(\zeta)$ in cells $i-2+n$ equal $\overline{a}_{i-2+n}$ for $n = 0, 1, \dots, 4$. The reconstructed states are then obtained by evaluating Eq.~(\ref{eq:PQH_a}) at $\zeta_{i-1/2}$ and $\zeta_{i+1/2}$, respectively. The resulting expressions are
		\begin{align}
			\begin{split}\label{eq:PQH_aL}
				a_{i-1/2,\mrm{R}} = \frac{1}{60}\bigg({}& -3\overline{a}_{i-2} + 27\overline{a}_{i-1} + 47\overline{a}_i - 13\overline{a}_{i+1} + 2\overline{a}_{i+2}\bigg),
			\end{split}\\
			\begin{split}\label{eq:PQH_aR}
				a_{i+1/2,\mrm{L}} = \frac{1}{60}\bigg({}& 2\overline{a}_{i-2} - 13\overline{a}_{i-1} + 47\overline{a}_i + 27\overline{a}_{i+1} - 3\overline{a}_{i+2}\bigg).
			\end{split}
		\end{align}
		The limiter we apply to passive scalars in the PQH method is the same as that in the PPH method, see above. The PQH method is exact wherever $a(\zeta)$ is locally a quartic polynomial, 5\textsuperscript{th}-order accurate for general but smooth functions $a(\zeta)$, and it requires three ghost cells at domain boundaries.
		
		We have only performed a small number of tests with these two methods. Specifically, we ran simulations of the Kelvin-Helmholtz problem (Sect.~\ref{sec:results_kh}) with the initial Mach number of $\mach_0 = 10^{-3}$ and the HLLC flux function on grids of $64 \times 32$ and $128 \times 64$ cells. There were no overshoots in the passive tracer in the results of these tests.
		
		\section{1D test cases}
		\label{sec:1D_test_cases}
		
		We use two simple 1D experiments -- linear advection and the propagation of a linear sound wave -- to compare the accuracy that the six reconstruction schemes can reach if not constrained by the 2\textsuperscript{nd}-order accuracy limit imposed by our multidimensional scheme.
		
		\subsection{Linear advection}
		\label{sec:linear_advection}
		
		The initial conditions in this experiment correspond to a right-going contact wave:
		\begin{align}
			\rho(x) &= \gamma \left[ 1 + 0.01 \sin\left(2 \pi x\right)\right], \label{eq:linear_advection_rho} \\
			u(x) &= 0.1, \\
			p(x) &= 1.
		\end{align}
		We consider the interval $0 \le x \le 1$ with periodic boundary conditions and the equation of state of an ideal gas with the ratio of specific heats $\gamma = 1.4$. The average speed of sound is unity and it varies by $0.5\%$ due to the density variation. The solution is sought at $t = 10$, when the sinusoid has been advected by one period and the analytic solution becomes identical to the initial condition. This makes the quantification of numerical errors trivial.
		
		We use the same code for this experiment as we do for all the other experiments reported in this work. Since some of the schemes tested exceed 2\textsuperscript{nd} order of accuracy, cell averages cannot be approximated by sampling the initial condition at cell centres. We initialise the discrete density profile using the formula
		\begin{align}
			\rho_i &= \gamma \left[ 1 + 0.01 \frac{\sin\left(\pi \Delta x\right)}{\pi \Delta x}\sin\left(2 \pi x_i\right)\right],  \label{eq:linear-advection-rhoi}  
		\end{align}
		where $\Delta x$ is the constant grid spacing and $x_i$ the location of \mbox{$i$-th} cell's centre. This formula is easily obtained by analytically averaging Eq.~(\ref{eq:linear_advection_rho}) in the interval $x_i - \Delta x / 2 \le x \le x_i + \Delta x / 2$. We suppress time-stepping errors by using the RK3 scheme with $\mrm{CFL} = 0.01$. Because the Mach number of the flow is $0.1$, this means that the wave is advected by only $10^{-3} \Delta x$ during every 3\textsuperscript{rd}-order-accurate time step. We use the \mbox{LHLLC} flux function for this experiment.
		
\begin{figure}
			\includegraphics[width=\linewidth]{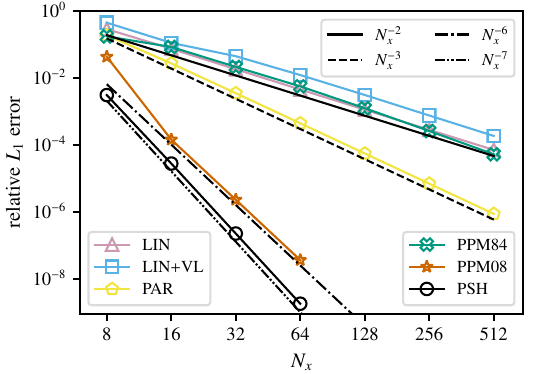}
			\caption{Relative $L_1$ errors obtained by advecting a sinusoid by one period on 1D grids with $N_x$ cells using different reconstruction schemes. A few scaling relations are shown to guide the eye.}
			\label{fig:linear-advection-L1}
		\end{figure}

		The $L^1$ errors we obtain on grids with different numbers of cells $N_x$ are shown in Fig.~\ref{fig:linear-advection-L1}. We test the LIN, LIN+VL, PAR, and PPM84 schemes on grids with 8 to 512 cells. We are forced to stop at grids of 64 cells for the most accurate schemes PPM08 and PSH, because the errors rapidly become dominated by the finite precision of floating-point arithmetic. The asymptotic orders of accuracy based on $L^1$ errors measured on the three finest grids available are reported in Table~\ref{tab:rec_methods}. Both LIN and LIN+VL reach the expected 2\textsuperscript{nd} order of accuracy. The absolute errors produced by LIN+VL are somewhat larger as compared with LIN because of the presence of the van Leer limiter. The PPM84 scheme is formally 4\textsuperscript{th}-order accurate but that only holds for monotonic solutions. PPM84 contains a limiter that flattens the slope at local extrema and that reduces the scheme's asymptotic order to $2.3$ in this test case. We have experimentally turned off both limiters present in PPM84, which increased the asymptotic order of accuracy to $4.0$, as expected. The PAR, PPM08, and PSH methods reach the asymptotic orders of $3.0$, $6.0$, and $7.0$, also matching theoretical expectations.
		
		\subsection{Sound wave}
		\label{sec:sound_wave}
		
		The initial conditions in this experiment correspond to a right-going sound wave:
		\begin{align}
			\rho(x) &= \rho_0 \left[ 1 + \gamma^{-1} \Psi(x)\right], \\
			u(x) &= \gamma^{-1} c_0 \Psi(x), \\
			p(x) &= p_0 \left[ 1 + \Psi(x)\right],
		\end{align}
		where
		\begin{align}
			\Psi(x) = \Psi_0 \sin\left(2 \pi x\right).
		\end{align}
		We use $\rho_0 = \gamma$, $p_0 = 1$ and the equation of state of an ideal gas with the ratio of specific heats $\gamma = 1.4$, so that the unperturbed speed of sound is $c_0 = 1$. We set the amplitude $\Psi_0 = 10^{-10}$ to suppress non-linear effects. This allows us to avoid accuracy constraints imposed by our 2\textsuperscript{nd}-order-accurate transformations between primitive and conserved variables. Round-off errors are suppressed by using 128-bit floating point numbers. In a way analogous to Eq.~(\ref{eq:linear-advection-rhoi}), we initialise the discrete wave such that the cell average of $\Psi(x)$ in cell $i$ is\footnote{We do not average the $\Psi^2$ term, which would appear in the product $\rho u$, because the term is only ${\approx}\,10^{-20}$ with $\Psi_0 = 10^{-10}$. This term must be kept small or other non-linear effects would also cease to be negligible in the range of relative $L_1$ errors we explore.}
		\begin{align}
			\Psi_i = \frac{\sin\left(\pi \Delta x\right)}{\pi \Delta x} \Psi(x_i).
		\end{align}
		
		The simulations are stopped at $t = 1$, when the sound wave has propagated by a single wavelength around the periodic domain $0 \le x \le 1$. Neglecting non-linear effects, which are of order $\Psi_0^2 = 10^{-20}$, the evolved solution is expected to be identical to the initial condition. Timestepping errors are suppressed by using the RK3 scheme with $\mrm{CFL} = 0.001$, i.e. the wave moves by only $10^{-3} \Delta x$ per time step. We use the LHLLC flux function for this experiment.
		
\begin{figure}
			\includegraphics[width=\linewidth]{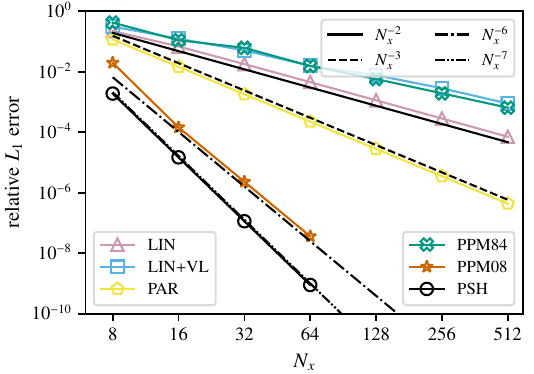}
			\caption{Relative $L_1$ errors in velocity $u$ obtained by propagating a sinusoidal sound wave by one period on 1D grids with $N_x$ cells using different reconstruction schemes. A few scaling relations are shown to guide the eye.}
			\label{fig:sound-wave-L1}
		\end{figure}

		We show $L^1$ errors in the velocity $u$ on grids with $8$ to $512$ cells in Fig.~\ref{fig:sound-wave-L1}. The simulation series with the most accurate PPM08 and PSH schemes are stopped at grids of $64$ cells, because the absolute errors rapidly approach $10^{-20}$, i.e. the magnitude of the residual non-linear effects. The asymptotic orders of accuracy, as defined by the $L^1$ errors measured on the three finest grids available, are $2.0$, $3.0$, $6.0$, and $7.0$ for the LIN, PAR, PPM08, and PSH methods, respectively, matching theoretical expectations. The LIN+VL and PPM84 methods only reach the order of $1.6$, which seems to be caused by the combination of the low-Mach flux function LHLLC with the limiters contained in these two methods. We have checked that both methods reach the order of $2.0$ with the RUSANOV and HLLC flux functions (PPM84 flattens the slope at local extrema, which reduces its order, see also Sect.~\ref{sec:linear_advection}).
		
\begin{figure}
			\includegraphics[width=\linewidth]{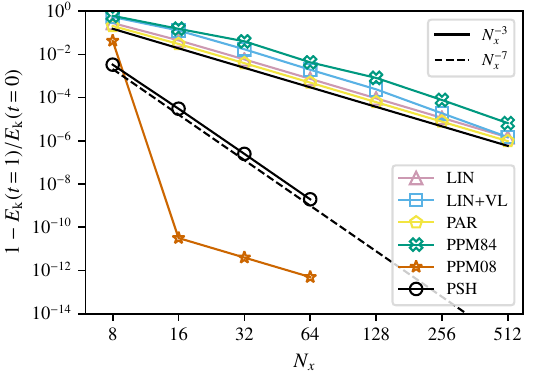}
			\caption{Relative decrease in kinetic energy after propagating a sinusoidal sound wave by one period on 1D grids with $N_x$ cells using different reconstruction schemes. Two scaling relations are shown to guide the eye.}
			\label{fig:sound-wave-damping}
		\end{figure}

		The $L^1$ errors discussed so far do not distinguish between amplitude and phase errors. We quantify amplitude errors by computing the relative loss of the total kinetic energy from $t = 0$ to $t = 1$, see Fig.~\ref{fig:sound-wave-damping}. Some kinetic energy is lost in all of the simulations, which is a sign of stability. The amount of energy dissipated decreases with the 3\textsuperscript{rd} power of the grid spacing for LIN, LIN+VL, PAR, and PPM84. It is not immediately clear why this is the case given that the methods have different orders of accuracy, but we did not investigate this further. In case of PSH, the dissipation rate decreases with the 7\textsuperscript{th} power of the grid spacing, matching the method's order of accuracy. PPM08 is a special case -- the method preserves extrema and sinusoids resolved by $16$ or more cells per wavelength turn out to be smooth enough not to trigger any of the method's limiters. PPM08 reduces to a simple interpolation function in this special case, eliminating jumps at all cell interfaces, see Sect.~\ref{sec:method_PPM08}. This, in turn, eliminates all explicit dissipative terms in the flux function. The dissipation rate drops by many orders of magnitude and time-stepping errors start to dominate.
		
		\section{Sound wave generation with a compact third order WENO scheme}\label{sec:sound-weno}
		
		The results presented in Sects.~\ref{sec:results_kh} and \ref{sec:code-comparison} reveal that nonlinear, switching reconstruction schemes (such as LIN+VL, PPM84, and PPM08) generate excess acoustic power which is not present in simulations run with unlimited (linear) reconstruction methods. The question naturally arises whether nonlinear but nonswitching schemes (e.g., WENO) also generate artificial sound waves in simulations of low-Mach-number flows. In this Appendix we try to answer this question using the compact third order (CTO) WENO scheme of \cite{kolb2014}. This scheme reconstructs the quantity $a$ within the cell $i$ using a weighted sum of three polynomials, $a_\mathrm{L,R,C}(\zeta)$. The ultimate expression for the underlying function $a(\zeta)$  reads
		\begin{equation}\label{eq:a_zeta}
			a(\zeta) = w_\mathrm{L} a_\mathrm{L}(\zeta) + w_\mathrm{C} a_\mathrm{C}(\zeta) + w_\mathrm{R} a_\mathrm{R}(\zeta).
		\end{equation}
		The weighting coefficients are computed as
		\begin{equation}
			w_m = \frac{\alpha_m}{\sum_{k\in\{\mathrm{L,C,R}\}}\alpha_k},
		\end{equation}
		where 
		\begin{equation}\label{eq:alphai}
			\alpha_m = \frac{c_m}{(\epsilon_\mathrm{WENO}+\mathrm{IS}_m)^p},
		\end{equation}
		$m \in \{\mathrm{L,C,R}\}$, $c_\mathrm{L}=c_\mathrm{R}=0.25$, $c_\mathrm{C}=0.5$, and $\mathrm{IS}_m$ are the smoothness indicators of the scheme\footnote{For a formal derivation of these indicators, see Sect. 2.1 of \cite{kolb2014}.},
		\begin{align}
			\mathrm{IS}_\mathrm{L} = & \ (\overline{a}_i-\overline{a}_{i-1})^2, \\
			\mathrm{IS}_\mathrm{R} = & \ (\overline{a}_{i+1}-\overline{a}_{i})^2, \\
			\mathrm{IS}_\mathrm{C} = & \ \frac{13}{12c_\mathrm{C}^2}(\overline{a}_{i+1}-2\overline{a}_{i}+\overline{a}_{i-1})^2+\frac{1}{4}(\overline{a}_{i+1}-\overline{a}_{i-1})^2.
		\end{align}
		In Eq.~(\ref{eq:alphai}), we set $p=2$. The functions $a(\zeta)_\mathrm{L,R}$ are one-sided linear reconstructions
		\begin{align}
			a_\mathrm{L}(\zeta) = & \ \overline{a}_i + (\overline{a}_{i}-\overline{a}_{i-1})(\zeta-\zeta_i), \\
			a_\mathrm{R}(\zeta) = & \ \overline{a}_i + (\overline{a}_{i+1}-\overline{a}_{i})(\zeta-\zeta_i),
		\end{align}
		whereas $a_\mathrm{C}(\zeta)$ is defined such that
		\begin{equation}
			a_\mathrm{opt}(\zeta) = c_\mathrm{L}a_\mathrm{L}(\zeta) + c_\mathrm{C}a_\mathrm{C}(\zeta)+c_\mathrm{R}a_\mathrm{R}(\zeta)
		\end{equation}
		is the unique parabola that conserves the cell volume averages $\overline{a}_{i-1}$, $\overline{a}_{i}$, and $\overline{a}_{i+1}$ over cells $i-1$, $i$, and $i+1$, respectively. This constraint implies
		\begin{equation}
			a_\mathrm{opt}(\zeta) = a_i + \left.\frac{\partial a}{\partial \zeta}\right\vert_{\zeta_i}(\zeta-\zeta_i) + \left.\frac{1}{2}\frac{\partial^2 a}{\partial \zeta^2}\right\vert_{\zeta_i}(\zeta-\zeta_i)^2,
		\end{equation}
		with 
		\begin{alignat}{2}
			&\ \ a_i   && =  \ \overline{a}_i - \frac{1}{24}(\overline{a}_{i+1} - 2 \overline{a}_i + \overline{a}_{i-1}), \\
			&\left.\frac{\partial a}{\partial \zeta}\right\vert_{\zeta_i} &&  =   \ \frac{\overline{a}_{i+1}-\overline{a}_{i-1}}{2}, \\
			&\left.\frac{\partial^2 a}{\partial \zeta^2}\right\vert_{\zeta_i}  && =   \ \overline{a}_{i+1}-2\overline{a}_{i}+\overline{a}_{i-1}.
		\end{alignat}
		The smoothness indicators $\mathrm{IS}_\mathrm{L,R,C}$ are such that the smoothest among the polynomials $a(\zeta)_\mathrm{L,R,C}$ has the largest weight in Eq.~(\ref{eq:a_zeta}). This feature allows CTO-WENO to achieve third-order spatial accuracy in smooth parts of the flow while at the same time it remains robust near discontinuities. The parameter $\epsilon_\mathrm{WENO}$ that appears in the denominator of $\alpha_m$ avoids division by zero in the case a smoothness indicator $\mathrm{IS}_m=0$. Thus, the oscillatory behavior of $a(\zeta)$ is also determined by the value of this parameter. In fact, when $\epsilon_\mathrm{WENO} \gg \mathrm{IS}_\mathrm{L,R,C}$, the weighting coefficients become almost equal and the scheme is unlimited even near discontinuities, making it similar to our PAR method. On the other hand, if $\epsilon_\mathrm{WENO} \ll \mathrm{IS}_m$, the weighting coefficients are only determined by $\mathrm{IS}_\mathrm{L,R,C}$ and the scheme becomes close to being TVD near discontinuities or in poorly resolved parts of the flow.
		
		Although CTO-WENO, unlike LIN+VL or the PPM methods, does not involve any conditional statements, the weighting coefficients that occur in $a(\zeta)$ can still abruptly change from time step to time step if a barely resolved wave or feature in the flow crosses that particular cell. Such a rapid change in the form of the polynomial $a(\zeta)$ can generate high-frequency perturbations in the state quantities on the grid scale and affect the propagation of sound waves. Therefore, we expect a version of CTO-WENO that uses small values of $\epsilon_\mathrm{WENO}$ to generate artificial acoustic noise, whereas large values of $\epsilon_\mathrm{WENO}$ should generate results closer to our unlimited reconstruction methods.
		
\begin{figure*}\label{fig:freq-weno}
			\centering
			\includegraphics[width=0.62\textwidth]{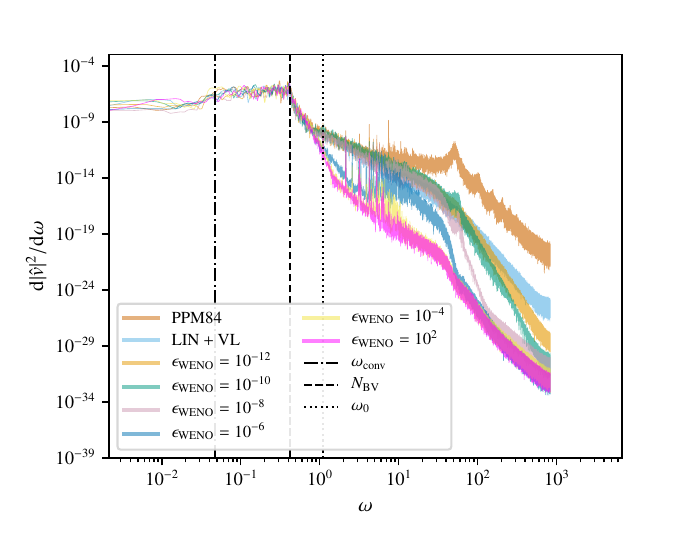}
			\caption{Frequency power spectrum of the vertical velocity component $v$ extracted in the middle of the stable layer at $(x,y,z)=(-0.2,2.5,-0.2)$ over the time series $t \in (10\tau_\mathrm{conv},32\tau_\mathrm{conv})$ in the problem of turbulent convection and wave excitation described in Sect.~\ref{sec:code-comparison}. These results are obtained on a grid with $128^3$ cells, using the $\mathrm{LHLLC}$ Riemann solver and the CTO-WENO scheme of \cite{kolb2014} for different values of the parameter $\epsilon_\mathrm{WENO}$ occurring in the smoothness indicators of the scheme (see Sect.~\ref{sec:sound-weno}), ranging from $10^{-12}$ to $10^2$. As a reference, power spectra obtained with the numerical options $\mathrm{LHLLC+PPM84}$ and $\mathrm{LHLLC+LIN+VL}$ are also shown. The convective turnover frequency ($\omega_\mathrm{conv}=2\pi/\tau_\mathrm{conv}$), the Brunt-V\"ais\"al\"a frequency at the location of the point probe ($N_\mathrm{BV}$), and the frequency of the fundamental oscillation mode of the cavity ($\omega_0=1.1$) are represented by the black dashed-dotted, dashed, and dotted lines, respectively.}
		\end{figure*}

		To test this hypothesis, we run a series of simulations of the setup involving turbulent convective flows and excitation of internal waves described in Sect.~\ref{sec:code-comparison}. We fix the grid resolution to $128^3$ cells and the Riemann solver is \mbox{LHLLC}. We run one simulation for each value of $\epsilon_\mathrm{WENO} \in (10^{-12},\ 10^{-10},\ 10^{-8},\ 10^{-6},\ 10^{-4},\ 10^{2})$, so that several intermediate cases between the two extreme behaviors of the scheme (close to being TVD and fully oscillatory) are considered. We extract the frequency spectrum of the vertical velocity component from the middle of the stable layer as done for the analysis in Sect.~\ref{sec:frequency-spectrum}. The results are shown in Fig.~\ref{fig:freq-weno}. As expected, the power stored in sound waves is considerably increased when very small values of $\epsilon_\mathrm{WENO}$ are used. For $\epsilon_\mathrm{WENO}=10^{-12}$, the power spectrum obtained using CTO-WENO resembles that produced by LIN+VL. Decreasing the value of $\epsilon_\mathrm{WENO}$ progressively reduces (in a monotonic way) the energy of the sound waves and eventually the continuum of the power spectrum converges for  $\epsilon_\mathrm{WENO}\gtrsim10^{-4}$. In the study performed by \cite{kolb2014}, the value of $\epsilon_\mathrm{WENO}$ that achieves the optimal order of accuracy lies within the range\footnote{Here we assume that the reconstructed variable $a$ is dimensionless.} $\Delta x^3 \lesssim \epsilon_\mathrm{WENO} \lesssim \Delta x^2$, which, for this setup, corresponds to $10^{-6} \lesssim \epsilon_\mathrm{WENO} \lesssim 10^{-4}$. However, we note that, in this test, the values of the primitive variables reconstructed at the grid cell interfaces span several orders of magnitude. Therefore, using a unique value of $\epsilon_\mathrm{WENO}$ could potentially result in different  oscillatory properties of the scheme depending on which variable is being reconstructed. One way to avoid this problem is to rescale the reconstructed quantity $a$ such that its mean value across the stencil is close to unity. However, we decide not to investigate this effect on the generation of artificial sound waves further. 
		
		\section{Additional plots for the Kelvin--Helmholtz problem}
		
		\begin{figure*}
			\includegraphics[width=\linewidth]{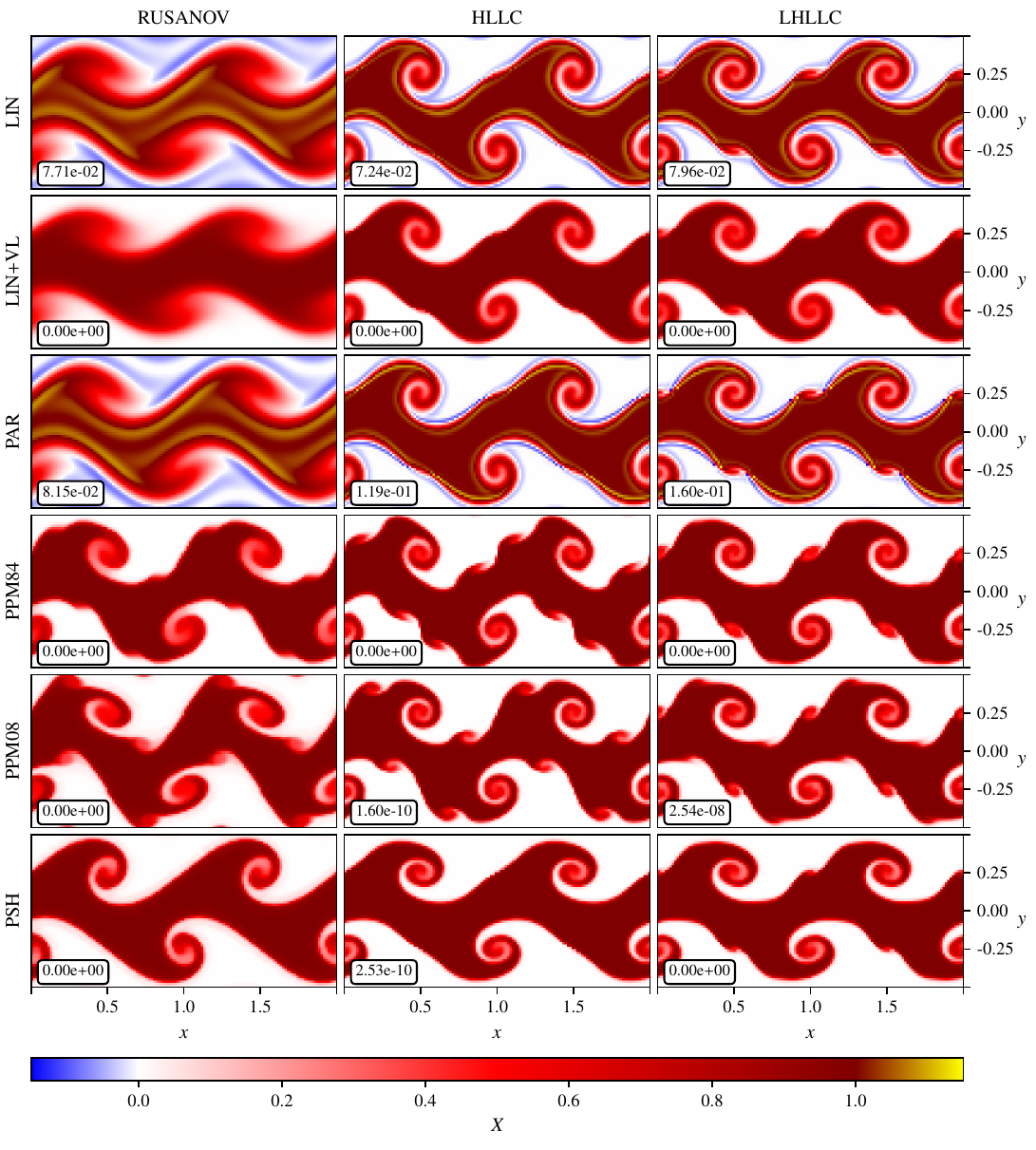}
			\caption{As Fig.~\ref{fig:kh2d-machx-1.000e-02-128x64-ps} but with the initial Mach number $\mach_0 = 10^{-1}$.}
			\label{fig:kh2d-machx-1.000e-01-128x64-ps}
		\end{figure*}
		
		\begin{figure*}
			\includegraphics[width=\linewidth]{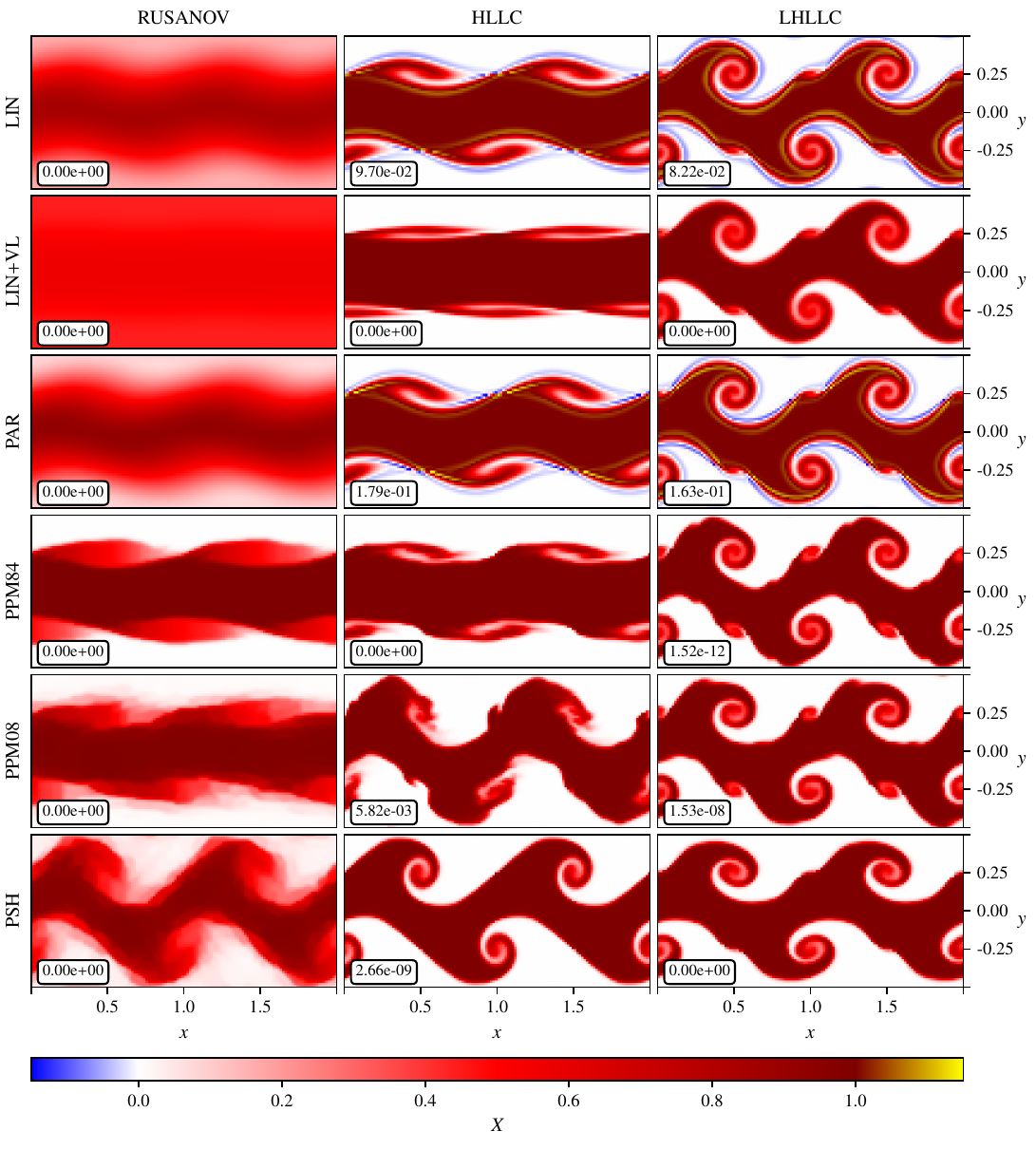}
			\caption{As Fig.~\ref{fig:kh2d-machx-1.000e-02-128x64-ps} but with the initial Mach number $\mach_0 = 10^{-3}$.}
			\label{fig:kh2d-machx-1.000e-03-128x64-ps}
		\end{figure*}
		
		\begin{figure*}
			\includegraphics[width=\linewidth]{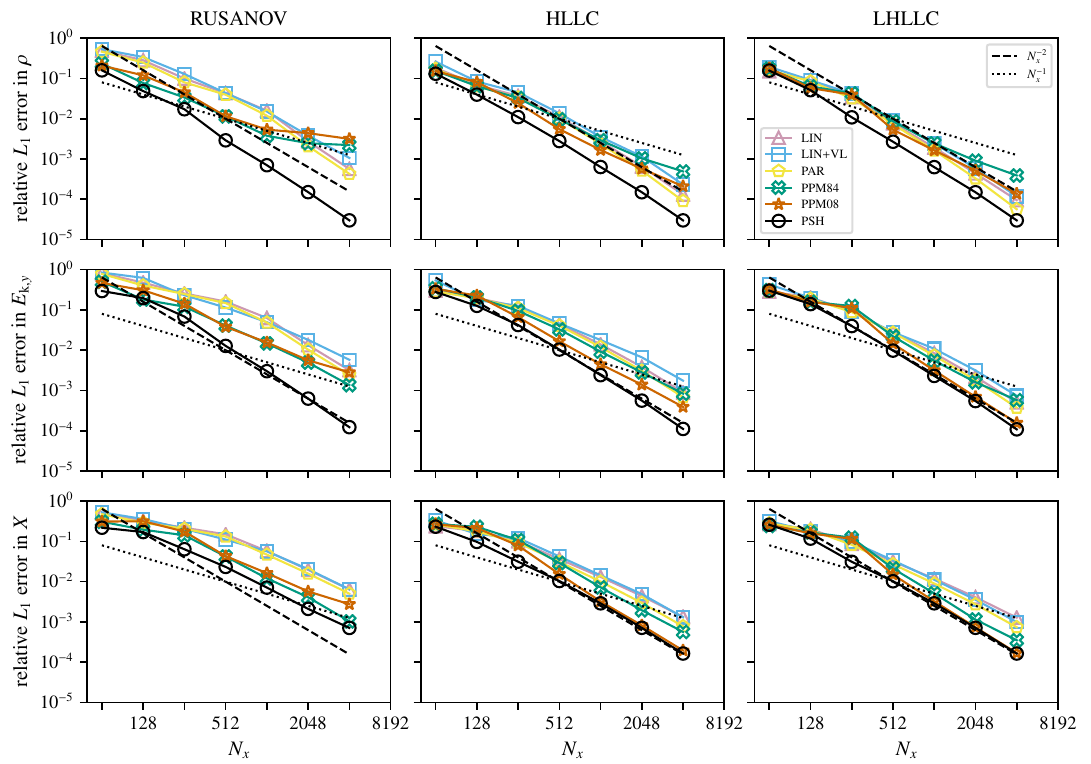}
			\caption{As Fig.~\ref{fig:kh2d-machx-1.000e-02-L1} but with the initial Mach number $\mach_0 = 10^{-1}$.}
			\label{fig:kh2d-machx-1.000e-01-L1}
		\end{figure*}
		
		\begin{figure*}
			\includegraphics[width=\linewidth]{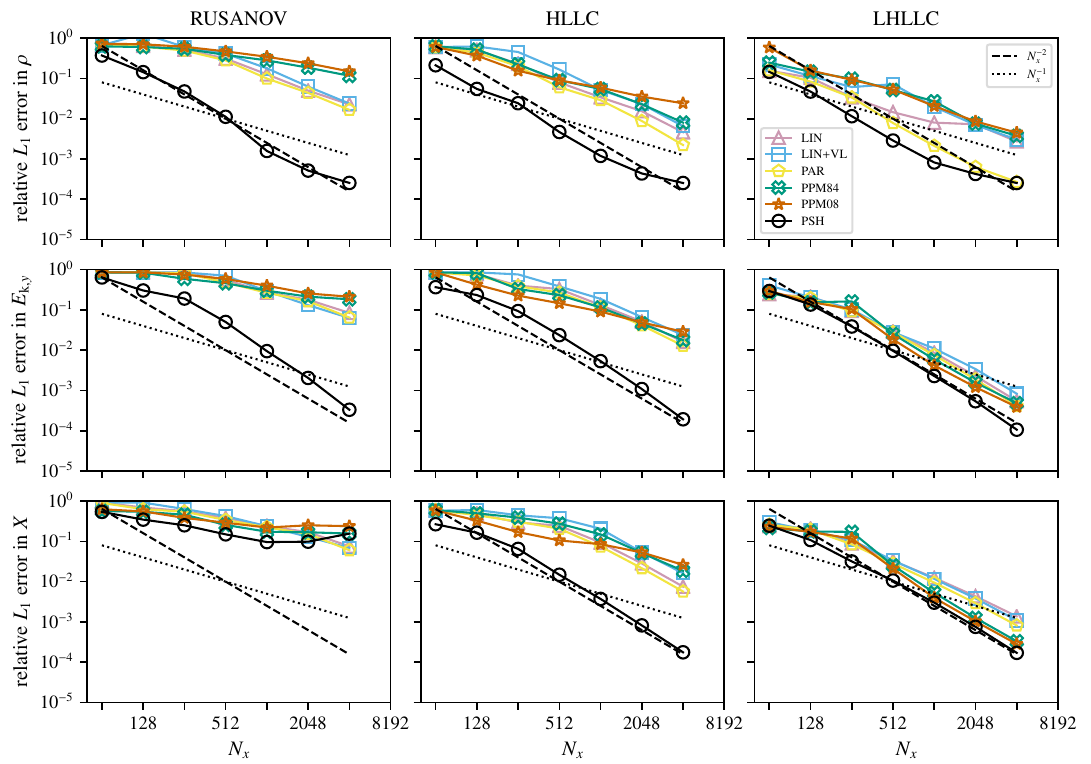}
			\caption{As Fig.~\ref{fig:kh2d-machx-1.000e-02-L1} but with the initial Mach number $\mach_0 = 10^{-3}$.}
			\label{fig:kh2d-machx-1.000e-03-L1}
		\end{figure*}
		
		\begin{figure*}
			\includegraphics[width=\linewidth]{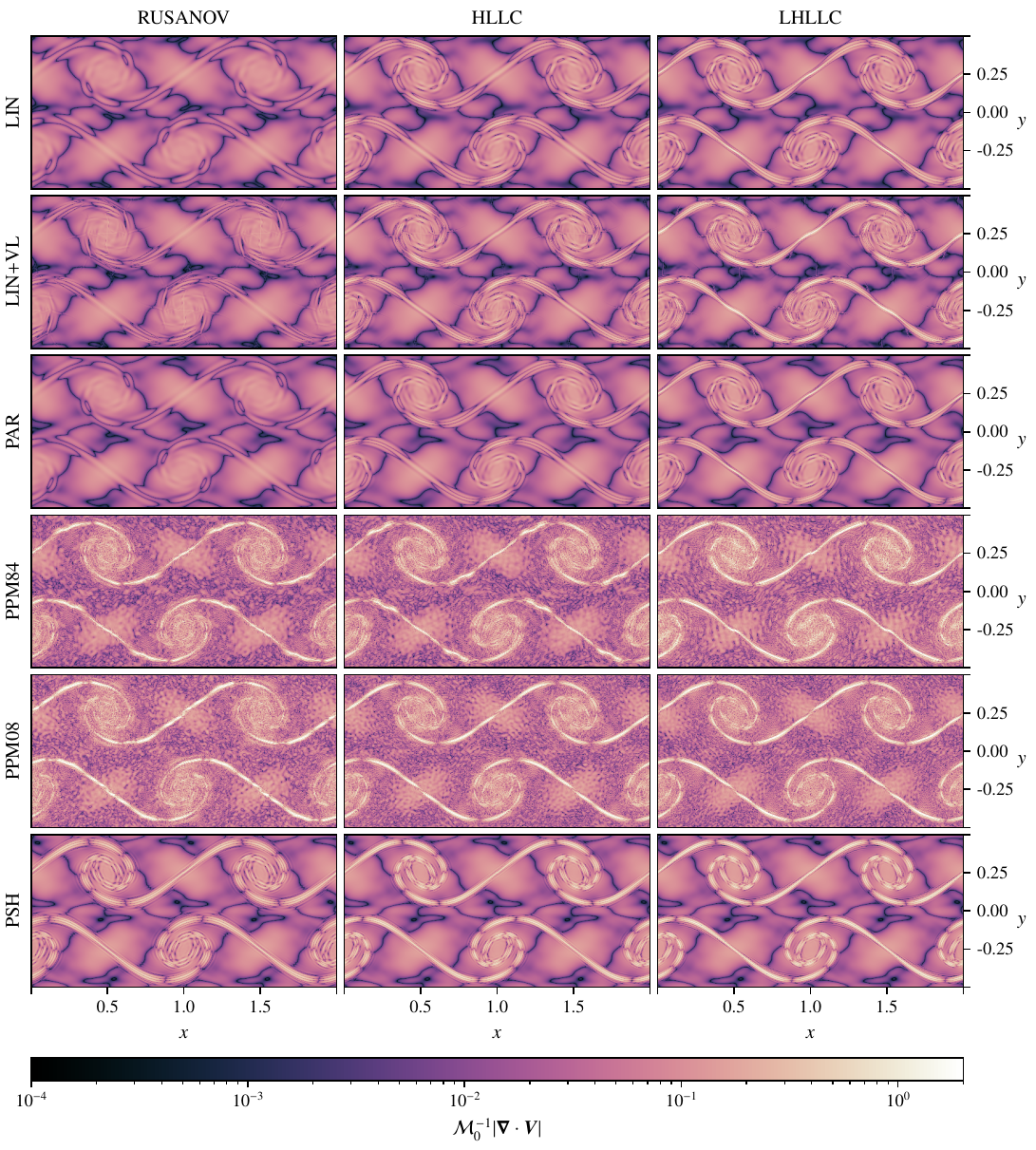}
			\caption{As Fig.~\ref{fig:kh2d-machx-1.000e-02-512x256-abs_div_u} but with the initial Mach number $\mach_0 = 10^{-1}$.}
			\label{fig:kh2d-machx-1.000e-01-512x256-abs_div_u}
		\end{figure*}
		
		\begin{figure*}
			\includegraphics[width=\linewidth]{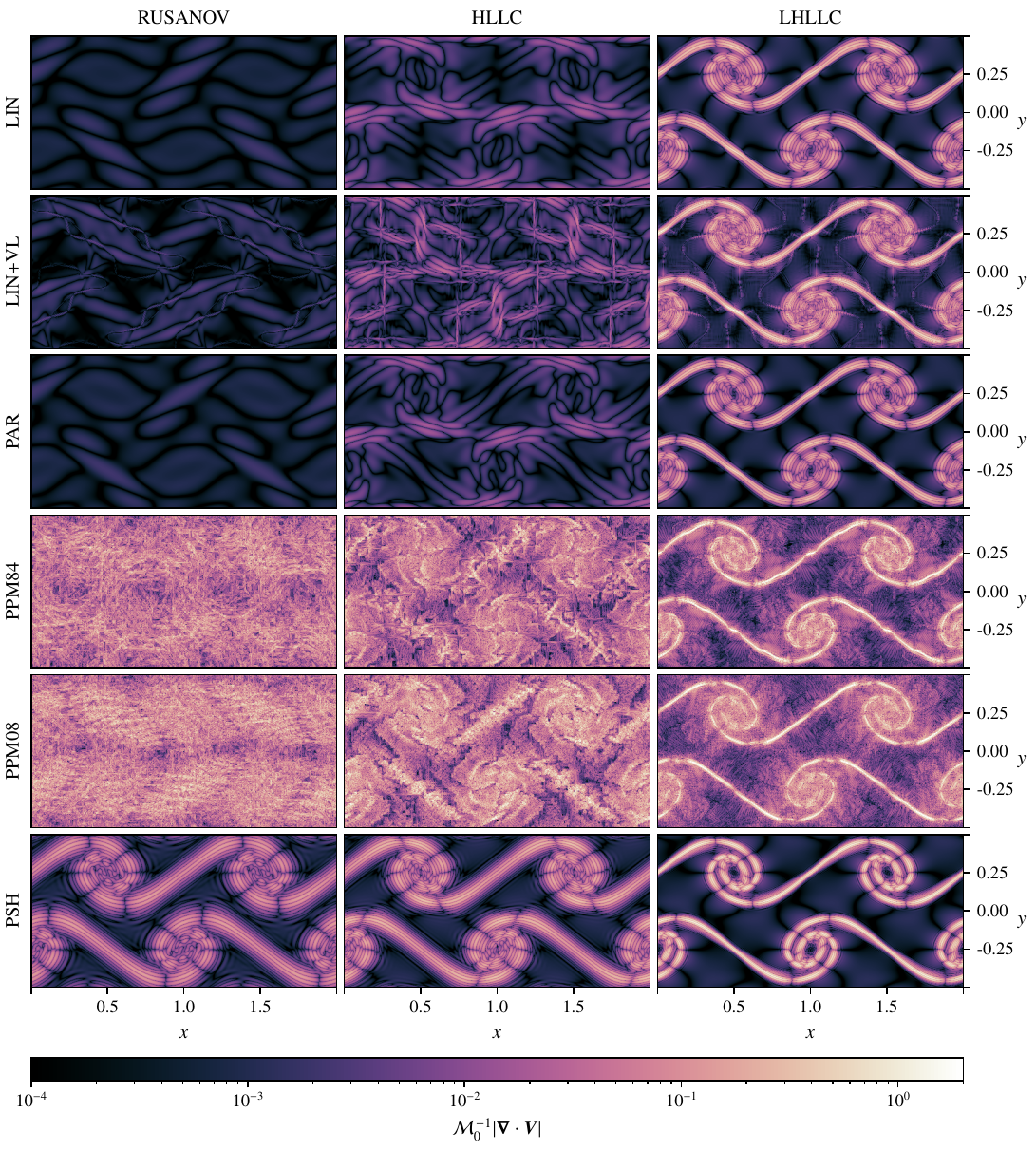}
			\caption{As Fig.~\ref{fig:kh2d-machx-1.000e-02-512x256-abs_div_u} but with the initial Mach number $\mach_0 = 10^{-3}$.}
			\label{fig:kh2d-machx-1.000e-03-512x256-abs_div_u}
		\end{figure*}
		
		\begin{figure*}
			\includegraphics[width=\linewidth]{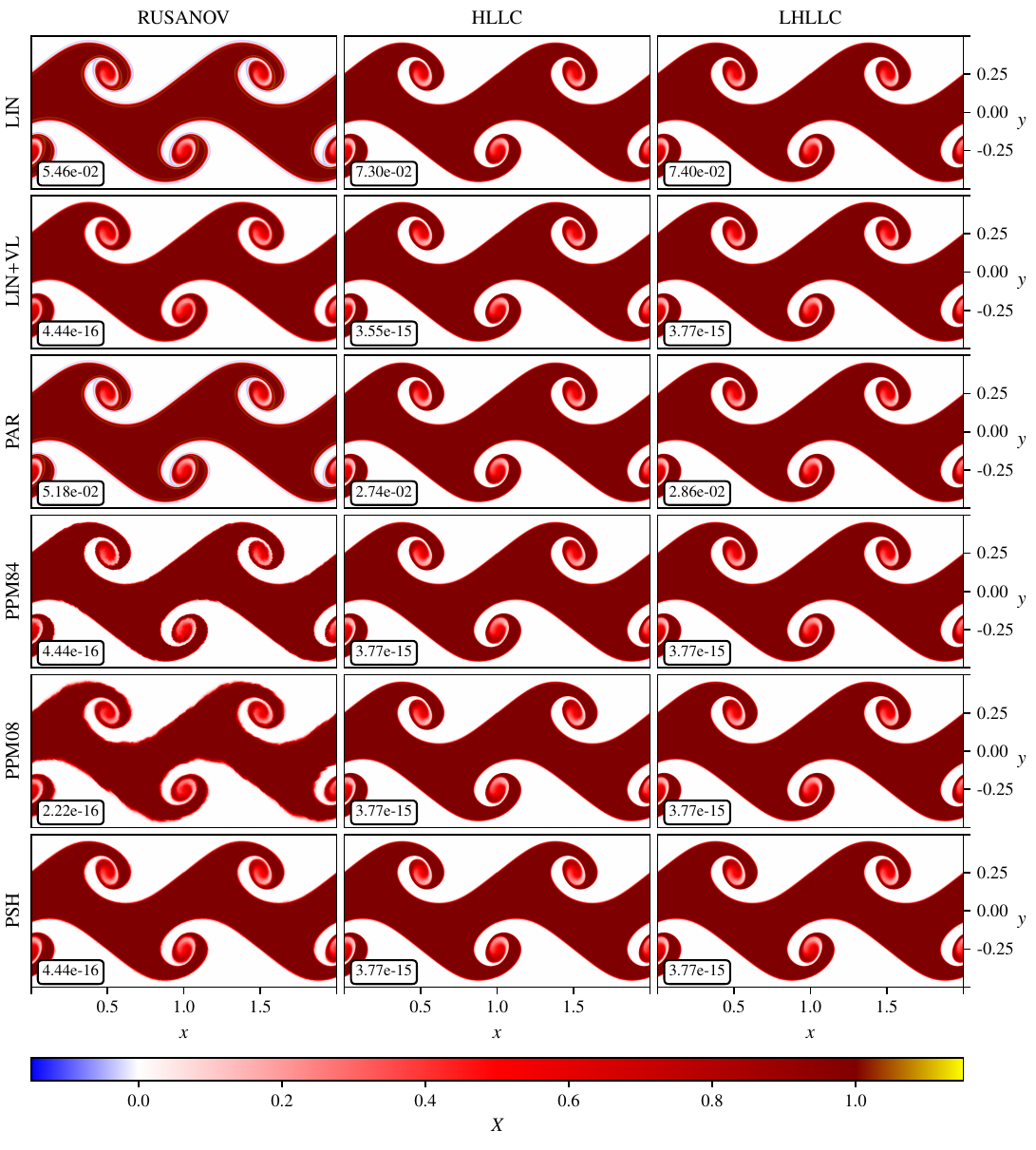}
			\caption{As Fig.~\ref{fig:kh2d-machx-1.000e-02-128x64-ps} but showing simulations computed on the $4096 \times 2048$ grid.}
			\label{fig:kh2d-machx-1.000e-02-4096x2048-ps}
		\end{figure*}
		
		\begin{figure*}
			\includegraphics[width=\linewidth]{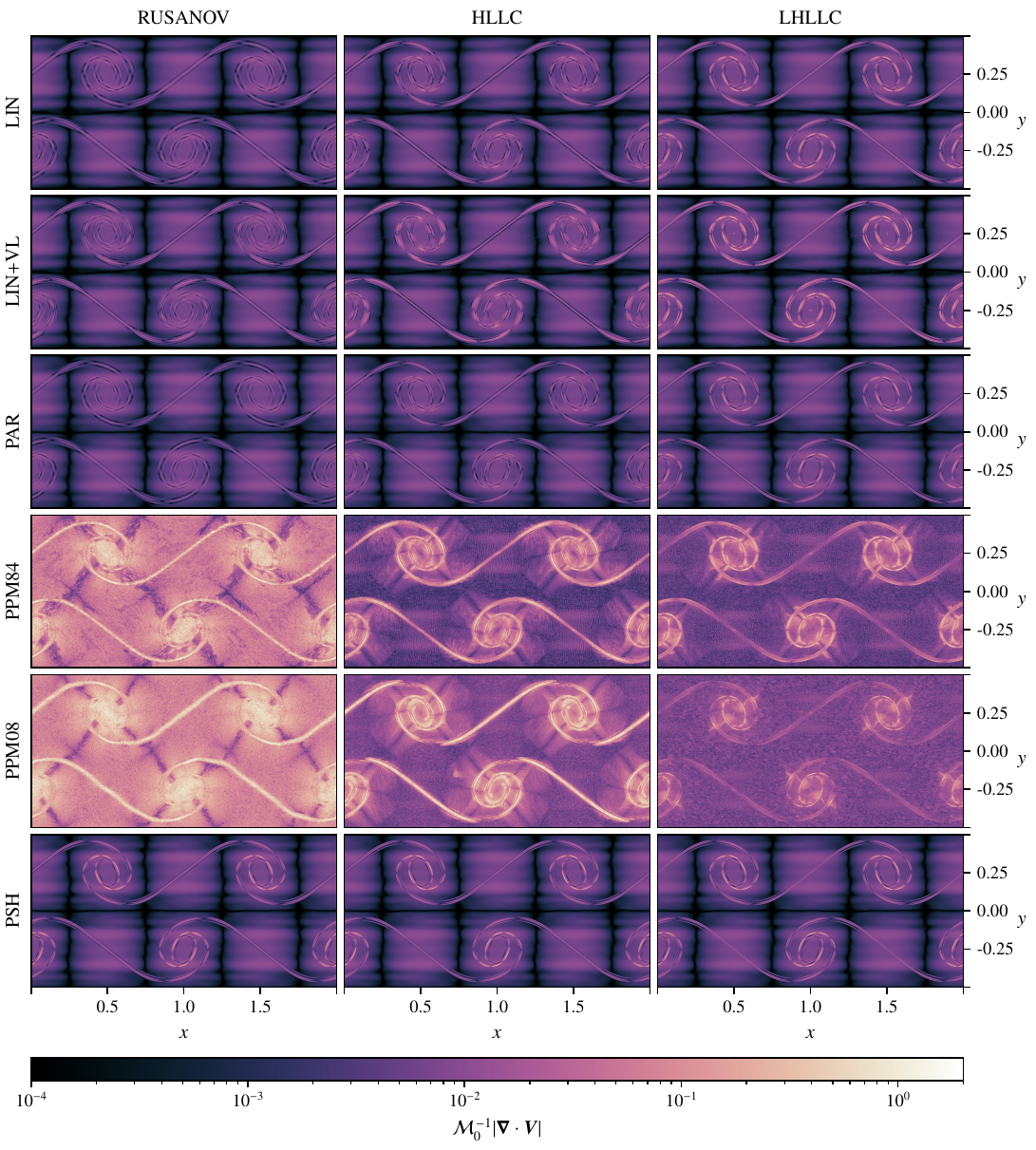}
			\caption{As Fig.~\ref{fig:kh2d-machx-1.000e-02-512x256-abs_div_u} but showing simulations computed on the $4096 \times 2048$ grid.}
			\label{fig:kh2d-machx-1.000e-02-4096x2048-abs_div_u}
		\end{figure*}
		
	\end{appendix}
\end{document}